\documentclass[11pt]{JHEP3}

\JHEPspecialurl{http://jhep.sissa.it/JOURNAL/JHEP3.tar.gz}

\usepackage{epsfig,multicol}
\usepackage{amsmath,amssymb}
\usepackage{mathrsfs}

\newcommand\fverb{\setbox\fverbbox=\hbox\bgroup\verb}
\newcommand\fverbdo{\egroup\medskip\noindent%
			\fbox{\unhbox\fverbbox}\ }
\newcommand\fverbit{\egroup\item[\fbox{\unhbox\fverbbox}]}
\newbox\fverbbox

\newcommand{\pslash}{p\kern-1ex /}
\newcommand{\qslash}{q\kern-1ex /}
\newcommand{\lslash}{l\kern-1ex /}
\newcommand{\sslash}{s\kern-1ex /}
\newcommand{\kaslash}{k_a\kern-2ex /}
\newcommand{\kbslash}{k_b\kern-2ex /}
\newcommand{\Dslash}{\mathcal{D}\kern-1.5ex /}

\newcommand{\beqa}{\begin{eqnarray}}
\newcommand{\eeqa}{\end{eqnarray}}

\newcommand{\ba}{\begin{eqnarray}}
\newcommand{\ea}{\end{eqnarray}}
\newcommand{\be}{\begin{equation}}

\newcommand{\alg}[1]{\mathfrak{#1}}

\title{Hybrid-NLIE for the AdS/CFT spectral problem}

\author{J\'anos Balog and \'Arp\'ad Heged\H us\\
Institute for Particle and Nuclear Physics, 
Wigner Research Centre for Physics,\\ 
H-1525 Budapest 114, P.O.B. 49, Hungary\\}

\received{\today} 		
\accepted{\today}	

\abstract{
Hybrid-NLIE equations, an alternative finite NLIE description for the
spectral problem of the super sigma model of AdS/CFT and 
its $\gamma$-deformations are derived by replacing the semi-infinite
SU(2) and SU(4) parts of the AdS/CFT TBA equations by a few appropriately
chosen complex NLIE variables, which are coupled among themselves and to the 
Y-functions associated to the remaining central nodes of the TBA diagram. 
The integral equations are written explicitly for the ground state of
the $\gamma$-deformed system. We linearize these NLIE equations, 
analytically calculate the first correction to the
asymptotic solution and find agreement with analogous results coming
from the original TBA formalism. 
Our equations differ substantially from the recently published 
finite FiNLIE formulation of the spectral problem.
}

\begin{document}


\newcommand{\con}{\,\star\hspace{-3.7mm}\bigcirc\,}
\newcommand{\convu}{\,\star\hspace{-3.1mm}\bigcirc\,}
\newcommand{\Eps}{\Epsilon}
\newcommand{\gM}{\mathcal{M}}
\newcommand{\dD}{\mathcal{D}}
\newcommand{\gG}{\mathcal{G}}
\newcommand{\pa}{\partial}
\newcommand{\eps}{\epsilon}
\newcommand{\La}{\Lambda}
\newcommand{\De}{\Delta}
\newcommand{\nonu}{\nonumber}
\newcommand{\beq}{\begin{eqnarray}}
\newcommand{\eeq}{\end{eqnarray}}
\newcommand{\ka}{\kappa}
\newcommand{\ee}{\end{equation}}
\newcommand{\an}{\ensuremath{\alpha_0}}
\newcommand{\bn}{\ensuremath{\beta_0}}
\newcommand{\dn}{\ensuremath{\delta_0}}
\newcommand{\al}{\alpha}
\newcommand{\bm}{\begin{multline}}
\newcommand{\fm}{\end{multline}}
\newcommand{\de}{\delta}


\section{Introduction}

One of the most important problems in testing the AdS/CFT
correspondence \cite{adscft} in the planar limit is to determine the
finite size spectrum of the $AdS_5 \times S^5$ superstring sigma
model \cite{AFrev}. After integrability was discovered in the string
worldsheet theory, the mirror Thermodynamic Bethe Ansatz (TBA)
technique was proposed \cite{AJK,AF07} to determine
nonperturbatively the spectrum of the string theory. The TBA
equations for AdS/CFT were first derived for the ground state
\cite{AF09a,Bombardelli:2009ns,AF09b,GKKV09,AF09d} and then using an
analytic continuation trick~\cite{DT} TBA equations were conjectured
for magnon excited states
\cite{GKKV09,GKV09b,Arutyunov:2009ax,BH-BJ} and bound states
\cite{AFT11}
 in the $\alg{sl}(2)$ and $\alg{su}(2)$ sectors of the theory.
The Y-system (and T-system) associated to the AdS/CFT problem were
proposed first in \cite{GKV09} and were later derived from the
mirror TBA equations.

Analogous progress has been made for the deformed, but still
integrable cousins of the superstring sigma model
\cite{Z27,Z28,Z29,Z30,Z32,FrolovTsT}. An important integrability
preserving deformation is the so-called $\gamma$-deformed theory
\cite{Z29} which has three deformation parameters. The implementation of
the deformations in the integrable context
\cite{Z34,Z36,Z33,Z35,Z37} led to the TBA formulation of the finite
size problem in the deformed theory \cite{TBA2ndLusch}. An
important property of the $\gamma$-deformed theory is that since
supersymmetry is not preserved, unlike in the undeformed case, the ground 
state energy is non-vanishing. Recently \cite{orbi} the most general integrable
deformations (obtained by orbifolding and TsT-transforming the original
string sigma model) were found and the corresponding TBA system was 
proposed.

In the undeformed theory the correctness of the (conjectured) TBA
equations has been nicely demonstrated by the convincing agreement
with gauge theory results \cite{Sieg,Vel,KL02,40_5} in the weak
coupling limit \cite{GKV09,AFS,BHxxx,BH-BJ} through the generalized
L\"uscher approach
\cite{JL07,BJ08,Bajnok:2008qj,BJ09,Lukowski:2009ce}, and with string
theory results in the strong coupling limit
\cite{GKV09b,Gromov,ujKazi,Frolnum,StrongKonishi}.

 In the $\gamma$-deformed theory the TBA and Y-system has been checked
\cite{Z36,TBA2ndLusch} in the large volume and small coupling
regimes against 1st \cite{BJ08} and 2nd \cite{TBA2ndLusch} order
L\"uscher formulae and direct field theory computations
\cite{betapert}.

In spite of the success of the TBA technique in AdS/CFT, it has some
obvious disadvantages as well. First of all, like all the TBA
equations of known sigma-models it contains infinitely many unknown
functions, which makes the study of their properties, both
analytically and numerically, difficult.

A possible way to give a simpler and finite formulation of the
spectral problem of an integrable model is the so-called NLIE
formulation, where only a few unknown functions appear in the
resulting set of nonlinear integral equations.

The NLIE approach for finite size physics of integrable models
originates from the paper \cite{KBP}, where it was discovered that
the basic objects of the integrable structure like T- and
Q-functions, their functional relations and analytic properties in
the complex rapidity plane allow one to set up, though in a
non-unique and non-trivial way, a compact set of non-linear integral
equations that governs the finite size spectrum of the theory. Later
this method has been developed further and extended to describe
lattice models \cite{KP91,Suz98,exs1,HubbardNLIE,RacsNLIEk,Klumpersl4} 
and quantum field theories
\cite{nlie1,nlie21,nlie22,nlie23,nlie24,nlie25,GKV08,KL10} as well.

In the case of quantum field theories the basic starting point for the
NLIE approach is the reformulation of the TBA equations through
functional relations \cite{KNS11}. It is known that the infinite
Y-system can be rephrased as a T-system,
a set of discrete Hirota equations, which can be solved in terms
of a few Q-functions. This offers the possibility to replace the
infinite set of TBA variables by a finite set of only a few variables.
In order to transform the functional relations into integral equations
the analyticity properties of the Y,~T,~and~Q-functions must be known.
In AdS/CFT the analyticity part of the problem is more
complicated than it is in the previously elaborated relativistically
invariant examples
\cite{nlie1,nlie21,nlie22,nlie23,nlie24,nlie25,KL10,GKV08} since
Y-functions are defined on an infinite genus Riemann surface, such
that the physical and mirror sheets are not equivalent and
Y-functions satisfy non-trivial discontinuity relations on the
mirror sheet \cite{Tateo1}. This is responsible for the non-local properties of
the mirror TBA. However it is possible to bring the TBA equations
into a quasi-local form \cite{BH10} which contains only next to nearest
neighbor interactions among Y-functions opening
 the way to apply techniques worked out for relativistic models
 \cite{nlie1,nlie21,nlie22,nlie23,nlie24,nlie25,KL10,GKV08}.

Recently it has been shown that there exists a \lq\lq magic" sheet with
short cuts where the T-functions have very simple discontinuity
structure and they admit a $\mathbb{Z}_4$ symmetry which
supplemented by analyticity requirements dictated by the asymptotic
solution and group-theoretical constraints
allows one to derive both the mirror TBA and its simplified
finite reformulation FiNLIE \cite{GKLV11} as well. This was the
first finite reformulation of the AdS/CFT spectral problem. In
\cite{GKLV11} left-right symmetric states ($\alg{sl}(2)$ or 
$\alg{su}(2)$ states) were considered and the
3 unknowns of the FiNLIE are discontinuities of
two T-functions along short cuts in the \lq\lq magic" sheet
and the discontinuity of a certain gauge transformation along
the real axis and they encode all information on the T- and
Y-systems of the AdS/CFT.

In this paper we will give another finite reformulation of the
spectral problem, but remaining on the mirror sheet. We follow the
approach of the very first NLIE paper \cite{KBP}, where the basic
objects of the integrable structure like T- and Q-functions are used
to build the NLIE unknowns and where the functional relations they
satisfy and their analyticity properties determine the actual form
of the NLIE. Our approach may be called hybrid-NLIE, since by
appropriate NLIEs we resum the semi-infinite 
SU$(2)$ and SU$(4)$ parts of the mirror TBA equations.
This approach was initiated in \cite{RyoHybrid} by resumming the two 
SU$(2)$ wings
similarly as it had been done in the case of the SU$(2)$ related
relativistic models \cite{nlie21,nlie22,nlie23,nlie24,nlie25}. The
basic idea of the construction as follows.
\begin{figure}[htb]
\begin{flushleft}
\hskip 15mm \leavevmode \epsfxsize=120mm \epsfbox{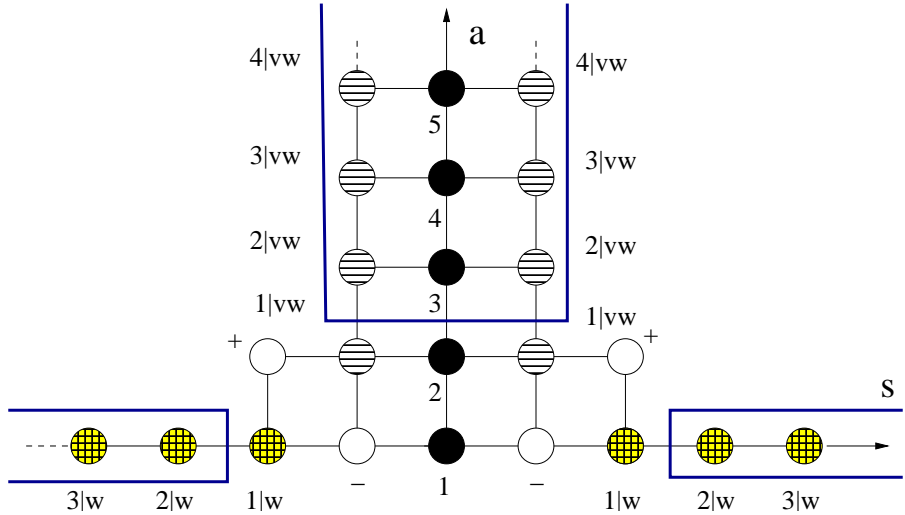}
\end{flushleft}
\caption{{\footnotesize Division of the AdS/CFT Y-system into
right-wing, left-wing, upper and central nodes. }} \label{ImageY}
\end{figure}

The set of Y-functions appearing in the AdS/CFT spectral problem
can be naturally grouped into four subsets (see Figure 1):

\begin{itemize}

\item
right-wing nodes: $Y^{(+)}_{m\vert w}$ for $m\geq2$.

\item
left-wing nodes: $Y^{(-)}_{m\vert w}$ for $m\geq2$.

\item
upper nodes: $Y^{(\pm)}_{m\vert vw}$ for $m\geq2$ and $Y_Q$ for
$Q\geq3$.

\item
central nodes: $Y_1,Y_2,Y^{(\pm)}_{1\vert w},Y^{(\pm)}_{1\vert vw}$
and $Y^{(\pm)}_\pm$.

\end{itemize}

The first step \cite{RyoHybrid} of the construction of the hybrid-NLIE
is to replace the right-wing SU$(2)$ type nodes by a complex NLIE
variable coupled to itself and to the central nodes\footnote{Note
that a resummation of the horizontal nodes, similarly to what is
used for the FiNLIE, was already proposed in \cite{talk}.}. 
The coupling of
the central nodes to $Y^{(+)}_{2\vert w}$ is replaced by coupling to
the NLIE variable. Similar considerations apply to the left-wing
part of the problem.

The second step is to replace the Y-functions corresponding to the
upper SU$(4)$ type nodes by 12 NLIE functions. These are coupled to
themselves as well as to the 1st row of upper nodes $(Y_3,Y_{2\vert
vw}^{(\pm)})$ while the 1st row of the upper nodes are coupled to
the central nodes and the upper NLIE functions.


Thus we can we replace the semi-infinite SU$(2)$ and SU$(4)$ parts of the
TBA diagram by two SU$(2)$ and an SU$(4)$ type relativistic NLIEs
which are sewn together by the quasi-local TBA equations for the
central nodes. We call the final equations hybrid-NLIE for AdS/CFT.
Though we have more unknown functions than in \cite{GKLV11}, our
NLIE equations are based on the B\"acklund transformations of the
corresponding T-system and can be straightforwardly generalized to a
wide range of relativistically invariant integrable models.
In this paper we will complete the derivation of our equations only for the ground state, but
in order to get non-trivial results, in the $\gamma$-deformed AdS/CFT
model, because in this case the ground state solution of
the Y-system is non-trivial and the ground state equations can be
checked by analytical computations using the 2nd order L\"uscher
formula \cite{TBA2ndLusch}.

Though at the level of equations we concentrate on the ground state,
all our considerations concerning the analyticity properties and the
construction of NLIE unknowns given in sections~4-6 are general 
and valid for excited states of the model as well.

The paper is organized as follows: in section~2 we give our
starting point by recalling the quasi-local form of the TBA
equations. In section~3 we review the B\"acklund transformations
and analyticity strips which form the basis of the construction of
our variables. In section~4 we construct the NLIE functions and the
functional relations they satisfy. In section~5 
the asymptotic solution is given at all levels of the
nesting for the upper nodes and in
section~6 the NLIE construction for the horizontal part is given.
Section~7 contains the derivation of the SU$(2)$ and
SU$(4)$ type NLIEs.
In section~8 we linearize our NLIE equations for the $\gamma$-deformed
ground state, analytically calculate the first correction to the
asymptotic solution and show the agreement with analogous results coming
from the original TBA formalism. In section~9 we summarize our results
by collecting all NLIE equations for the description of the $\gamma$-deformed
ground state.
Appendix~A contains our basic notations and the kernels of the TBA.
We collected the basic building blocks of the asymptotic solution in
appendix~B. The paper is closed by appendix~C where the kernels of
the SU$(4)$ type NLIE are listed.

\section{Quasi-local twisted TBA}

In this section we review the quasi-local form of the mirror TBA for
the three parameter $\gamma$-deformed AdS/CFT theories
\cite{Z27,Z28,Z29,Z30,Z31}. Contrary to the undeformed or
$\beta$-deformed theories, in the most general case no
supersymmetry is preserved and the ground state energy is non-zero.
This allows us to test our ideas on the ground state equations
directly. For this reason and for the sake of simplicity at the level of
equations we will restrict our attention to the ground state of the
model though as it has already been mentioned in the introduction
all of our equations can be extended to excited states without
difficulties.

At the level of the worldsheet scattering theory the $\gamma$-deformation 
can be implemented in two ways: either imposing
operatorial (particle number dependent) twisted boundary 
conditions \cite{Z33}, or by imposing a (c-number) twisted boundary 
condition and considering a
twisted scattering matrix for the excitations \cite{Z34,Z35}.

The ground state TBA equations of the $\gamma$-deformed
theory have been derived in \cite{TBA2ndLusch} and it was found that
the twist parameters enter the so-called canonical version of the
equations as if they were Y-system preserving
chemical potentials in the undeformed theory \cite{Tateo1,AF09b}.
Thus the Y-system for the $\gamma$-deformed theory is
identical to that of the undeformed theory.
On the other hand the twist parameters cancel from the simplified
version of the twisted TBA and they re-appear as parameters of
the boundary conditions imposed on the Y-functions at large $u$.
These canonical and simplified twisted TBA equations can be
reformulated in a quasi-local form \cite{BH10} which we review now
for the ground state.

The quasi-local formulation of the mirror TBA is possible since all kernels
${\cal K}_Q \equiv {\cal K}_Q(u,v)$ entering the
TBA equations satisfy  the important identity:
\begin{equation}
{\cal K}_Q-s \star ({\cal K}_{Q-1}+{\cal K}_{Q+1})=\delta {\cal
K}_Q, \qquad{\cal K}_0 \equiv 0, \qquad Q=1,2,... \label{kerid0}
\end{equation}
with $\delta {\cal K}_Q$ vanishing with the exception of a few values of the 
index $Q$. Explicit formulas are given at the end of appendix~A.
In order to compactly present the quasi-local TBA equations we introduce the
notations:
\begin{equation}
L_Q=\ln(1+Y_Q), \qquad  {\cal R}_Q=\ln\left(1+\frac{1}{Y_Q}\right)
\qquad Q=1,2,... \nonumber
\end{equation}
\begin{equation}
 r_m^{(\alpha)}=\ln(1+Y_{m\vert vw}^{(\alpha)}), \qquad   
{\cal L}_\pm^{(\alpha)}=\ln\left(-\left(1-\frac{1}{Y_{\pm}^{(\alpha)}} 
\right)\right), \qquad  m=1,2,\dots, \quad \alpha=\pm
\label{jeloles}
\end{equation}
\begin{equation}
\ H^{(\alpha)}=\ln\left(\frac{1-Y_-^{(\alpha)}}
{1-Y_+^{(\alpha)}}\right) \qquad \alpha=\pm. \nonumber
 \end{equation}
Next we introduce the linear functional{\footnote{The expression
$\Omega({\cal K}_Q)=\sum\limits_{Q=1}^{\infty} L_Q \star
 {\cal K}_Q$ in the non-local versions of the mirror TBA.}}
of the vector kernel ${\cal K}_ Q$  with the definition:
\begin{equation}
\begin{split}
\Omega({\cal K}_Q)=&{\cal R}_2\star(\delta{\cal K}_2+
2s_{1/2}\star\delta{\cal K}_2)+{\cal R}_2\star\sigma_{1/2}\star{\cal
K}_1
-{\cal R}_1\star\sigma_{1/2}\star{\cal K}_2\\
+& \sum\limits_{\alpha=\pm} \left(  \,
r_1^{(\alpha)}\star\sigma_{1/2}\star\delta{\cal K}_2 +
r_1^{(\alpha)} \star s_{1/2}\star{\cal
K}_1-H^{(\alpha)}\,\hat\star\, s_{1/2}\star{\cal K}_2 \right).
\end{split}
\label{final}
\end{equation}
We use the kernels
\begin{equation}
s(u)=\frac{g}{4\cosh\frac{\pi gu}{2}}, \qquad 
s_{1/2}(u)=\frac{1}{2}s\left(\frac{u}{2}\right),
\qquad \sigma_{1/2}(u)=\frac{g}{2\sqrt{2}}\,\frac{\cosh\frac{\pi gu}{4}}
{\cosh\frac{\pi gu}{2}},
\end{equation}
and $\star$ and $\hat{\star}$ denote convolutions running on 
$\mathbb{R}$ and $[-2,2] \subset \mathbb{R}$  respectively. 
See appendix~A.

The quasi-local TBA equations for the ground state are composed 
of two groups of equations. Equations in the first group follow
from the Y-system relations, they are local and their form is 
the same as in the simplified version of the equations:
\begin{eqnarray}
Y^{(\alpha)}_{m\vert vw}&=&\exp\left\{
\ln\left[\frac{(1+Y^{(\alpha)}_{m+1\vert
vw})(1+Y^{(\alpha)}_{m-1\vert vw})}
{(1+Y_{m+1})}\right]\star s\right\},\qquad m\geq2,\label{TBAmvw}\\
Y^{(\alpha)}_{1\vert vw}&=&\exp\left\{
\ln\left[\frac{(1+Y^{(\alpha)}_{2\vert vw})}{(1+Y_2)}\right]\star s
+\ln\left[\frac{1-Y_-^{(\alpha)}}{1-Y_+^{(\alpha)}}\right]\
\hat\star\ s
\right\},\label{TBA1vw}\\
Y^{(\alpha)}_{m\vert w}&=&\exp\left\{
\ln\left[(1+Y^{(\alpha)}_{m+1\vert w})(1+Y^{(\alpha)}_{m-1\vert
w})\right]
\star s\right\},\qquad m\geq2,\label{TBAmw}\\\
Y^{(\alpha)}_{1\vert w}&=&\exp\left\{
\ln\left[1+Y^{(\alpha)}_{2\vert w}\right]\star s+
\ln\left[\frac{1-\frac{1}{Y_-^{(\alpha)}}}{1-\frac{1}{Y_+^{(\alpha)}}}\right]
\ \hat\star\ s \right\}\,,\label{TBA1w}\\
Y_Q&=&\exp\left\{\ln\left[ \frac{Y_{Q+1}\,Y_{Q-1}
(1+Y^{(+)}_{Q-1\vert vw})(1+Y^{(-)}_{Q-1\vert vw})}
{Y^{(+)}_{Q-1\vert vw}Y^{(-)}_{Q-1\vert vw}
(1+Y_{Q+1})(1+Y_{Q-1})}\right]\star s \right\},\quad
Q\geq2.\label{TBAQ}
\end{eqnarray}
The second group consists of quasi-local (next to nearest neighbor interacting)
central node equations and they take the form:
\begin{equation}
\frac{Y_-^{(\alpha)}}{Y_+^{(\alpha)}}= \exp\left\{-L_1 \star K_{1y}
-\Omega(K_{Qy}) \right\}, \label{YmperYp}
\end{equation}
\begin{equation}
\begin{split}
Y^{(\alpha)}_+Y^{(\alpha)}_-= &\exp\Bigg\{
2\ln\left[\frac{1+Y^{(\alpha)}_{1\vert vw}}{1+Y^{(\alpha)}_{1\vert
w}} \right]\star s + L_1\star\left[
-K_1+2K^{11}_{xv}\star s\right]  \\
& -\Omega(K_Q)+2 \, \Omega(K^{Q1}_{xv}\star s) \Bigg\},
\end{split} \label{YpxYm}
\end{equation}
\begin{equation}
\begin{split}
\ln Y_1=&-L\tilde{\cal E}_1+
\sum_{\alpha=\pm} \, r_1^{(\alpha)}\star s\ \hat\star\ K_{y1}\\
&- \sum_{\alpha=\pm} \left( \ln\left[\frac{1-Y_-^{(\alpha)}}
{1-Y_+^{(\alpha)}}\right]\ \hat\star\ s\star K^{11}_{vwx}
+{\mathscr L}_-^{(\alpha)}\ \hat\star\ K^{y1}_-+{\mathscr L}_+^{(\alpha)}
\ \hat\star\ K^{y1}_+ \right)  \\
&\qquad+ L_1\star K^{11}_{{\alg{sl}(2)}}
+\Omega(K^{Q1}_{{\alg{sl}(2)}})+2 \,\Omega(s\star K^{Q-1\,1}_{vwx}).
\end{split}
\label{hybrid}
\end{equation}
These equations are parameter-free, but the $\gamma$-deformation parameters
enter the conditions determining the asymptotic large $u$ behavior of
Y-functions:
\begin{equation} \label{largeuvw}
Y_{m|vw}^{(\alpha)}\to m \, (m+2), \qquad Y_{m|w}^{(+)}\to [m]_q \, [m+2]_q,
\qquad
Y_{m|w}^{(-)}\to [m]_{\dot{q}} \, [m+2]_{\dot{q}}, \qquad m=1,2,...
\end{equation}
and
\begin{equation} \label{largeupm}
Y_{\pm}^{(+)} \to \frac{2}{[2]_q}, \qquad Y_{\pm}^{(-)} 
\to \frac{2}{[2]_{\dot{q}}},
\end{equation}
where $[m]_q$ denotes the $q$-number $[m]_q=\frac{q^m-q^{-m}}{q-q^{-1}}$.
They correspond to the (c-number) boundary conditions in the description of
\cite{Z34,Z35} and they are expressed by the deformation parameters
of the $\gamma$-deformed theory  as follows: 
$q=e^{i \frac{\gamma_3+\gamma_2}{2}L},
\quad \dot{q}=e^{i \frac{\gamma_3-\gamma_2}{2}L}$.
The parameter $\gamma_1$ does not enter the ground state equations since it 
corresponds to the twist parameter of the S-matrix.

To complete the quasi-local TBA description of the $\gamma$-deformed AdS/CFT 
we should supplement the integral equations (\ref{TBAmvw}-\ref{hybrid}) with 
the energy formula. In the quasi-local description the energy expression is a
function of the central nodes only:
\begin{equation}
\begin{split}
{\cal E}=&L_1\star \tilde{J}_1+ ({\cal R}_2-{\cal
R}^o_2+L^o_2)\star\sigma_{1/2}\star \tilde{J}_1
-({\cal R}_1-{\cal R}^o_1+L^o_1)\star\sigma_{1/2}\star \tilde{J}_2\\
+& \sum_{\alpha=\pm} \left( (r_1^{(\alpha)}-r^{(\alpha)o}_1
)\star s_{1/2}\star \tilde{J}_1 -
(H^{(\alpha)}-H^{(\alpha)o})\,\hat\star\, s_{1/2}\star
\tilde{J}_2 \right),
\end{split}
\label{finalenergy}
\end{equation}
where we introduced the notation $\tilde{J}_Q(u)=-\frac{1}{2 \,
\pi}\frac{{\rm d}\tilde p^Q}{{\rm d}u}$ and the upper index
$^o$ means that the corresponding expression should be taken at
the asymptotic solution. This representation is necessary 
for all integrals in (\ref{finalenergy}) to converge. In the
$\gamma$-deformed model the asymptotic solutions are identical with
the large $u$ limits for Y-functions appearing in (\ref{largeuvw}), 
(\ref{largeupm}) while the asymptotic solution for the momentum carrying 
nodes takes the form 
\begin{equation}
Y_Q^o \sim (2-[2]_q) \, (2-[2]_{\dot{q}}) \, Q^2 \, e^{-L \,
\tilde{\cal E}_Q}.
\end{equation}

The quasi-local TBA equations presented in this section are the starting
point for our NLIE description. We will transform the semi-infinite
set of TBA equations (\ref{TBAmw}) and (\ref{TBAmvw}), (\ref{TBAQ})
to NLIE equations of SU$(2)$ and SU$(4)$ type, respectively by
(the nested hierarchy of) B\"acklund transformations. This is described
in the next sections.

\section{Hierarchy of B\"acklund transformations and analyticity strips}

In this paper we will denote the AdS/CFT Y-functions in the index
conventions\footnote{The precise relation between our Y-functions and those
of ref.~\cite{GKV09} is $Y_{a,s}(u)={\mathbf y}_{a,s}(-2u/g)$.} of
\cite{GKV09} by ${\mathbf y}_{a,s}$ and the corresponding T-system elements
by ${\mathbf t}_{a,s}$. They satisfy the usual Y-T relations
\begin{equation}
{\mathbf y}_{a,s}=\frac{{\mathbf t}_{a,s+1}\,{\mathbf t}_{a,s-1}}
{{\mathbf t}_{a+1,s}\,{\mathbf t}_{a-1,s}}
\label{YT}
\end{equation}
and the T-system equations
\begin{equation}
{\mathbf t}^+_{a,s}\,{\mathbf t}^-_{a,s}=
{\mathbf t}_{a+1,s}\,{\mathbf t}_{a-1,s}+
{\mathbf t}_{a,s+1}\,{\mathbf t}_{a,s-1}.
\label{TT}
\end{equation}

The T-functions ${\mathbf t}_{a,s}$ for $s=0,1,2$ have been constructed 
explicitly in \cite{BH9} in a particular gauge called the BA (Bethe Ansatz) 
gauge. It was found that their analytic properties can be summarized as:

${\mathbf t}_{a,0}$ is of type $(-1-a,a+1)$, \ \ $a\geq1$,

${\mathbf t}_{a,1}$ is of type $(-a,a)$, \ \ \ $a\geq1$,

${\mathbf t}_{a,2}$ is of type $(1-a,a-1)$, \ \ $a\geq2$.

\noindent
A function $f(u)$ is called of type $(c,d)$ if it is meromorphic in the strip
$c/g<{\rm Im}u<d/g$.

We can extend the above solution of T-functions for $s=-1,-2$ by the Y-T 
relations (\ref{YT}). Using the fact that ${\mathbf y}_{a,0}=Y_a$ is of type
$(-a,a)$ and the relation
\begin{equation}
{\mathbf t}_{a,-1}={\mathbf y}_{a,0}\,\frac{{\mathbf t}_{a+1,0}\,
{\mathbf t}_{a-1,0}}{{\mathbf t}_{a,1}}
\end{equation}
we see that

${\mathbf t}_{a,-1}$ is of type $(-a,a)$, \ \ \ $a\geq1$,

\noindent
and similarly from
\begin{equation}
{\mathbf t}_{a,-2}={\mathbf y}_{a,-1}\,\frac{{\mathbf t}_{a+1,-1}\,
{\mathbf t}_{a-1,-1}}{{\mathbf t}_{a,0}}
\end{equation}
and using the fact that ${\mathbf y}_{a,-1}=Y^{(-)}_{a-1\vert vw}$ 
is of type $(1-a,a-1)$ we find that

${\mathbf t}_{a,-2}$ is of type $(1-a,a-1)$, \ \ \ $a\geq2$.

The AdS/CFT T-system satisfies the boundary condition 
${\mathbf t}_{a,\pm3}=0$, $a\geq3$ and hence the boundary T-functions
${\mathbf t}_{a,\pm2}$, $a\geq2$ are solutions of the discrete Laplace
equation. These are parametrized by the four functions ${\cal A}$, ${\cal B}$,
${\cal C}$ and ${\cal D}$: 
\begin{equation}
{\mathbf t}_{a,2}={\cal A}^{[a]}\,{\cal B}^{[-a]},\qquad\qquad
{\mathbf t}_{a,-2}={\cal C}^{[a]}\,{\cal D}^{[-a]}.
\label{ABCD}
\end{equation}
This implies
\begin{eqnarray}
\frac{{\mathbf t}_{a+1,2}}{{\mathbf t}^+_{a,2}}&=&B^{[-a]},\qquad\quad
B=\frac{{\cal B}^-}{{\cal B}^+},\label{B-1}\\
\frac{{\mathbf t}_{a+1,2}}{{\mathbf t}^-_{a,2}}&=&A^{[a]},\qquad\quad
\ \ A=\frac{{\cal A}^+}{{\cal A}^-},\label{A-1}\\
\frac{{\mathbf t}_{a+1,-2}}{{\mathbf t}^+_{a,-2}}&=&D^{[-a]},\qquad\quad
D=\frac{{\cal D}^-}{{\cal D}^+},\label{D-1}\\
\frac{{\mathbf t}_{a+1,-2}}{{\mathbf t}^-_{a,-2}}&=&C^{[a]},\qquad\quad
\ \ C=\frac{{\cal C}^+}{{\cal C}^-}.\label{C-1}
\end{eqnarray}
From (\ref{B-1}) we obtain that $B$ is of type $(-2a,-2)$ and since this
is true for any $a\geq2$ finally we can conclude that $B$ is of type
$(-\infty,-2)$. Similarly we find that $D$ is also of type $(-\infty,-2)$
and $A$ and $C$ are of type $(2,\infty)$.

\subsection{Principal chiral model conventions}

In this paper we will also use an alternative notation for the Y and 
T-functions corresponding to the upper nodes ($a\geq2$) of the AdS/CFT TBA 
diagram. Using this new notation this part becomes identical to the 
corresponding functional relations of an SU(4) principal chiral model. 
In this latter model, for general SU($k$), there are T-functions $T_{a,s}$  
$a=0,1,\dots,k$ satisfying the T-system equations
\begin{equation}
T^+_{a,s}\,T^-_{a,s}=T_{a+1,s}\,T_{a-1,s}+T_{a,s+1}\,T_{a,s-1}
\label{TT2}
\end{equation}
and the boundary conditions 
\begin{equation}
T_{-1,s}=T_{k+1,s}=0.
\label{boundaryT}
\end{equation}
The mapping of the AdS/CFT T-functions to the SU(4) principal model variables
is given by
\begin{equation}
{\mathbf t}_{a,2}=T_{0,a},\quad
{\mathbf t}_{a,1}=T_{1,a},\quad
{\mathbf t}_{a,0}=T_{2,a},\quad
{\mathbf t}_{a,-1}=T_{3,a},\quad
{\mathbf t}_{a,-2}=T_{4,a}
\label{iden}
\end{equation}
and we can summarize the analytic properties of the T-functions
in this notation as 
\begin{equation}
T_{4,s}\ {\rm and \ }T_{0,s}\ {\rm are\ of\ type\ }(1-s,s-1),\qquad
T_{3,s}\ {\rm and \ }T_{1,s}\ {\rm are\ of\ type\ }(-s,s),
\nonumber
\end{equation}
and $T_{2,s}$ is of type $(-1-s,s+1)$.
The principal chiral model Y-T relations are of the same form as (\ref{YT}): 
\begin{equation}
y_{a,s}=\frac{T_{a,s+1}\,T_{a,s-1}}{T_{a+1,s}\,T_{a-1,s}},\quad
Y_{a,s}=1+y_{a,s}=\frac{T^+_{a,s}\,T^-_{a,s}}{T_{a+1,s}\,T_{a-1,s}},\quad
a=1,\dots,k-1.
\label{YT2}
\end{equation}
Note that the identification (\ref{iden}) implies the exchange of the indices
$a\leftrightarrow s$ and consequently the relation among the Y-functions
is given by 
\begin{equation}
Y_{1,s}=1+\frac{1}{{\mathbf y}_{s,1}}=1+Y^{(+)}_{s-1\vert vw},\quad
Y_{2,s}=1+\frac{1}{{\mathbf y}_{s,0}}=1+\frac{1}{Y_s},\quad
Y_{3,s}=1+\frac{1}{{\mathbf y}_{s,-1}}=1+Y^{(-)}_{s-1\vert vw}.
\label{iden2}
\end{equation}

\subsection{B\"acklund transformations}

The advantage of using the principal model conventions is that in this language
it is easy to formulate the hierarchy of B\"acklund transformations that will
play an important role in our considerations. In addition, while the SU(4) 
case is relevant for the upper nodes, similar considerations, but for the 
SU(2) case, are relevant for the right-wing nodes and similarly for the 
left-wing nodes.

Given a set of T-functions satisfying the T-system equations (\ref{TT2})
and boundary conditions (\ref{boundaryT}) we can find the set of F-functions,
$F_{a,s}$, $a=0,\dots,k-1$ with boundary condition
\begin{equation}
F_{-1,s}=F_{k,s}=0
\label{boundaryF}
\end{equation}
by solving the equations \cite{KLWZ96}
(B\"acklund transformation) for $a=0,1,\dots,k-1$
\begin{eqnarray}
T_{a+1,s+1}\,F_{a,s}&=&T_{a+1,s}^+\,F_{a,s+1}^- + T_{a,s}\,F_{a+1,s+1},
\label{BT1}\\
T^+_{a,s+1}\,F_{a,s}&=&T_{a,s}\,F_{a,s+1}^+ + T^+_{a+1,s}\,F_{a-1,s+1}.
\label{BT2}
\end{eqnarray}
It can be shown \cite{KLWZ96} that the F-functions also satisfy the T-system 
relations (\ref{TT2}) (with $k-1$) thus the mapping $T\rightarrow F$ is from 
a solution of the SU($k$) T-problem to a solution of the SU($k-1$) T-problem.
Hence it is natural to define the analogs of the Y-functions corresponding 
to the F-functions:
\begin{equation}
w_{a,s}=\frac{F_{a,s+1}\,F_{a,s-1}}{F_{a+1,s}\,F_{a-1,s}},\quad
W_{a,s}=1+w_{a,s}=\frac{F^+_{a,s}\,F^-_{a,s}}{F_{a+1,s}\,F_{a-1,s}},\quad
a=1,\dots,k-2.
\label{WF}
\end{equation}
Let us recall that the boundary components satisfy the discrete Laplace
equation and can be factorized as
\begin{equation}
T_{0,s}={\cal A}^{[s]}\,{\cal B}^{[-s]},\qquad\qquad
T_{k,s}={\cal C}^{[s]}\,{\cal D}^{[-s]}
\end{equation}
and define the ratios
\begin{equation}
A=\frac{{\cal A}^+}{{\cal A}^-},\quad\qquad
B=\frac{{\cal B}^-}{{\cal B}^+},\quad\qquad
C=\frac{{\cal C}^+}{{\cal C}^-},\quad\qquad
D=\frac{{\cal D}^-}{{\cal D}^+}.
\end{equation}
Similarly for the F-functions we have
\begin{equation}
F_{0,s}=\alpha^{[s]}\,\beta^{[-s]},\qquad\qquad
F_{k-1,s}=\gamma^{[s]}\,\delta^{[-s]}
\end{equation}
and
\begin{equation}
a=\frac{\alpha^+}{\alpha^-},\quad\qquad
b=\frac{\beta^-}{\beta^+},\quad\qquad
c=\frac{\gamma^+}{\gamma^-},\quad\qquad
d=\frac{\delta^-}{\delta^+}.
\end{equation}
It is easy to see from the B\"acklund transformations (\ref{BT1}) and
(\ref{BT2}) that one of the boundary ratios is preserved on both boundary 
lines:
\begin{equation}
a=A,\qquad\quad d=D^+.
\label{AD}
\end{equation}
By studying the analytic properties of the T-functions appearing in the
B\"acklund transformations and assuming maximal possible analyticity
(meromorphicity) strips for the resulting F-functions we find that ($s\geq2$):

$F_{0,s}$ is of type $(1-s,s-1)$, 

$F_{1,s}$ is of type $(-s,s)$,

$F_{2,s}$ is of type $(-1-s,s-1)$, 

$F_{3,s}$ is of type $(-s,s-2)$. 

\noindent
The T-system equations are invariant under the gauge transformations
\begin{equation}
T_{a,s}\rightarrow \tilde T_{a,s}=\lambda_{a,s}\,T_{a,s},
\label{gau1}
\end{equation}
where the gauge function is of the form
\begin{equation}
\lambda_{a,s}=f_1^{[-s-a]}\,f_2^{[s-a]}\,f_3^{[a-s]}\,f_4^{[s+a]}.
\label{gau2}
\end{equation}
The Y-functions defined by (\ref{YT2}) are gauge invariant and if we define
the gauge transformation of the F-functions by
\begin{equation}
F_{a,s}\rightarrow \tilde F_{a,s}=\omega_{a,s}\,F_{a,s},
\end{equation}
with
\begin{equation}
\omega_{a,s}=f_1^{[-s-a]}\,f_2^{[s-a]}\,f_3^{[2+a-s]}\,f_4^{[s+a]}
\end{equation}
then also the B\"acklund transformations remain invariant. Similarly the
W-functions defined by (\ref{WF}) are also gauge invariant. 

We have seen that the properties of the F-functions are very similar
to those of the T-functions apart from the reduction $k\rightarrow k-1$.
We can now formulate a B\"acklund transformation starting from the F-functions
playing the role of the T-functions and by repeating the steps build the
chain of B\"acklund transformations corresponding to
\begin{equation}
{\rm SU}(k)\rightarrow {\rm SU}(k-1)\rightarrow {\rm SU}(k-2)\dots
\end{equation}

To unify and simplify the notation we now introduce the family of T-functions 
$ T^{(r)}_{a,s}$ where $r=1,\dots,k$ indicates the B\"acklund level,
the range of the index $a$ is $a=0,\dots,r$ and the boundary conditions are
\begin{equation}
T^{(r)}_{-1,s}=T^{(r)}_{r+1,s}=0.
\end{equation}
With this notation
\begin{equation}
T^{(k)}_{a,s}=T_{a,s},\qquad\qquad T^{(k-1)}_{a,s}=F_{a,s}
\end{equation}
and so on. The boundary factorization can be written as
\begin{equation}
T^{(r)}_{0,s}={\cal A}^{(r)[s]}\,{\cal B}^{(r)[-s]},\qquad\qquad
T^{(r)}_{r,s}={\cal C}^{(r)[s]}\,{\cal D}^{(r)[-s]}
\end{equation}
and we can identify
\begin{equation}
{\cal A}^{(k)}={\cal A},\qquad
{\cal B}^{(k)}={\cal B},\qquad
{\cal C}^{(k)}={\cal C},\qquad
{\cal D}^{(k)}={\cal D}
\end{equation}
and
\begin{equation}
{\cal A}^{(k-1)}=\alpha,\qquad
{\cal B}^{(k-1)}=\beta,\qquad
{\cal C}^{(k-1)}=\gamma,\qquad
{\cal D}^{(k-1)}=\delta.
\end{equation}

\noindent
Further we define
\begin{equation}
A^{(r)}=\frac{{\cal A}^{(r)+}}{{\cal A}^{(r)-}},\quad\qquad
B^{(r)}=\frac{{\cal B}^{(r)-}}{{\cal B}^{(r)+}},\quad\qquad
C^{(r)}=\frac{{\cal C}^{(r)+}}{{\cal C}^{(r)-}},\quad\qquad
D^{(r)}=\frac{{\cal D}^{(r)-}}{{\cal D}^{(r)+}}.
\end{equation}
Here
\begin{equation}
A^{(k)}=A,\quad
B^{(k)}=B,\quad
C^{(k)}=C,\quad
D^{(k)}=D,\quad
\end{equation}
\begin{equation}
A^{(k-1)}=a,\quad
B^{(k-1)}=b,\quad
C^{(k-1)}=c,\quad
D^{(k-1)}=d
\end{equation}
and so on. Continuing (\ref{AD}) we have 
\begin{equation}
A^{(k)}=A^{(k-p)},\qquad\quad D^{(k)}=D^{(k-p)[-p]},\qquad p=1,\dots,k-1.
\label{AD2}
\end{equation}

\noindent
Let us also introduce the family of Y-functions using the notation
\begin{equation}
y^{(r)}_{a,s},\qquad Y^{(r)}_{a,s}=1+y^{(r)}_{a,s},\quad a=1,\dots,r,
\end{equation}
where $r=1,\dots,k-1$ and
\begin{equation}
y^{(k-1)}_{a,s}=y_{a,s},\qquad y^{(k-2)}_{a,s}=w_{a,s},
\end{equation}
and so on.

Returning to the SU(4) special case the analyticity strips of 
$T^{(4)}_{a,s}=T_{a,s}$ and $T^{(3)}_{a,s}=F_{a,s}$ are already determined
above and using the lower level B\"acklund equations can be established 
also for $T^{(2)}_{a,s}$ and $T^{(1)}_{a,s}$. We find that

$T^{(2)}_{0,s}$ is of type $(1-s,s-1)$, 

$T^{(2)}_{1,s}$ is of type $(-s,s-2)$,

$T^{(2)}_{2,s}$ is of type $(-1-s,s-3)$, 

$T^{(1)}_{0,s}$ is of type $(1-s,s-1)$, 

$T^{(1)}_{1,s}$ is of type $(-2-s,s-4)$.

For the boundary ratios we find that

$A^{(4)},A^{(3)},A^{(2)},A^{(1)}$ are of type $(2,\infty)$, 

$B^{(4)},B^{(3)},B^{(2)},B^{(1)}$ are of type $(-\infty,-2)$, 

$C^{(4)},C^{(3)-},C^{(2)--},C^{(1)---}$ are of type $(2,\infty)$, 

$D^{(4)},D^{(3)-},D^{(2)--},D^{(1)---}$ are of type $(-\infty,-2)$. 

For completeness we also list the analyticity strips for the Y-functions:

$y^{(3)}_{1,s}=y_{1,s}$ is of type $(1-s,s-1)$,

$y^{(3)}_{2,s}=y_{2,s}$ is of type $(-s,s)$,

$y^{(3)}_{3,s}=y_{3,s}$ is of type $(1-s,s-1)$,

$y^{(2)}_{1,s}=w_{1,s}$ is of type $(1-s,s-1)$,

$y^{(2)}_{2,s}=w_{2,s}$ is of type $(-s,s-2)$,

$y^{(1)}_{1,s}$ is of type $(1-s,s-3)$.

\section{NLIE variables and functional equations}

The gauge invariant Y-functions can be obtained from the T-system equations
by dividing the equations by one of the terms. Similarly we can form gauge 
invariant ratios by dividing the B\"acklund equations by one of the three 
terms. These gauge invariant ratios will be the variables in our NLIE 
equations.

Using the first B\"acklund equation (\ref{BT1}) we can form the ratios
\begin{equation}
b_{a,s}=\frac{T^+_{a,s}\,F^-_{a-1,s+1}}{T_{a-1,s}\,F_{a,s+1}},\qquad
B_{a,s}=\frac{T_{a,s+1}\,F_{a-1,s}}{T_{a-1,s}\,F_{a,s+1}},\qquad
a=1,\dots,k-1.
\label{bB}
\end{equation}
In terms of these variables the (\ref{BT1}) equations simply become
\begin{equation}
B_{a,s}=1+b_{a,s},\qquad a=1,\dots,k-1.
\end{equation}
Similary we obtain from the second B\"acklund equation (\ref{BT2})
\begin{equation}
d_{a,s}=\frac{T^-_{a,s}\,F_{a,s+1}}{T_{a+1,s}\,F^-_{a-1,s+1}},\qquad
D_{a,s}=\frac{T_{a,s+1}\,F^-_{a,s}}{T_{a+1,s}\,F^-_{a-1,s+1}},\qquad
a=1,\dots,k-1.
\label{dD}
\end{equation}
In terms of these variables the (\ref{BT2}) equations simplify to
\begin{equation}
D_{a,s}=1+d_{a,s},\qquad a=1,\dots,k-1.
\end{equation}
Since we know the analyticity strips of the T and F-functions we can see that 
for our SU(4) case the NLIE functions are of type:

$b_{1,s}$:\ \ $(1-s,s-1)$, \hspace{4.7cm} $d_{1,s}$:\ \ $(1-s,s+1)$,

$b_{2,s}$:\ \ $(-s,s)$, \hspace{5.7cm} $d_{2,s}$:\ \ $(-s,s)$,

$b_{3,s}$:\ \ $(-1-s,s-1)$, \hspace{4.4cm} $d_{3,s}$:\ \ $(1-s,s-1)$.

\noindent
Generalizing the definitions (\ref{bB}) and (\ref{dD}) to all B\"acklund
levels we can introduce the corresponding NLIE functions:
\begin{equation}
b^{(r)}_{a,s},\quad
d^{(r)}_{a,s},\quad
B^{(r)}_{a,s}=1+b^{(r)}_{a,s},\quad
D^{(r)}_{a,s}=1+d^{(r)}_{a,s},\quad
a=1,\dots,r,\qquad r=1,\dots,k-1,
\end{equation}
where
\begin{equation}
b^{(k-1)}_{a,s}=b_{a,s},\qquad\qquad d^{(k-1)}_{a,s}=d_{a,s}
\end{equation}
and the analyticity strips are

$b^{(2)}_{1,s}$:\ \ $(1-s,s-1)$, \hspace{4.7cm} 
$d^{(2)}_{1,s}$:\ \ $(1-s,s-1)$,

$b^{(2)}_{2,s}$:\ \ $(-s,s-2)$, \hspace{5.1cm} 
$d^{(2)}_{2,s}$:\ \ $(-s,s-2)$,

$b^{(1)}_{1,s}$:\ \ $(1-s,s-3)$, \hspace{4.7cm} 
$d^{(1)}_{1,s}$:\ \ $(1-s,s-3)$.

\noindent
It is easy to verify using the definitions (\ref{bB}) and (\ref{dD}) that
the NLIE functions satisfy the functional equations
\begin{eqnarray}
b_{a,s}d_{a,s}&=&Y_{a,s},\qquad\quad a=1,\dots,k-1,\label{FE1}\\
d^+_{a,s}b_{a+1,s}&=&W_{a,s+1},\qquad\quad a=1,\dots,k-2,\label{FE2}\\
\beta_{a,s}D_{a,s}\delta^+_{a+1,s}B^+_{a+1,s}&=&W_{a,s},
\qquad\quad a=1,\dots,k-2,\label{FE3}\\
B^-_{a,s}D^+_{a,s}\beta_{a+1,s}\delta_{a-1,s}&=&Y_{a,s+1},
\qquad\quad a=1,\dots,k-1,\label{FE4}\\
\nonumber
\end{eqnarray}
where
\begin{equation}
\beta_{a,s}=\frac{b_{a,s}}{B_{a,s}},\qquad a=1,\dots,k-1,\qquad \beta_{k,s}=1
\end{equation}
and
\begin{equation}
\delta_{a,s}=\frac{d_{a,s}}{D_{a,s}},\qquad a=1,\dots,k-1,\qquad 
\delta_{0,s}=1.
\end{equation}
Again, we can write analogous relations for all B\"acklund levels:
\begin{eqnarray}
b^{(r)}_{a,s}d^{(r)}_{a,s}&=&Y^{(r)}_{a,s},
\qquad\quad a=1,\dots,r,\label{FEr1}\\
d^{(r)+}_{a,s}b^{(r)}_{a+1,s}&=&Y^{(r-1)}_{a,s+1},
\qquad\quad a=1,\dots,r-1,\label{FEr2}\\
\beta^{(r)}_{a,s}D^{(r)}_{a,s}\delta^{(r)+}_{a+1,s}B^{(r)+}_{a+1,s}
&=&Y^{(r-1)}_{a,s},
\qquad\quad a=1,\dots,r-1,\label{FEr3}\\
B^{(r)-}_{a,s}D^{(r)+}_{a,s}\beta^{(r)}_{a+1,s}\delta^{(r)}_{a-1,s}
&=&Y^{(r)}_{a,s+1},
\qquad\quad a=1,\dots,r\label{FEr4}\\
\nonumber
\end{eqnarray}
with
\begin{equation}
\beta^{(r)}_{a,s}=\frac{b^{(r)}_{a,s}}{B^{(r)}_{a,s}},
\qquad a=1,\dots,r,\qquad \beta^{(r)}_{r+1,s}=1
\end{equation}
and
\begin{equation}
\delta^{(r)}_{a,s}=\frac{d^{(r)}_{a,s}}{D^{(r)}_{a,s}},
\qquad a=1,\dots,r,\qquad 
\delta^{(r)}_{0,s}=1.
\end{equation}
Using the functional equations (\ref{FEr1}-\ref{FEr4}) we can extend the 
analyticity strips for some NLIE functions. For example, the relation
\begin{equation}
B^{(3)-}_{1,s}D^{(3)+}_{1,s}\beta^{(3)}_{2,s}=Y^{(3)}_{1,s+1}
\end{equation}
should be valid in the strip $(-s,s)$ and allows us to extend the analyticity
strip of $b^{(3)}_{1,s}$ to $(-1-s,s-1)$. Similarly we can make the extensions
to:

$d^{(3)}_{3,s}$:\ \ $(1-s,s+1)$,

$b^{(2)}_{1,s}$:\ \ $(-1-s,s-1)$,

$d^{(2)}_{2,s}$:\ \ $(-s,s)$,

$b^{(1)}_{1,s}$:\ \ $(-1-s,s-3)$,

$d^{(1)}_{1,s}$:\ \ $(1-s,s-1)$.

\noindent
Actually, the NLIE functions $b^{(1)}_{1,2}$ and $d^{(1)}_{1,2}$ had no
range at all before this extension and even more seriously the $s=2$
functional equations (\ref{FEr3}) for $r=2$ and $a=1$ and (\ref{FEr1}) for 
$r=1$ and $a=1$ have no range (even after the extensions). We conclude that
our system of funcional equations is meaningful for $s\geq3$ only.

For completeness, we here summarize the extended analyticity strips of our NLIE
functions:

$b^{(3)}_{1,s}$:\ \ $(-1-s,s-1)$, \hspace{4.7cm} 
$d^{(3)}_{1,s}$:\ \ $(1-s,s+1)$,

$b^{(3)}_{2,s}$:\ \ $(-s,s)$, \hspace{6.0cm} 
$d^{(3)}_{2,s}$:\ \ $(-s,s)$,

$b^{(3)}_{3,s}$:\ \ $(-1-s,s-1)$, \hspace{4.7cm} 
$d^{(3)}_{3,s}$:\ \ $(1-s,s+1)$,

$b^{(2)}_{1,s}$:\ \ $(-1-s,s-1)$, \hspace{4.7cm} 
$d^{(2)}_{1,s}$:\ \ $(1-s,s-1)$,

$b^{(2)}_{2,s}$:\ \ $(-s,s-2)$, \hspace{5.4cm} 
$d^{(2)}_{2,s}$:\ \ $(-s,s)$,

$b^{(1)}_{1,s}$:\ \ $(-1-s,s-3)$, \hspace{4.7cm} 
$d^{(1)}_{1,s}$:\ \ $(1-s,s-1)$.

We will see that the system of functional equations (\ref{FEr1}-\ref{FEr4})
cannot be translated to a closed set of NLIE integral equations. They have
to be completed with further relations, which we will call the 
\lq\lq half-plane'' functional relations (because their building blocks have 
good analyticity properties either in the upper or the lower half
plane). They can also be obtained from the definitions (\ref{bB}) and 
(\ref{dD}) and are of the form 
\begin{equation}
\beta_{1,s}=\frac{T^+_{1,s}}{T_{1,s+1}}\,b^{[-1-s]},\qquad
\delta_{k-1,s}=\frac{T^-_{k-1,s}}{T_{k-1,s+1}}\,c^{[s]}.
\label{halfplane}
\end{equation}
While the NLIE functional equations (\ref{FEr1}-\ref{FEr4}) are written in 
terms of gauge invariant variables this is apparently not the case for
the half-plane functional equations (\ref{halfplane}). We can however 
reformulate them such that they contain explicitly gauge invariant combinations
only. To find such a form, we first have to express the T-functions 
$T_{a,s}$, $a=1,\dots,k-1$ in terms
of the gauge invariant Y-functions and the boundary (factorized) variables
$T_{0,s}$ and $T_{k,s}$. We start by writing the logarithmic derivative of the
Y-T relations (\ref{YT2}) as
\begin{equation}
dl\,T^+_{a,s}+dl\,T^-_{a,s}-\sum_{b=1}^{k-1}\,\Delta_{ab}\,dl\,T_{b,s}
=dl\,Y_{a,s}+\delta_{a1}dl\,T_{0,s}+\delta_{a\,k-1}dl\,T_{k,s},
\label{Delt}
\end{equation}
where
\begin{equation}
\Delta_{ab}=\delta_{a\,b+1}+\delta_{a\,b-1},\qquad\quad
a,b=1,\dots,k-1
\end{equation}
and we introduce the notation
\begin{equation}
dl F(u)=\frac{{\rm d}}{{\rm d}u}\ln F(u)=\frac{F^\prime(u)}{F(u)}
\end{equation}
for any function $F(u)$.

Using Fourier transformation techniques, we can calculate the inverse
of the linear operator appearing on the LHS of the equation (\ref{Delt}).
We will denote this inverse operator\footnote{It will be 
explicitly given in section 7.} by $M_{ab}$ and write the solution as
\begin{equation}
dl\,T_{a,s}=\sum_{b=1}^{k-1}\,M_{ab}\star dl\, Y_{b,s}+M_{a1}\star dl\,T_{0,s}+
M_{a\,k-1}\star dl\,T_{k,s},\qquad a=1,\dots,k-1.
\label{sol}
\end{equation}
(\ref{sol}) in this form is valid only up to source terms, i.e. the 
contribution of pointlike singularities within the analyticity 
(meromorphicity) strips.
Using this result, we can write (again up to source terms)
\begin{equation}
\begin{split}
dl\,\beta_{1,s}&=\sum_{b=1}^{k-1}\,M_{1b}\star dl\,\frac{Y^+_{b,s}}
{Y_{b,s+1}}-M_{11}\star dl\,B^{[-s]}
-M_{1\,k-1}\star dl\,D^{[-s]}+dl\, b^{[-1-s]},\\
dl\,\delta_{k-1,s}&=\sum_{b=1}^{k-1}\,M_{k-1\,b}\star dl\,\frac{Y^-_{b,s}}
{Y_{b,s+1}}-M_{k-1\,1}\star dl\,A^{[s]}
-M_{k-1\,k-1}\star dl\,C^{[s]}+dl\, c^{[s]}.
\end{split}
\end{equation}
Using this result, we now write the \lq\lq half-plane'' equations for all
($r=1,2,3$) levels of our SU(4) problem: 
\begin{equation}
\begin{split}
dl\,\beta^{(r)}_{1,s}&=\sum_{b=1}^r\,M^{(r)}_{1b}\star 
dl\,\frac{Y^{(r)+}_{b,s}}
{Y^{(r)}_{b,s+1}}+\dots,\\
dl\,\delta^{(r)}_{r,s}&=\sum_{b=1}^r\,M^{(r)}_{rb}\star 
dl\,\frac{Y^{(r)-}_{b,s}}
{Y^{(r)}_{b,s+1}}+\dots,
\end{split}
\label{half12}
\end{equation}
where $M^{(r)}_{ab}$ $a,b=1,\dots,r$ is the $r\times r$ matrix kernel at
B\"acklund level $r$ and the dots indicate that the equations are valid up to
source terms and also terms vanishing (after Fourier transformation),
similarly to what is explained later before the equations (\ref{half1})
and (\ref{half2}), for negative and positive frequencies, for the upper
and lower equations, respectively.

\section{Asymptotic solution of the B\"acklund hierarchy}

In this section we calculate the asymptotic solution of the whole B\"acklund 
hierarchy and of the corresponding NLIE variables and lower level Y-functions.
All asymptotic functions are indicated by an upper index $^o$. The asymptotic
hierarchy of B\"acklund T-functions is given by the Bethe Ansatz and we start
with recalling the Bethe Ansatz solution for the SU$(2\vert2)$ fat-hook 
\cite{KSZ07}.

\subsection{Bethe Ansatz solution of the SU$(2\vert2)$ fat-hook}

To construct the hierarchy of T-functions relevant for the upper part of
our AdS/CFT T-system we will use the solution of the analogous problem
for the SU$(2\vert2)$ fat-hook. We will denote this set of T-functions by
\begin{equation}
T^{(k,m)}(a,s,u),\qquad\quad  k=0,1,2,\qquad m=0,1,2.
\end{equation}
(For a general SU$(K\vert M)$ fat-hook one has $k=0,\dots K$, $m=0,\dots,M$.)
The solution will be given \cite{KSZ07} in terms of 9 Q-functions
\begin{equation}
Q^{(k,m)}(u),\qquad\quad  k,m=0,1,2
\end{equation}
listed in appendix \ref{AsySol}.
These are not all independent, they satisfy a number of quadratic 
QQ-relations \cite{KSZ07}. The meaning of these Q-functions is that in terms 
of these the boundary values of the T-functions are given as
\begin{eqnarray}
T^{(k,m)}(0,s,u)&=&Q^{(k,m)[-s]}(u),\qquad\quad -\infty<s<\infty,\label{TQ1}\\
T^{(k,m)}(a,0,u)&=&Q^{(k,m)[a]}(u),\qquad\quad 0\leq a<\infty,\\
T^{(k,m)}(k,s,u)&=&Q^{(k,0)[s+k]}(u)Q^{(0,m)[-s-k]}(u),\qquad
m\leq s<\infty,\\
T^{(k,m)}(a,m,u)&=&(-1)^{m(a-k)}Q^{(k,0)[a+m]}(u)
Q^{(0,m)[-a-m]}(u),\qquad k\leq a<\infty.\label{TQ4}
\end{eqnarray} 
We also introduce the simplified notation
\begin{equation}
T^{(2,2)}(a,s,u)=T(a,s,u),\qquad\quad T^{(2,1)}(a,s,u)=F(a,s,u)
\end{equation}
for the most important members of the hierarchy.

The B\"acklund transformation we need is similar to (but not identical with)
the transformations of section 3:
\begin{equation}
\begin{split}
F(a,s+1,u)T(a-1,s,u)-F(a-1,s,u)&T(a,s+1,u)\\
&+F^+(a,s,u)T^-(a-1,s+1,u)=0,
\end{split}
\label{bl1}
\end{equation}
\begin{equation}
\begin{split}
-F^-(a,s,u)T(a-1,s,u)+F(a-1,s,u)&T^-(a,s,u)\\
&+F(a,s-1,u)T^-(a-1,s+1,u)=0.
\end{split}
\label{bl2}
\end{equation}
Before this Bethe Ansatz solution can be used for our purposes, we have to go 
to the (1,1) gauge, where the T-functions are equal to unity along the left 
and lower boundaries of the fat-hook. Denoting the (1,1) gauge T-functions by 
$\hat T$, the relation between this solution and the asymptotic limit of our 
T-functions (both in the original conventions and in the principal chiral 
model conventions) is given as
\begin{equation}
\hat T(a,s,u)=\frac{T(a,s,u)}{Q^{(2,2)[a-s]}(u)}={\mathbf t}^o_{a,s}(u).
\label{5.10}
\end{equation}
In particular
\begin{eqnarray}
\hat T(0,s,u)&=&{\mathbf t}^o_{0,s}(u)=1,\label{def1}\\
\hat T(a,0,u)&=&{\mathbf t}^o_{a,0}(u)=T^o_{2,a}(u)=1,\label{def2}\\
\hat T(a,1,u)&=&{\mathbf t}^o_{a,1}(u)=T^o_{1,a}(u)=t_a(u),\label{def3}\\
\hat T(a,2,u)&=&{\mathbf t}^o_{a,2}(u)=T^o_{0,a}(u)=
{\cal A}^{o[a]}(u){\cal B}^{o[-a]}(u).\label{def4}
\end{eqnarray}
Here
\begin{equation}
{\cal A}^o=\frac{Q^{(2,0)++}}{Q^{(2,2)--}},\qquad
{\cal B}^o=Q^{(0,2)--}.
\label{5.15}
\end{equation}
Introducing the function
\begin{equation}
E(u)={\rm e}^{\frac{gu\pi}{2}},\qquad\quad E^+=iE,\quad
E^{[2\sigma]}=(-1)^\sigma E
\label{defE}
\end{equation}
and defining
\begin{equation}
\beta^o=E^-Q^{(0,1)},\qquad\quad \gamma^o=E\frac{Q^{(2,1)+}}{Q^{(2,2)-}}
\label{beto}
\end{equation}
the B\"acklund transformation (\ref{bl1}) with $s=0$ can be rewritten as
\begin{equation}
\gamma^{o[a-1]}t_a^+-\gamma^{o[a+1]}t_{a-1}=
{\cal A}^{o[a]}\beta^{o[-a]}
\label{rew1}
\end{equation}
and the B\"acklund transformation (\ref{bl2}) with $s=1$ as
\begin{equation}
\beta^{o[1-a]}t_a-\beta^{o[-1-a]}t^+_{a-1}=
\gamma^{o[a]}{\cal B}^{o[1-a]}.
\label{rew2}
\end{equation}

\subsection{Asymptotic solution of the AdS/CFT T-system}

In this subsection we recall the asymptotic solution of the AdS/CFT
Y-system and T-system, which is given by two (left and right) copies
of the SU$(2\vert2)$ fat-hook. We will use the notations of \cite{BH9}.
Let us denote these two copies by
\begin{equation}
{\mathbf t}^X_{a,s},\qquad s=0,1,2,\qquad X=L,R. 
\end{equation}
Further definitions and relations are:
\begin{eqnarray}
{\mathbf t}^X_{0,s}&=&{\mathbf t}^X_{a,0}=1,\qquad\quad {\mathbf t}^R_{a,1}=
r_a,\qquad {\mathbf t}^L_{a,1}=\ell_a,\\
{\mathbf t}^X_{a,2}&=&{\cal A}^{[a]}_X{\cal B}_X^{[-a]},\qquad a\geq2
\end{eqnarray}
and we also define $\beta_X$, $\gamma_X$ for $X=L,R$ using (\ref{beto}).

The asymptotic solution for the massive nodes on the AdS/CFT Y-system is
given by
\begin{equation}
{\mathbf y}^o_{a,0}=\eta_a{\mathbf t}^R_{a,1}{\mathbf t}^L_{a,1}=
\eta_ar_a\ell_a,
\end{equation}
where the prefactor $\eta_a$ is given in (\ref{B.25}), (\ref{B.26}).
The prefactor satisfies the discrete Laplace equation
\begin{equation}
\eta_a^+\eta_a^-=\eta_{a+1}\eta_{a-1}\qquad(\eta_0=1)
\end{equation}
and actually it can be written in the form 
\begin{equation}
\eta_a=\frac{\psi^{[a]}}{\psi^{[-a]}},
\label{eta}
\end{equation}
where $\psi$ has cuts along the real axis and is exponentially small,
${\cal O}(\varepsilon)$, in the upper half plane and is exponentially
large, ${\cal O}(1/\varepsilon)$, in the lower half plane. Thus the ratio
(\ref{eta}) is ${\cal O}(\varepsilon^2)$ for $-a/g<{\rm Im}u<a/g$.

The asymptotic solution of the complete AdS/CFT T-system is now constructed as
follows. The right half of the upper part is
\begin{eqnarray}
{\mathbf t}^o_{a,2}&=&{\mathbf t}^R_{a,2}={\cal A}^{[a]}_R{\cal B}^{[-a]}_R,\\
{\mathbf t}^o_{a,1}&=&{\mathbf t}^R_{a,1}=r_a,\\
{\mathbf t}^o_{a,0}&=&1
\end{eqnarray}
and using the relation
\begin{equation}
\frac{{\mathbf t}^o_{a,-1}{\mathbf t}^o_{a,1}}
{{\mathbf t}^o_{a+1,0}{\mathbf t}^o_{a-1,0}}={\mathbf y}^o_{a,0}=
\eta_a{\mathbf t}^R_{a,1}{\mathbf t}^L_{a,1}
\end{equation}
we obtain
\begin{equation}
{\mathbf t}^o_{a,-1}=\eta_a{\mathbf t}^L_{a,1}=\eta_a\ell_a
\end{equation}
and similarly we get
\begin{equation}
{\mathbf t}^o_{a,-2}=\eta_a^+\eta_a^-{\mathbf t}^L_{a,2}=
\eta_a^+\eta_a^-{\cal A}^{[a]}_L{\cal B}^{[-a]}_L.
\end{equation}

Comparing this to (\ref{ABCD}) we find
\begin{equation}
{\cal A}^o={\cal A}_R,\qquad
{\cal B}^o={\cal B}_R,\qquad
{\cal C}^o=\psi^+\psi^-{\cal A}_L,\qquad
{\cal D}^o=\frac{{\cal B}_L}{\psi^+\psi^-}
\end{equation}
and the asymptotic limit of the ratios (\ref{B-1}-\ref{C-1}) are given by 
\begin{eqnarray}
A^o&=&\frac{{\cal A}^{o+}}{{\cal A}^{o-}}=\frac{{\cal A}^+_R}{{\cal A}^-_R}
=A_R,\\
B^o&=&\frac{{\cal B}^{o-}}{{\cal B}^{o+}}=\frac{{\cal B}^-_R}{{\cal B}^+_R}
=B_R,\\
C^o&=&\frac{{\cal C}^{o+}}{{\cal C}^{o-}}=\frac{\psi^{++}}{\psi^{--}}
\frac{{\cal A}^+_L}{{\cal A}^-_L}=\eta_2A_L,\\
D^o&=&\frac{{\cal D}^{o-}}{{\cal D}^{o+}}=\frac{\psi^{++}}{\psi^{--}}
\frac{{\cal B}^-_L}{{\cal B}^+_L}=\eta_2B_L.
\end{eqnarray}
All four functions $A^o$, $B^o$, $C^o$, $D^o$ have cuts at $\pm 2i/g$.
For completeness we record that in the principal chiral model conventions
\begin{equation}
T^o_{0,s}={\mathbf t}^R_{s,2},\quad
T^o_{1,s}={\mathbf t}^R_{s,1},\quad
T^o_{2,s}=1,\quad
T^o_{3,s}=\eta_s{\mathbf t}^L_{s,1},\quad
T^o_{4,s}=\eta_s^+\eta_s^-{\mathbf t}^L_{s,2}.
\label{To}
\end{equation}

We now discuss a special case of the gauge transformations defined by
(\ref{gau1}-\ref{gau2}). Let us choose
\begin{equation}
f_1=\frac{1}{f_2},\qquad f_3=f_2^{[-4]},\qquad f_4=\frac{1}{f_3},
\end{equation}
where $f_2$ is the solution of
\begin{equation}
\frac{f_2^+}{f_2^-}=\psi^{++}.
\end{equation}
This means that
\begin{equation}
\lambda_{a,s}=\frac{f_2^{[a-s-4]}f_2^{[s-a]}}
{f_2^{[-s-a]}f_2^{[s+a-4]}}
\end{equation}
and in particular
\begin{equation}
\lambda_{0,s}=\eta^+_s\eta^-_s,\quad
\lambda_{1,s}=\eta_s,\quad
\lambda_{2,s}=1,\quad
\lambda_{3,s}=\frac{1}{\eta_s},\quad
\lambda_{4,s}=\frac{1}{\eta^+_s\eta^-_s}.
\label{lambdaas}
\end{equation}
After this gauge transformation we have
\begin{equation}
\tilde T^o_{0,s}=\eta_s^+\eta_s^-
{\mathbf t}^R_{s,2},\quad
\tilde T^o_{1,s}=\eta_s{\mathbf t}^R_{s,1},\quad
\tilde T^o_{2,s}=1,\quad
\tilde T^o_{3,s}={\mathbf t}^L_{s,1},\quad
\tilde T^o_{4,s}={\mathbf t}^L_{s,2}.
\label{tildeTo}
\end{equation}
This means that before the gauge transformation the components of
$T^o_{a,s}$ behave as

$T^o_{a,s}$:\ ${\cal O}(1,1,1,\varepsilon^2,\varepsilon^4)$
\ for $a=0,1,2,3,4$, respectively,

\noindent
and after the gauge transformation we have

$\tilde T^o_{a,s}$:\ ${\cal O}(\varepsilon^4,\varepsilon^2,1,1,1)$
\ for $a=0,1,2,3,4$, respectively.

Note that we have chosen the gauge transformation (\ref{lambdaas}) such that 
in the new gauge the asymptotic solution is the mirror image (under the 
exchange of left and right) of the original. This implies (among other things)
the above mirror symmetric ${\cal O}(\varepsilon)$ behaviour of the asymptotic
solution in the two gauges.  

Here are the gauge functions corresponding to the lower B\"acklund levels:
\begin{equation}
\omega_{0,s}=\frac{\psi^{[s+1]}\psi^{[s-1]}}
{\psi^{[1-s]}},
\quad
\omega_{1,s}=\psi^{[s]},\quad
\omega_{2,s}=\psi^{[1-s]},\quad
\omega_{3,s}=\frac{\psi^{[2-s]}\psi^{[-s]}}{\psi^{[s]}},
\label{omega}
\end{equation}
\begin{equation}
\nu_{0,s}=\psi^{[s+1]}\psi^{[s-1]},\qquad
\nu_{1,s}=\psi^{[2-s]}\psi^{[s]},\qquad
\nu_{2,s}=\psi^{[3-s]}\psi^{[1-s]},\qquad
\label{nu}
\end{equation}
\begin{equation}
\rho_{0,s}=\psi^{[3-s]}\psi^{[s+1]}\psi^{[s-1]},\qquad
\qquad\qquad
\rho_{1,s}=\psi^{[s]}\psi^{[4-s]}\psi^{[2-s]}.
\label{rho}
\end{equation}
The gauge functions $\omega_{a,s}$, $\nu_{a,s}$, $\rho_{a,s}$
correspond to the gauge transformation of the lower level T-functions
$T^{(r)}_{a,s}$ for $r=3,2,1$, respectively.

So far we have constructed the asymptotic solution of the T-functions
$T^o_{as}=T^{(4)o}_{a,s}$ at the highest level of the B\"acklund
hierarchy. We now proceed to the T-functions at lower levels. We start with
$F^o_{a,s}=T^{(3)o}_{a,s}$. We assume the pattern

$F^o_{a,s}$:\ ${\cal O}(1,1,\varepsilon,\varepsilon^3)$
\ for $a=0,1,2,3$, respectively,

which implies, using the formulae (\ref{omega})

$\tilde F^o_{a,s}$:\ ${\cal O}(\varepsilon^3,\varepsilon,1,1)$
\ for $a=0,1,2,3$, respectively.

\noindent
The explanation of this behaviour is as follows. It is easy to see that the
first two ($a=0,1$) B\"acklund equations are naturally solved by ${\cal O}(1)$
F-functions since the T-functions occurring in them are also ${\cal O}(1)$.
Similarly after gauge transformation the last two ($a=2,3$) F-components
are naturally ${\cal O}(1)$. Together with (\ref{omega}) they already fix
the above pattern. According to this pattern 
the B\"acklund equations take the following asymptotic form:
\begin{eqnarray}
T^o_{1,s+1}F^o_{0,s}&=&T^{o+}_{1,s}F^{o-}_{0,s+1}+
T^o_{0,s}F^o_{1,s+1},\label{asyTF1}\\
T^o_{2,s+1}F^o_{1,s}&=&T^{o+}_{2,s}F^{o-}_{1,s+1},\label{asyTF2}\\
\tilde T^o_{3,s+1}\tilde F^o_{2,s}&=&\tilde T^{o+}_{3,s}\tilde F^{o-}_{2,s+1}
+\tilde T^o_{2,s}\tilde F^o_{3,s+1},\\
\tilde T^o_{4,s+1}\tilde F^o_{3,s}&=&
\tilde T^{o+}_{4,s}\tilde F^{o-}_{3,s+1}\label{asyTF4}
\end{eqnarray}
and
\begin{eqnarray}
T^{o+}_{0,s+1}F^o_{0,s}&=&T^o_{0,s}F^{o+}_{0,s+1},\label{asyTF5}\\
T^{o+}_{1,s+1}F^o_{1,s}&=&T^o_{1,s}F^{o+}_{1,s+1}+T^{o+}_{2,s}F^o_{0,s+1},\\
\tilde T^{o+}_{2,s+1}\tilde F^o_{2,s}&=&
\tilde T^o_{2,s}\tilde F^{o+}_{2,s+1},\label{asyTF7}\\
\tilde T^{o+}_{3,s+1}\tilde F^o_{3,s}&=&\tilde T^o_{3,s}\tilde F^{o+}_{3,s+1}
+\tilde T^{o+}_{4,s}\tilde F^o_{2,s+1}.\label{asyTF8}
\end{eqnarray}
Here some ${\cal O}(\varepsilon)$ terms were omitted from (\ref{asyTF2}) and
(\ref{asyTF7}) and then
all equations are written in terms of ${\cal O}(1)$ variables. Using the
asymptotic solution (\ref{To}) and (\ref{tildeTo}) and also the identities
(\ref{rew1}) and (\ref{rew2}) we find that the solution
of (\ref{asyTF1}-\ref{asyTF4}) and (\ref{asyTF5}-\ref{asyTF8}) is given by
\begin{equation}
F^o_{0,s}={\cal A}^{[s]}_R\beta^{[-s]}_R,\qquad
F^o_{1,s}=\gamma^{[s]}_R,\qquad
\tilde F^o_{2,s}=\beta^{[-s]}_L,\qquad
\tilde F^o_{3,s}={\cal B}^{[1-s]}_L\gamma^{[s]}_L.
\end{equation}
Next we solve the asymptotic equations for $G^o_{a,s}=T^{(2)o}_{a,s}$. We
assume that

$G^o_{a,s}$:\ ${\cal O}(1,\varepsilon,\varepsilon^2)$
and
$\tilde G^o_{a,s}$:\ ${\cal O}(\varepsilon^2,\varepsilon,1)$
\ for $a=0,1,2$, respectively.

\noindent
Using this pattern (and the previous one for $F^o_{a,s}$) we have
\begin{eqnarray}
F^o_{1,s+1}G^o_{0,s}&=&F^{o+}_{1,s}G^{o-}_{0,s+1},\label{asyFG1}\\
\tilde F^o_{2,s+1}\tilde G^o_{1,s}&=&\tilde F^{o+}_{2,s}\tilde G^{o-}_{1,s+1}
+\tilde F^o_{1,s}\tilde G^o_{2,s+1},\\
\tilde F^o_{3,s+1}\tilde G^o_{2,s}&=&\tilde F^{o+}_{3,s}\tilde G^{o-}_{2,s+1}
\label{asyFG3}
\end{eqnarray}
and
\begin{eqnarray}
F^{o+}_{0,s+1}G^o_{0,s}&=&F^o_{0,s}G^{o+}_{0,s+1},\label{asyFG4}\\
F^{o+}_{1,s+1}G^o_{1,s}&=&F^o_{1,s}G^{o+}_{1,s+1}+F^{o+}_{2,s}G^o_{0,s+1},\\
\tilde F^{o+}_{2,s+1}\tilde G^o_{2,s}&=&
\tilde F^o_{2,s}\tilde G^{o+}_{2,s+1}.\label{asyFG6}
\end{eqnarray}
The solution of (\ref{asyFG1}-\ref{asyFG3}) and 
(\ref{asyFG4}-\ref{asyFG6}) is
\begin{equation}
G^o_{0,s}={\cal A}^{[s]}_R,\qquad
G^o_{1,s}=\frac{1}{\psi^{[2-s]}}\gamma^{[s]}_R\beta^{[1-s]}_L\left(
w^{[s]}+y^{[2-s]}\right),\qquad
\tilde G^o_{2,s}={\cal B}^{[2-s]}_L,
\label{Go}
\end{equation}
where the functions $w$ and $y$ are the solutions of
\begin{equation}
w^--w^+=\frac{{\cal A}_R}{\gamma^+_R\gamma^-_R},\qquad\quad
y^+-y^-=\frac{{\cal B}_L}{\beta_L\beta^{--}_L}.
\end{equation}
We note that (\ref{Go}) is actually an exact solution of the SU(2) T-system
for $T^{(2)o}_{a,s}=G^o_{a,s}$. Completed with
\begin{equation}
T^{(1)o}_{0,s}=\frac{{\cal A}^{[s]}_R\beta^{[2-s]}_L}{\psi^{[3-s]}},\qquad
T^{(1)o}_{1,s}=-\frac{{\cal B}^{[3-s]}_L\gamma^{[s]}_R}
{\psi^{[2-s]}\psi^{[4-s]}}
\end{equation}
they form an exact solution of the SU(2) B\"acklund system. The lowest level
solution shows the following pattern.

$T^{(1)o}_{a,s}$:\ ${\cal O}(\varepsilon,\varepsilon^2)$
and
$\tilde T^{(1)o}_{a,s}$:\ ${\cal O}(\varepsilon^2,\varepsilon)$
\ for $a=0,1$, respectively.

\subsection{Asymptotic solution for Y-functions and NLIE variables}

For completeness we here summarize the asymptotic solution for the gauge
invariant objects: Y-functions and NLIE variables for all B\"acklund
levels.
\begin{eqnarray}
y^{(3)o}_{1,s}&=&\frac{r_{s+1}r_{s-1}}{{\cal A}^{[s]}_R{\cal B}^{[-s]}_R},
\qquad\quad
Y^{(3)o}_{1,s}=\frac{r^+_sr^-_s}{{\cal A}^{[s]}_R{\cal B}^{[-s]}_R},\\
y^{(3)o}_{2,s}&=&Y^{(3)o}_{2,s}=\frac{1}{\eta_sr_s\ell_s},\\
y^{(3)o}_{3,s}&=&\frac{\ell_{s+1}\ell_{s-1}}
{{\cal A}^{[s]}_L{\cal B}^{[-s]}_L},
\qquad\quad
Y^{(3)o}_{3,s}=\frac{\ell^+_s\ell^-_s}{{\cal A}^{[s]}_L{\cal B}^{[-s]}_L},\\
y^{(2)o}_{1,s}&=&Y^{(2)o}_{1,s}=\left(\frac{\psi^+}
{\beta_L\beta_R}\right)^{[-s]}\frac{1}{w^{[s-1]}-w^{[s+1]}},\\
y^{(2)o}_{2,s}&=&Y^{(2)o}_{2,s}=\frac{1}{(\psi\gamma_L\gamma_R)^{[s]}}
\frac{1}{y^{[2-s]}-y^{[-s]}},\\
y^{(1)o}_{1,s}&=&\frac{
\left(w^{[s+1]}+y^{[1-s]}\right)
\left(w^{[s-1]}+y^{[3-s]}\right)}
{\left(w^{[s-1]}-w^{[s+1]}\right)
\left(y^{[3-s]}-y^{[1-s]}\right)},\\
Y^{(1)o}_{1,s}&=&\frac{
\left(w^{[s+1]}+y^{[3-s]}\right)
\left(w^{[s-1]}+y^{[1-s]}\right)}
{\left(w^{[s-1]}-w^{[s+1]}\right)
\left(y^{[3-s]}-y^{[1-s]}\right)}.
\end{eqnarray}

\begin{eqnarray}
b^{(3)o}_{1,s}&=&r^+_s\frac{\beta^{[-2-s]}_R}
{{\cal B}^{[-s]}_R\gamma^{[s+1]}_R},\qquad
B^{(3)o}_{1,s}=r_{s+1}\frac{\beta^{[-s]}_R}
{{\cal B}^{[-s]}_R\gamma^{[s+1]}_R},\\
b^{(3)o}_{2,s}&=&B^{(3)o}_{2,s}=\psi^{[-s]}\frac{\gamma^{[s]}_R}
{r_s\beta^{[-1-s]}_L},\\
b^{(3)o}_{3,s}&=&\ell^+_s\frac{\beta^{[-2-s]}_L}
{{\cal B}^{[-s]}_L\gamma^{[s+1]}_L},\qquad
B^{(3)o}_{3,s}=\ell_{s+1}\frac{\beta^{[-s]}_L}
{{\cal B}^{[-s]}_L\gamma^{[s+1]}_L},\\
d^{(3)o}_{1,s}&=&r^-_s\frac{\gamma^{[s+1]}_R}
{{\cal A}^{[s]}_R\beta^{[-2-s]}_R},\qquad
D^{(3)o}_{1,s}=r_{s+1}\frac{\gamma^{[s-1]}_R}
{{\cal A}^{[s]}_R\beta^{[-2-s]}_R},\\
d^{(3)o}_{2,s}&=&D^{(3)o}_{2,s}=\frac{1}{\psi^{[s]}}\frac{\beta^{[-1-s]}_L}
{\ell_s \gamma^{[s]}_R},\\
d^{(3)o}_{3,s}&=&\ell^-_s\frac{\gamma^{[s+1]}_L}
{{\cal A}^{[s]}_L\beta^{[-2-s]}_L},\qquad
D^{(3)o}_{3,s}=\ell_{s+1}\frac{\gamma^{[s-1]}_L}
{{\cal A}^{[s]}_L\beta^{[-2-s]}_L}.
\end{eqnarray}

\begin{eqnarray}
b^{(2)o}_{1,s}&=&B^{(2)o}_{1,s}=\left(\frac{\psi^+}
{\beta_R\beta_L}\right)^{[-s]}\frac{1}{w^{[s+1]}+y^{[1-s]}},\\
b^{(2)o}_{2,s}&=&\frac{w^{[s]}+y^{[-s]}}{y^{[2-s]}-y^{[-s]}},\qquad\quad
\ \ B^{(2)o}_{2,s}=\frac{w^{[s]}+y^{[2-s]}}{y^{[2-s]}-y^{[-s]}},\\
d^{(2)o}_{1,s}&=&\frac{w^{[s+1]}+y^{[1-s]}}{w^{[s-1]}-w^{[s+1]}},\qquad\quad
D^{(2)o}_{1,s}=\frac{w^{[s-1]}+y^{[1-s]}}{w^{[s-1]}-w^{[s+1]}},\\
d^{(2)o}_{2,s}&=&D^{(2)o}_{2,s}=\frac{1}
{(\psi\gamma_R\gamma_L)^{[s]}}\,\frac{1}{w^{[s]}+y^{[-s]}}.
\end{eqnarray}

\begin{eqnarray}
b^{(1)o}_{1,s}&=&\frac{w^{[s+1]}+y^{[3-s]}}{y^{[1-s]}-y^{[3-s]}},
\qquad\qquad
B^{(1)o}_{1,s}=\frac{w^{[s+1]}+y^{[1-s]}}{y^{[1-s]}-y^{[3-s]}},\\
d^{(1)o}_{1,s}&=&\frac{w^{[s-1]}+y^{[1-s]}}{w^{[s+1]}-w^{[s-1]}},
\qquad\qquad
D^{(1)o}_{1,s}=\frac{w^{[s+1]}+y^{[1-s]}}{w^{[s+1]}-w^{[s-1]}}.
\end{eqnarray}

\subsection{Identification of $Q^{(k,m)}$}

We will specify our building blocks $Q^{(k,m)}$ completely in two cases.
We first discuss the case of a general (excited) state in the original, 
non-deformed model and later in a separete sub-subsection for the ground state
of the $\gamma$-deformed model. 

The building blocks, in terms of which the asymptotic solutions are expressed 
are $r_s$, ${\cal A}_R$, ${\cal B}_R$, $\beta_R$, $\gamma_R$, $w$, $\ell_s$,
${\cal A}_L$, ${\cal B}_L$, $\beta_L$, $\gamma_L$ and $y$. For each of the
Bethe Ansatz solutions (either $R$ or $L$) we thus need to know $t_s$, 
${\cal A}^o$, ${\cal B}^o$, $\beta^o$ and $\gamma^o$. $w^o$ and $y^o$
can then be determined from the relations
\begin{equation}
w^{o-}-w^{o+}=\frac{{\cal A}^o}{\gamma^{o+}\gamma^{o-}},\qquad\qquad
y^{o+}-y^{o-}=\frac{{\cal B}^o}{\beta^o\beta^{o--}}.
\label{wyo}
\end{equation}
In terms of the 9 Q-functions $Q^{(k,m)}$, $k,m=0,1,2$, the building blocks
are expressed as
\begin{equation}
{\cal A}^o=\frac{Q^{(2,0)++}}{Q^{(2,2)--}},\qquad
{\cal B}^o=Q^{(0,2)--},\qquad \beta^o=E^-Q^{(0,1)},\qquad
\gamma^o=E\,\frac{Q^{(2,1)+}}{Q^{(2,2)-}},
\end{equation}
where $E$ was defined in (\ref{defE}).
It is also useful to know the building blocks as expressed in terms of
the Bethe Ansatz solution of \cite{GKV09} 
(see appendix B and \cite{BH9}):
\begin{eqnarray}
{\cal A}^o&=&-{\cal F}^{(0)}\,\frac{R^-_m}{R^-_p}\,\frac{1}{Q^{--}}\,
\frac{Q^-_1}{Q^-_3},\qquad\quad \beta^o=E^-Q^+_3,\\
{\cal B}^o&=&-{\cal G}^{(0)}\,\frac{R^+_m}{R^+_p}\,Q^{++}\,
\frac{Q^+_3}{Q^+_1},\qquad\quad \ \gamma^o=E\,\frac{Q_1}{Q^-}\,\frac{R_m}{R_p}.
\end{eqnarray}
The expression of the T-system elements $t_s$ in terms of the same Bethe
Ansatz functions is also given in appendix B.

The combination necessary to calculate the function $w$ becomes
\begin{equation}
\frac{{\cal A}^o}{\gamma^{o+}\gamma^{o-}}=-\frac{Q}{E^2}\left\{
\frac{Q^{++}_2}{Q_2(Q_1Q_3)^+}+
\frac{Q^{--}_2}{Q_2(Q_1Q_3)^-}-
\frac{R^-_m}{R^-_p(Q_1Q_3)^-}-
\frac{R^+_p}{R^+_m(Q_1Q_3)^+}\right\},
\end{equation}
which, for the case of an even number of particles in the
 ${\, \alg{sl}(2) \,}$ sector, simplifies to
\begin{equation}
\frac{{\cal A}^o}{\gamma^{o+}\gamma^{o-}}=-\frac{1}{E^2}\left\{
2Q-R^-_mB^-_p-R^+_pB^+_m\right\}.
\label{sl2A}
\end{equation}
Similarly we have
\begin{equation}
\frac{{\cal B}^o}{\beta^o\beta^{o--}}=-\frac{Q}{E^2}\left\{
\frac{Q^{++}_2}{Q_2(Q_1Q_3)^+}+
\frac{Q^{--}_2}{Q_2(Q_1Q_3)^-}-
\frac{B^-_m}{B^-_p(Q_1Q_3)^-}-
\frac{B^+_p}{B^+_m(Q_1Q_3)^+}\right\},
\end{equation}
and for the case of an even number of particles in the  ${\, \alg{sl}(2) \,}$ 
sector 
\begin{equation}
\frac{{\cal B}^o}{\beta^o\beta^{o--}}=-\frac{1}{E^2}\left\{
2Q-R^-_pB^-_m-R^+_mB^+_p\right\}.
\label{sl2B}
\end{equation}

Using (\ref{sl2A}) and (\ref{sl2B}) for the case of the Konishi operator
corresponding to two particles with rapidities $u_1=-u_2=\omega$ and
\begin{equation}
x_1^+=x_s\left(\omega+\frac{i}{g}\right)=\xi,\qquad
x_2^+=-\xi^*,\qquad x_1^-=\xi^*,\qquad x^-_2=-\xi
\end{equation}
we can explicitly solve (\ref{wyo}): 
\begin{eqnarray}
w^o(u)&=&\frac{1}{E^2}\left\{(h-h^*)\left(x-\frac{1}{x}\right)
+iguH+w_c\right\}=w(u),\\
y^o(u)&=&\frac{1}{E^2}\left\{(h-h^*)\left(x-\frac{1}{x}\right)
-iguH+y_c\right\}=y(u),
\end{eqnarray}
where
\begin{equation}
h=\xi-\frac{1}{\xi},\qquad\qquad
H=4+hh^*+\frac{1}{g^2}-\omega^2
\end{equation}
and $w_c$, $y_c$ are \lq\lq integration'' constants. (Only their sum enters the
asymptotic expressions.)

\subsubsection{$\gamma$-deformed constant solution}

We also give here the identification of the Q-functions in the case of the
constant (vacuum) solution for the $\gamma$-deformed model. Following 
ref. \cite{TBA2ndLusch}, the right-wing part of the problem
is charaterized by the constant $q$ and the analogous left-wing part by the
constant $\dot q$. To distinguish it from the asymptotic solution discussed 
above (valid for arbitrary excited state in the undeformed
model) the Q-functions for the deformed vacuum will be denoted by lower case
$q^{(k,m)}$. For the right-wing case we have
\begin{equation}
\begin{split}
q^{(2,2)}&=1,\\
q^{(1,2)}&=S_o,\\
q^{(0,2)}&=\left(1-\frac{1}{q}\right)^2,
\end{split}
\qquad\qquad
\begin{split}
q^{(2,1)}&=1-q,\\ 
q^{(1,1)}&=S_o,\\
q^{(0,1)}&=\left(1-\frac{1}{q}\right),
\end{split}
\qquad\qquad
\begin{split}
q^{(2,0)}&=(1-q)^2,\\
q^{(1,0)}&=S_o,\\
q^{(0,0)}&=1,
\end{split}
\label{constant}
\end{equation}
where $S_o$ is the solution of
\begin{equation}
S_o^+=qS^-_o.
\end{equation}
The solution (\ref{constant}) can be obtained from the results of 
ref.~\cite{Z36}. In this case the building blocks are
\begin{equation}
t_s=(-1)^{s+1}s\frac{(1-q)^2}{q}
\end{equation}
and
\begin{equation}
{\cal A}^o=(1-q)^2,\qquad {\cal B}^o=\left(1-\frac{1}{q}\right)^2,\qquad
\beta^o=E^-\left(1-\frac{1}{q}\right),\qquad \gamma^o=E(1-q).
\end{equation}
It is easy to solve (\ref{wyo}):
\begin{equation}
w^o(u)=-\frac{ig}{2E^2}(u+w_c),\qquad\qquad y^o(u)=\frac{ig}{2E^2}(u-y_c).
\end{equation}

The asymptotic solution of the gauge invariant functions is as follows.
\begin{eqnarray}
y^{(3)o}_{1,s}&=&y^{(3)o}_{3,s}=s^2-1,\qquad\quad
Y^{(3)o}_{1,s}=Y^{(3)o}_{3,s}=s^2,\\
y^{(3)o}_{2,s}&=&Y^{(3)o}_{2,s}=\frac{1}{\eta_s}\frac{1}{s^2}
\frac{q\dot q}{(1-q)^2(1-\dot q)^2},\\
y^{(2)o}_{1,s}&=&Y^{(2)o}_{1,s}=-\psi^{[1-s]}
\,\frac{q\dot q}{(1-q)(1-\dot q)},\\
y^{(2)o}_{2,s}&=&Y^{(2)o}_{2,s}=-\frac{1}{\psi^{[s]}}
\,\frac{1}{(1-q)(1-\dot q)},\\
y^{(1)o}_{1,s}&=&(s-\Delta)(s-2-\Delta),\qquad\qquad
Y^{(1)o}_{1,s}=(s-1-\Delta)^2,
\end{eqnarray}
where the constant $\Delta$ is a multiple of the sum of the arbitrary
constants $w_c$ and $y_c$.
\begin{eqnarray}
b^{(3)o}_{1,s}&=&b^{(3)o}_{3,s}=d^{(3)o}_{1,s}=d^{(3)o}_{3,s}=s,\\
B^{(3)o}_{1,s}&=&B^{(3)o}_{3,s}=D^{(3)o}_{1,s}=D^{(3)o}_{3,s}=s+1,\\
b^{(3)o}_{2,s}&=&B^{(3)o}_{2,s}=-\psi^{[-s]}\,
\frac{q\dot q}{s(1-q)(1-\dot q)},\\
d^{(3)o}_{2,s}&=&D^{(3)o}_{2,s}=-\frac{1}{\psi^{[s]}}\,
\frac{1}{s(1-q)(1-\dot q)},\\
b^{(2)o}_{1,s}&=&B^{(2)o}_{1,s}=\frac{\psi^{[1-s]}}{s-\Delta}\,
\frac{q\dot q}{(1-q)(1-\dot q)},\\
b^{(2)o}_{2,s}&=&d^{(2)o}_{1,s}=\Delta-s,\qquad\qquad
B^{(2)o}_{2,s}=D^{(2)o}_{1,s}=\Delta+1-s,\\
d^{(2)o}_{2,s}&=&D^{(2)o}_{2,s}=\frac{1}{\psi^{[s]}}\,
\frac{1}{s-\Delta}\,
\frac{1}{(1-q)(1-\dot q)},\\
b^{(1)o}_{1,s}&=&d^{(1)o}_{1,s}=s-1-\Delta,\qquad\qquad
B^{(1)o}_{1,s}=D^{(1)o}_{1,s}=s-\Delta.
\end{eqnarray}

\section{The horizontal SU$(2)$ problem}

In this section we study the problem of finding the NLIE description of the 
TBA system corresponding to the right-wing nodes. (The problem of the 
left-wing nodes is completely analogous.) Our discussion of the B\"acklund
transformation, the construction of the NLIE variables and the NLIE functional
equations and the asymptotic solution of these variables will be very 
similar to, but considerably simpler than, what has been discussed in the 
preceeding three sections. The construction here is based on 
\cite{Suz98}, \cite{nlie25} and \cite{RyoHybrid}. 

We introduce the notation  
\begin{equation}
Y^{(+)}_{m\vert w}={\mathbf y}_{1,m+1}=x_m,\qquad X_m=1+x_m.
\end{equation}
These functions are of type $(-m,m)$ and satisfy the SU(2) Y-system equations
\begin{equation}
x_m^+x_m^-=X_{m+1}X_{m-1},\qquad\quad m=2,3,\dots.
\end{equation}
It is easy to construct the corresponding T-system as follows. It will turn
out to be convenient to work in a gauge different from what has been used
so far. In this gauge we start with
the construction of the first two T-functions $\tau_1$, $\tau_2$ from
\begin{equation}
\tau_m^+\tau_m^-=X_m,\qquad m=1,2.
\end{equation}
The solution of this type of functional equations is the basic problem in 
the theory of TBA integral equations \cite{KPtanh}. (See also the TBA lemmas
of \cite{BH9}.) Next we define 
\begin{equation}
\tau_3=\frac{x_2}{\tau_1}
\end{equation}
satisfying
\begin{equation}
\tau_3^+\tau_3^-=\frac{x_2^+x_2^-}{\tau_1^+\tau_1^-}=\frac{X_1X_3}{X_1}=X_3.
\end{equation}
Proceeding similarly we can construct $\tau_m$ for all $m=1,2,\dots$ satisfying
\begin{equation}
\tau_{m+1}\tau_{m-1}=x_m,\qquad m=2,3,\dots,\qquad\qquad
\tau_m^+\tau_m^-=X_m,\qquad m=1,2,\dots
\end{equation}
and also the T-system equations of the form
\begin{equation}
\tau^+_m\tau^-_m=1+\tau_{m+1}\tau_{m-1},\qquad m=2,3,\dots
\label{tau}
\end{equation}
The $\tau_m$ functions constructed this way are of type $(-1-m,m+1)$.
The SU(2) T-system (\ref{tau}) is in the gauge
\begin{equation}
t_{2,m}=t_{0,m}=1,\qquad t_{1,m}=\tau_m.
\end{equation}

Next we consider the SU(2) B\"acklund transformations. Specifying (\ref{BT1})
for $a=0,1$ we get the two equations 
\begin{equation}
\tau_{m+1}f_{0,m}=\tau_m^+f^-_{0,m+1}+f_{1,m+1},\qquad
f_{1,m}=f^-_{1,m+1}.
\label{bt1}
\end{equation}
The second one implies that there exists some function $Q$ such that 
\begin{equation}
f_{1,m}=Q^{[m+1]}.
\end{equation}
Similarly from
\begin{equation}
f_{0,m}=f^+_{0,m+1},\qquad
\tau^+_{m+1}f_{1,m}=\tau_mf^+_{1,m+1}+f_{0,m+1},
\label{bt2}
\end{equation}
which are the $a=0,1$ components of (\ref{BT2}), it follows that with some
function $\widetilde Q$
\begin{equation}
f_{0,m}=\widetilde Q^{[-m]}. 
\end{equation}
Rewriting (\ref{bt1}) and (\ref{bt2}) in terms of $Q$ and $\widetilde Q$, we 
have
\begin{eqnarray}
\tau_{m+1}\widetilde Q^{[-m]}&=&\tau^+_m\widetilde Q^{[-m-2]}+Q^{[m+2]},
\label{Baxter1}\\
\tau^+_{m+1}Q^{[m+1]}&=&\tau_m Q^{[m+3]}+\widetilde Q^{[-m-1]},\label{Baxter2}
\end{eqnarray}
which form a variant of Baxter's famous TQ-relations. From these equations
we find that $Q$ is of type $(0,2m+2)$ and $\widetilde Q$ is of type
 $(-2m-2,0)$.
Since $m$ is arbitrary, $Q$ is free of cuts in the entire upper half plane
and $\widetilde Q$ is free of cuts in the lower half plane.

Having found the solution of the B\"acklund system we can construct the
NLIE variables in analogy to (\ref{bB}) and (\ref{dD}):
\begin{equation}
\begin{split}
b_m&=\frac{\tau_m^+\widetilde Q^{[-m-2]}}{Q^{[m+2]}},\\
B_m&=\frac{\tau_{m+1}\widetilde Q^{[-m]}}{Q^{[m+2]}},
\end{split}
\qquad\quad
\begin{split}
d_m&=\frac{\tau_m^-Q^{[m+2]}}{\widetilde Q^{[-m-2]}},\\
D_m&=\frac{\tau_{m+1} Q^{[m]}}{\widetilde Q^{[-m-2]}}.
\end{split}
\label{bBdD}
\end{equation}
Baxter's equations are equivalent to the relations
\begin{equation}
B_m=1+b_m,\qquad\qquad D_m=1+d_m
\end{equation}
and the analogues of the NLIE functional equations (\ref{FE1}), (\ref{FE4}) are
\begin{equation}
b_md_m=X_m,\qquad\qquad B_m^-D_m^+=X_{m+1}.
\label{fe}
\end{equation}
The NLIE functions $b_m$ and $d_m$ are of type $(-m-2,m)$ and $(-m,m+2)$,
respectively.

We have \lq\lq half-plane'' relations in this case as well. We write the ratios
\begin{equation}
\beta_m=\frac{b_m}{B_m}=\frac{\tau_m^+}{\tau_{m+1}}\,b_o^{[-m-1]}
\end{equation}
and
\begin{equation}
\delta_m=\frac{d_m}{D_m}=\frac{\tau_m^-}{\tau_{m+1}}\,c_o^{[m]}
\end{equation}
and since
\begin{equation}
b_o=\frac{\widetilde Q^-}{\widetilde Q^+}\qquad\quad{\rm and}\qquad\quad
c_o=\frac{Q^{++}}{Q}
\end{equation}
are of type $(-\infty,-1)$ and $(0,\infty)$, respectively, the second terms
on the right hand side of the first and second equations
\begin{eqnarray}
dl\,\frac{b_m}{B_m}&=&M^{(1)}_{11}\star dl\,\frac{X^+_m}{X_{m+1}}+dl\,
b_o^{[-m-1]},\label{half1}\\
dl\,\frac{d_m}{D_m}&=&M^{(1)}_{11}\star dl\,\frac{X^-_m}{X_{m+1}}+dl\,
c_o^{[m]}\label{half2}
\end{eqnarray}
do not contribute (after Fourier transformation) for negative and
positive frequencies, respectively. $M^{(1)}_{11}$ was defined in section~4
and its Fourier transform is given by (\ref{tildes}).

Next we discuss the asymptotic solution for the NLIE functions introduced
above. We go back to the Bethe Ansatz solution discussed in the previous
section and introduce the notation 
\begin{equation}
T^{(2,2)}(a,s,u)=t(a,s,u),\qquad T^{(1,2)}(a,s,u)=f(a,s,u).
\end{equation}
In this section we will make use the set of B\"acklund transformations 
\cite{KSZ07} (similar to (\ref{BT1}-\ref{BT2}))
\begin{equation}
\begin{split}
t(a+1,s,u)f^+(a,s,u)-t^+(a,s,u)&f(a+1,s,u)\\
&-t^+(a+1,s-1,u)f(a,s+1,u)=0,
\end{split}
\label{bak1}
\end{equation}
\begin{equation}
\begin{split}
t^+(a,s+1,u)f(a,s,u)-t(a,s,u)&f^+(a,s+1,u)\\
&-t^+(a+1,s,u)f(a-1,s+1,u)=0.
\end{split}
\label{bak2}
\end{equation}
Using (\ref{bak1}) for $a=0$ and (\ref{bak2}) for $a=1$ and the boundary
relations
\begin{eqnarray}
t(0,m,u)&=&Q^{(2,2)[-m]}(u),\\
f(0,m,u)&=&Q^{(1,2)[-m]}(u),\\
t(2,m,u)&=&Q^{(2,0)[m+2]}(u)\,Q^{(0,2)[-m-2]}(u),\\
f(1,m,u)&=&Q^{(1,0)[m+1]}(u)\,Q^{(0,2)[-m-1]}(u)
\end{eqnarray}
we can identify the asymptotic solution of the building blocks $\tau^o_m$,
$Q^o$ and $\widetilde Q^o$:
\begin{equation}
Q^o=\frac{Q^{(1,0)+}}{k_2},\qquad\quad
\widetilde Q^o=k_1^{--}\frac{Q^{(1,2)-}}{Q^{(0,2)---}},
\end{equation}
\begin{equation}
\tau^o_m=\frac{t(1,m+1,u)}{k_2^{[m+1]}k_1^{[-m-1]}Q^{(2,2)[-m]}}.
\end{equation}
Here and below in this section the argument of all functions is $u$. 
The factors $k_1$ and $k_2$ are the solutions of the relations
\begin{equation}
k_1^+k_1^-=\frac{Q^{(0,2)--}}{Q^{(2,2)++}},\qquad\quad
k_2^+k_2^-=Q^{(2,0)++}.
\end{equation}

We now write down the asymptotic form of the NLIE functions. In these
formulas we use the asymptotic T-functions in the (1,1) gauge used
in the previous section.
\begin{eqnarray}
b^o_m&=&\frac{\hat T^+(1,m+1,u)Q^{(2,2)[1-m]}Q^{(1,2)[-3-m]}}
{Q^{(1,0)[m+3]}Q^{(0,2)[-3-m]}Q^{(2,2)[-1-m]}},\\
B^o_m&=&\frac{\hat T(1,m+2,u)Q^{(1,2)[-1-m]}}
{Q^{(1,0)[m+3]}Q^{(0,2)[-3-m]}},\\
d^o_m&=&\frac{\hat T^-(1,m+1,u)Q^{(2,2)[-1-m]}Q^{(1,0)[m+3]}}
{Q^{(2,0)[m+3]}Q^{(1,2)[-3-m]}},\\
D^o_m&=&\frac{\hat T(1,m+2,u)Q^{(2,2)[-1-m]}Q^{(1,0)[m+1]}}
{Q^{(2,0)[m+3]}Q^{(1,2)[-3-m]}}.
\end{eqnarray}

We know that the right-wing part of the asymptotic solution, in particular
the T-functions $\hat T(1,s,u)={\mathbf t}^o_{1,s}(u)$ in the (1,1) gauge we
are using are stained with several cuts close to the real axis. Most of these
cuts are spurious and are not present in the gauge invariant Y-functions. This
is also the case for the gauge invariant NLIE functions (\ref{bBdD}). We can 
see this by considering the \lq\lq cut-free'' representation of the building
blocks. Using the definition 
\begin{equation}
\frac{R_p}{R_m}=\frac{\Omega^+}{\Omega^-}
\end{equation}
we can write
\begin{equation}
\hat T(2,s,u)=\tilde t^2_s\left(\frac{\Omega^{[2-s]}}
{\Omega^{[s]}}\right)^2,
\end{equation}
where $\tilde t^2_s$ has cuts only at $\pm\frac{i}{g}(s\pm1)$ and
\begin{equation}
\hat T(1,s,u)=\tilde t^1_s\,\frac{\Omega^{[1-s]}}
{\Omega^{[s-1]}},
\end{equation}
where $\tilde t^1_s$ has cuts only at $\pm\frac{is}{g}$.
The Y-functions are given by
\begin{equation}
x^o_m=\frac{\hat T(1,m+2,u)\hat T(1,m,u)}{\hat T(2,m+1,u)}=
\frac{\tilde t^1_{m+2}\tilde t^1_m}{\tilde t^2_{m+1}}
\left(\frac{R_p}{R_m}\right)^{[m]}\left(\frac{R_m}{R_p}\right)^{[-m]},
\end{equation}
which has cuts only at $\pm\frac{im}{g}$ and $\pm(m+2)\frac{i}{g}$.
Similarly we have
\begin{equation}
X^o_m=\frac{\hat T^+(1,m+1,u)\hat T^-(1,m+1,u)}{\hat T(2,m+1,u)}=
\frac{\tilde t^{1+}_{m+1}\tilde t^{1-}_{m+1}}{\tilde t^2_{m+1}}
\left(\frac{R_p}{R_m}\right)^{[m]}\left(\frac{R_m}{R_p}\right)^{[-m]}.
\end{equation}
We further define the functions $\widetilde Q^{(k,m)}$, which have better
analytic behaviour than the corresponding $Q^{(k,m)}$. We write
\begin{eqnarray}
Q^{(2,2)}&=&\frac{\widetilde Q^{(2,2)}}{\Omega^2},\qquad\quad 
\widetilde Q^{(2,2)}\ {\rm has\ no\ cuts\   },\\
Q^{(1,2)}&=&\frac{\widetilde Q^{(1,2)}}{\Omega^{++}},\qquad\quad 
\widetilde Q^{(1,2)}\ {\rm has\ cuts\ only\ at\ }\frac{-i}{g},\\
Q^{(1,0)}&=&\frac{\widetilde Q^{(1,0)}}{\Omega^{--}},\qquad\quad 
\widetilde Q^{(1,0)}\ {\rm has\ cuts\ only\ at\ }\frac{i}{g},\\
Q^{(0,2)}&=&\widetilde Q^{(0,2)},\qquad\quad 
\widetilde Q^{(0,2)}\ {\rm has\ cuts\ only\ at\ }\frac{i}{g},\frac{3i}{g},\\
Q^{(2,0)}&=&\frac{\widetilde Q^{(2,0)}}{\Omega^2},\qquad\quad 
\widetilde Q^{(2,0)}\ {\rm has\ cuts\ only\ at\ }\frac{-i}{g},\frac{-3i}{g}.
\end{eqnarray}

Finally the \lq\lq cut-free'' representation of the NLIE functions is as 
follows.
\begin{eqnarray}
b^o_m&=&\left(\frac{R_m}{R_p}\right)^{[-m]}\frac{\tilde t^{1+}_{m+1}
\widetilde Q^{(2,2)[1-m]}\widetilde Q^{(1,2)[-3-m]}}
{\widetilde Q^{(1,0)[m+3]}\widetilde Q^{(0,2)[-3-m]}
\widetilde Q^{(2,2)[-1-m]}},\\
B^o_m&=&\left(\frac{R_m}{R_p}\right)^{[-m]}\frac{\tilde t^1_{m+2}
\widetilde Q^{(1,2)[-1-m]}}
{\widetilde Q^{(1,0)[m+3]}\widetilde Q^{(0,2)[-3-m]}}
\end{eqnarray}
and
\begin{eqnarray}
d^o_m&=&\left\{\left(\frac{R_p}{R_m}\right)^{[m+2]}\right\}^2
\left(\frac{R_p}{R_m}\right)^{[m]}\frac{\tilde t^{1-}_{m+1}
\widetilde Q^{(2,2)[-1-m]}\widetilde Q^{(1,0)[m+3]}}
{\widetilde Q^{(2,0)[m+3]}\widetilde Q^{(1,2)[-3-m]}},\\
D^o_m&=&\left\{\left(\frac{R_p}{R_m}\right)^{[m+2]}\right\}^2
\left(\frac{R_p}{R_m}\right)^{[m]}\frac{\tilde t^1_{m+2}
\widetilde Q^{(2,2)[-1-m]}\widetilde Q^{(1,0)[m+1]}}
{\widetilde Q^{(2,0)[m+3]}\widetilde Q^{(1,2)[-3-m]}}.
\end{eqnarray}

We end this section by giving the asymptotic solution corresponding
to the deformed ground state using the building blocks (\ref{constant}).
The T-functions in the (1,1) gauge are
\begin{equation}
\hat T(1,s,u)=\frac{1-q}{1+q}\,q^{-s}(1-q^{2s}),\qquad\quad
\hat T(2,s,u)=\frac{(1-q)^4}{q^2}.
\end{equation}
Further we have
\begin{equation}
\tau^o_m=\frac{q^{-m}}{1-q^2}(1-q^{2m+2}),\qquad\qquad
Q^o=\widetilde Q^o=\frac{S_o^+}{1-q}
\end{equation}
and
\begin{equation}
\begin{split}
b^o_m&=\frac{q^{-2m-2}-1}{1-q^2},\\
d^o_m&=\frac{q^2-q^{2m+4}}{1-q^2},
\end{split}
\qquad\quad
\begin{split}
B^o_m&=\frac{q^{-2m-2}-q^2}{1-q^2},\\
D^o_m&=\frac{1-q^{2m+4}}{1-q^2}.
\end{split}
\label{6.54}
\end{equation}

\section{NLIE for the ground state}

In this section we will obtain the NLIE integral equations and thus reduce the
system of integral equations to a finite set. We will transform the functional
equations (\ref{FEr1}-\ref{FEr4}) into integral equations which together with 
(\ref{halfplane}) form a complete set of NLIE integral equations equivalent to
the TBA integral equations for the ($a > 3$ part of the) upper nodes. But we 
start by considering the analogous but much simpler (and already solved) 
problem corresponding to the right-wing (and left-wing) nodes. We will see
that the two constructions proceed along very similar lines.  

In this paper we will consider the NLIE integral equations only up to
source terms, i.e. the NLIE for the ground state problem. Although the
addition of source terms is in principle straightforward, we leave the
elaboration of the excited state problem to future work.

Before we start the construction we fix some notations and conventions.
We will denote the Fourier transform of the function $f(u)$ by 
$\tilde f(\omega)$ and use the definition
\begin{equation}
\tilde f(\omega)=\int_{-\infty}^\infty{\rm d}u\,{\rm e}^{i\omega u}f(u).
\end{equation}
In particular, the Fourier transform of a logarithmic derivative will be 
denoted by
\begin{equation}
\widetilde{dl}f(\omega)=\int_{-\infty}^\infty{\rm d}u\,{\rm e}^{i\omega u}
\frac{f^\prime(u)}{f(u)}.
\end{equation}
We note that if we introduce the shorthand notation 
\begin{equation}
p={\rm e}^{\frac{\omega}{g}}
\end{equation}
we can write
\begin{equation}
\widetilde{f^{[\gamma]}}(\omega)=p^\gamma\,\tilde f(\omega)
\end{equation}
and in particular
\begin{equation}
\widetilde{f^+}(\omega)=p\tilde f(\omega),\qquad\qquad
\widetilde{f^-}(\omega)=\frac{1}{p}\tilde f(\omega).
\end{equation}
We also note that if $f(u)$ is analytic in the upper complex $u$ plane then
$\tilde f(\omega)=0$ for $\omega>0$ and likewise if $f(u)$ is analytic in the 
lower complex $u$ plane then $\tilde f(\omega)=0$ for $\omega<0$.

\subsection{NLIE integral equations for the right-wing nodes}

The NLIE integral equations for the one-dimensional SU(2) TBA chain
are well known \cite{KBP,nlie1,Suz98}. 
Here we will follow the construction in \cite{nlie25}
and \cite{RyoHybrid}. We start by rewriting (\ref{fe}) in Fourier space as
\begin{equation}
\widetilde{dl}\, b_m+\widetilde{dl} \,d_m=\widetilde{dl}\, X_m,\qquad\quad
\frac{1}{p}\,\widetilde{dl} \,B_m+p\,\widetilde{dl} \,D_m
=\widetilde{dl} X_{m+1}.
\end{equation}
This has to be supplemented by the Fourier space version of (\ref{half1})
and (\ref{half2}). According to what we noted above, there is no contribution
coming from the second term on the right hand side of (\ref{half1}) and
(\ref{half2}) for negative and positive frequencies, respectively. For the
first one we have
\begin{equation}
\widetilde{dl}\,b_m-\widetilde{dl}\,B_m=\tilde s
(p\widetilde{dl} X_m-\widetilde{dl} X_{m+1}),
\qquad p<1,
\end{equation}
where
\begin{equation}
\tilde s(\omega)=\frac{1}{p+\frac{1}{p}}=\tilde M^{(1)}_{11}(\omega)
\label{tildes}
\end{equation}
and for the second
\begin{equation}
\widetilde{dl}\,d_m-\widetilde{dl}\,D_m=\tilde s(\frac{1}{p}
\widetilde{dl} X_m -\widetilde{dl} X_{m+1}), \qquad p>1.
\end{equation}
Eliminating $\widetilde{dl} X_{m+1}$ from the equations we get	
\begin{equation}
\begin{split}
\widetilde{dl}\,b_m&=\frac{\widetilde{dl}\,B_m-\widetilde{dl}\,D_m+
p^2\widetilde{dl}\,X_m}{1+p^2},\\
\widetilde{dl}\,d_m&=\frac{\widetilde{dl}\,D_m-\widetilde{dl}\,B_m+
\widetilde{dl}\,X_m}{1+p^2},
\end{split}
\qquad\quad p>1,
\label{ps1}
\end{equation}
\begin{equation}
\begin{split}
\widetilde{dl}\,b_m&=\frac{p^2\widetilde{dl}\,B_m-p^2\widetilde{dl}\,D_m+
p^2\widetilde{dl}\,X_m}{1+p^2},\\
\widetilde{dl}\,d_m&=\frac{p^2\widetilde{dl}\,D_m-p^2\widetilde{dl}\,B_m+
\widetilde{dl}\,X_m}{1+p^2},
\end{split}
\qquad\quad p<1.
\label{pg1}
\end{equation}
$\widetilde{dl}\,X_{m+1}$ is determined from
\begin{equation}
\widetilde{dl} X_{m+1}=
\frac{1}{p}\,\widetilde{dl} \,B_m+p\,\widetilde{dl} \,D_m.
\label{Xm1}
\end{equation}

Having solved the problem in Fourier space we now return to rapidity space.
In order that the kernels appearing in (\ref{ps1}) and (\ref{pg1})
can be transformed (back) to rapidity space, we have to shift the argument
of the unknowns $b_m(u)$ and $d_m(u)$ in the imaginary direction. To emphasize
that our unknowns in the NLIE integral equations are functions after these 
shifts, we introduce the notation
\begin{equation}
\begin{split}
\mathfrak{b}_m(u)&=b^{[-\eta]}_m(u)=b_m\left(u-i\frac{\eta}{g}\right),\\
\mathfrak{d}_m(u)&=d^{[\eta]}_m(u)=d_m\left(u+i\frac{\eta}{g}\right),
\end{split}\qquad\quad
\begin{split}
\mathfrak{B}_m(u)&=B^{[-\eta]}_m(u)=B_m\left(u-i\frac{\eta}{g}\right),\\
\mathfrak{D}_m(u)&=D^{[\eta]}_m(u)=D_m\left(u+i\frac{\eta}{g}\right),
\end{split}
\end{equation}
which corresponds to the Fourier space relations
\begin{equation}
\begin{split}
\widetilde{dl}\,\mathfrak{b}_m&=p^{-\eta}\,\widetilde{dl}\,b_m,\\
\widetilde{dl}\,\mathfrak{d}_m&=p^{\eta}\,\widetilde{dl}\,d_m,
\end{split}\qquad\qquad
\begin{split}
\widetilde{dl}\,\mathfrak{B}_m&=p^{-\eta}\,\widetilde{dl}\,B_m,\\
\widetilde{dl}\,\mathfrak{D}_m&=p^{\eta}\,\widetilde{dl}\,D_m.
\end{split}
\end{equation}
Besides the basic TBA kernel function $s(u)$, whose Fourier transform
is given by (\ref{tildes}) the kernel $H(u)$ with Fourier transform
\begin{equation}
\tilde H(\omega)=\tilde s(\omega)\,{\rm e}^{-\frac{\vert\omega\vert}{g}}
\label{7.15}
\end{equation}
enters the rapidity space version of the NLIE equations:
\begin{eqnarray}
dl\,\mathfrak{b}_m&=&H\star dl\,\mathfrak{B}_m-H^{[-2\eta]}\star
dl\,\mathfrak{D}_m+s^{[1-\eta]}\star dl\,X_m,\\
dl\,\mathfrak{d}_m&=&H\star dl\,\mathfrak{D}_m-H^{[2\eta]}\star
dl\,\mathfrak{B}_m+s^{[\eta-1]}\star dl\,X_m.
\end{eqnarray}
Since $s(u)$ is analytic in the strip $(-1,1)$ and $H(u)$ in the strip
$(-2,2)$, the above NLIE equations are well defined if the parameter
$\eta$ is chosen in the range $0<\eta<1$. Integrating the equations once,
we obtain the final form of the ground state NLIE equations for the right-wing
nodes:
\begin{eqnarray}
\ln \mathfrak{b}_m&=&H\star \ln\mathfrak{B}_m-H^{[-2\eta]}\star
\ln\mathfrak{D}_m+s^{[1-\eta]}\star \ln X_m+ Cb_m,\label{7.18}\\
\ln\mathfrak{d}_m&=&H\star \ln\mathfrak{D}_m-H^{[2\eta]}\star
\ln\mathfrak{B}_m+s^{[\eta-1]}\star \ln X_m+ Cd_m,\label{7.19}
\end{eqnarray}
where the integration constants $Cb_m$, $Cd_m$ can be calculated using the
large $u$ asymptotics of the functions ${\mathfrak b}_m$, ${\mathfrak d}_m$
and $X_m$.
Finally we note that we are going to apply this construction with a fixed
value of $m$. $X_{m+1}$ only appears in the equation for the node $x_m$ 
in the form $s\star\ln X_{m+1}$. We can write this combination using
(\ref{Xm1}) as 
\begin{equation}
s\star\ln X_{m+1}=s^{[\eta-1]}\star\ln\mathfrak{B}_m+
s^{[1-\eta]}\star\ln\mathfrak{D}_m.
\end{equation}
Thus all Y-functions with index larger than $m$ are replaced by the two
NLIE variables ${\mathfrak b}_m$, ${\mathfrak d}_m$ and the set of 
integral equations for this truncated set is closed.

\subsection{NLIE equations for the upper nodes}

In this subsection we derive the NLIE equations for the upper nodes.
This will be done in the same spirit as in ref. \cite{Klumpersl4}.
The setting up of the equations is based on the same principles, but
the resulting set of equations (even the number of NLIE
variables) is completely different.
We start by rewriting the functional relations in terms of logarithmic 
derivatives in Fourier space. From (\ref{FEr1}) we obtain
\begin{eqnarray}
\widetilde{dl}\,b^{(3)}_{1,s}+\widetilde{dl}\,d^{(3)}_{1,s}&=&
\widetilde{dl}\,Y^{(3)}_{1,s},\label{aa1}\\
\widetilde{dl}\,b^{(3)}_{2,s}+\widetilde{dl}\,d^{(3)}_{2,s}&=&
\widetilde{dl}\,Y^{(3)}_{2,s},\label{aa2}\\
\widetilde{dl}\,b^{(3)}_{3,s}+\widetilde{dl}\,d^{(3)}_{3,s}&=&
\widetilde{dl}\,Y^{(3)}_{3,s},\label{aa3}\\
\widetilde{dl}\,b^{(2)}_{1,s}+\widetilde{dl}\,d^{(2)}_{1,s}&=&
\widetilde{dl}\,Y^{(2)}_{1,s},\label{aa4}\\
\widetilde{dl}\,b^{(2)}_{2,s}+\widetilde{dl}\,d^{(2)}_{2,s}&=&
\widetilde{dl}\,Y^{(2)}_{2,s},\label{aa5}\\
\widetilde{dl}\,b^{(1)}_{1,s}+\widetilde{dl}\,d^{(1)}_{1,s}&=&
\widetilde{dl}\,Y^{(1)}_{1,s}.\label{aa6}
\end{eqnarray}
From (\ref{FEr2}) we get
\begin{eqnarray}
\widetilde{dl}\,b^{(3)}_{2,s}+p\,\widetilde{dl}\,d^{(3)}_{1,s}&=&
\widetilde{dl}\,Y^{(2)}_{1,s+1},\label{bb1}\\
\widetilde{dl}\,b^{(3)}_{3,s}+p\,\widetilde{dl}\,d^{(3)}_{2,s}&=&
\widetilde{dl}\,Y^{(2)}_{2,s+1},\label{bb2}\\
\widetilde{dl}\,b^{(2)}_{2,s}+p\,\widetilde{dl}\,d^{(2)}_{1,s}&=&
\widetilde{dl}\,Y^{(1)}_{1,s+1}.\label{bb3}
\end{eqnarray}
From (\ref{FEr3}) we get
\begin{eqnarray}
\widetilde{dl}\,b^{(3)}_{1,s}-\widetilde{dl}\,B^{(3)}_{1,s}+
\widetilde{dl}\,D^{(3)}_{1,s}+
p\,\widetilde{dl}\,d^{(3)}_{2,s}-p\,\widetilde{dl}\,D^{(3)}_{2,s}+
p\,\widetilde{dl}\,B^{(3)}_{2,s}&=&\widetilde{dl}\,Y^{(2)}_{1,s},\label{cc1}\\
\widetilde{dl}\,b^{(3)}_{2,s}-\widetilde{dl}\,B^{(3)}_{2,s}+
\widetilde{dl}\,D^{(3)}_{2,s}+
p\,\widetilde{dl}\,d^{(3)}_{3,s}-p\,\widetilde{dl}\,D^{(3)}_{3,s}+
p\,\widetilde{dl}\,B^{(3)}_{3,s}&=&\widetilde{dl}\,Y^{(2)}_{2,s},\label{cc2}\\
\widetilde{dl}\,b^{(2)}_{1,s}-\widetilde{dl}\,B^{(2)}_{1,s}+
\widetilde{dl}\,D^{(2)}_{1,s}+
p\,\widetilde{dl}\,d^{(2)}_{2,s}-p\,\widetilde{dl}\,D^{(2)}_{2,s}+
p\,\widetilde{dl}\,B^{(2)}_{2,s}&=&\widetilde{dl}\,Y^{(1)}_{1,s}.\label{cc3}
\end{eqnarray}
Finally we have from (\ref{FEr4})
\begin{eqnarray}
\frac{1}{p}\widetilde{dl}\,B^{(3)}_{2,s}+p\,\widetilde{dl}\,D^{(3)}_{2,s}+
\widetilde{dl}\,b^{(3)}_{3,s}-
\widetilde{dl}\,B^{(3)}_{3,s}+\widetilde{dl}\,d^{(3)}_{1,s}-
\widetilde{dl}\,D^{(3)}_{1,s}&=&\widetilde{dl}\,Y^{(3)}_{2,s+1},\label{dd1}\\
\frac{1}{p}\widetilde{dl}\,B^{(3)}_{1,s}+p\,\widetilde{dl}\,D^{(3)}_{1,s}+
\widetilde{dl}\,b^{(3)}_{2,s}-
\widetilde{dl}\,B^{(3)}_{2,s}&=&\widetilde{dl}\,Y^{(3)}_{1,s+1},\label{dd2}\\
\frac{1}{p}\widetilde{dl}\,B^{(3)}_{3,s}+p\,\widetilde{dl}\,D^{(3)}_{3,s}+
+\widetilde{dl}\,d^{(3)}_{2,s}-
\widetilde{dl}\,D^{(3)}_{2,s}&=&\widetilde{dl}\,Y^{(3)}_{3,s+1},\label{dd3}\\
\frac{1}{p}\widetilde{dl}\,B^{(2)}_{1,s}+p\,\widetilde{dl}\,D^{(2)}_{1,s}+
\widetilde{dl}\,b^{(2)}_{2,s}-
\widetilde{dl}\,B^{(2)}_{2,s}&=&\widetilde{dl}\,Y^{(2)}_{1,s+1},\label{dd4}\\
\frac{1}{p}\widetilde{dl}\,B^{(2)}_{2,s}+p\,\widetilde{dl}\,D^{(2)}_{2,s}+
+\widetilde{dl}\,d^{(2)}_{1,s}-
\widetilde{dl}\,D^{(2)}_{1,s}&=&\widetilde{dl}\,Y^{(2)}_{2,s+1},\label{dd5}\\
\frac{1}{p}\widetilde{dl}\,B^{(1)}_{1,s}+p\,\widetilde{dl}\,D^{(1)}_{1,s}+
&=&\widetilde{dl}\,Y^{(1)}_{1,s+1}.\label{dd6}
\end{eqnarray}

The above set of equations has to be completed by the ones following from
the \lq\lq halfplane'' relations (\ref{half12}). Here we have to use the
\lq\lq halfplane'' properties of the boundary ratios discussed in section~3
to obtain for $p<1$ ($\omega<0$) 
\begin{equation}
\begin{split}
\widetilde{dl}\,b^{(3)}_{1,s}-\widetilde{dl}\,B^{(3)}_{1,s}=
\frac{1}{\Sigma^3-2\Sigma}&\Big\{
(\Sigma^2-1)(p\,\widetilde{dl}\,Y^{(3)}_{1,s}-
\widetilde{dl}\,Y^{(3)}_{1,s+1})\\
&\Sigma(p\,\widetilde{dl}\,Y^{(3)}_{2,s}-\widetilde{dl}\,Y^{(3)}_{2,s+1})+
(p\,\widetilde{dl}\,Y^{(3)}_{3,s}-\widetilde{dl}\,Y^{(3)}_{3,s+1})\Big\},
\end{split}
\label{ee1}
\end{equation}
\begin{equation}
\widetilde{dl}\,b^{(2)}_{1,s}-\widetilde{dl}\,B^{(2)}_{1,s}=
\frac{1}{\Sigma^2-1}\Big\{
\Sigma(p\,\widetilde{dl}\,Y^{(2)}_{1,s}-\widetilde{dl}\,Y^{(2)}_{1,s+1})+
(p\,\widetilde{dl}\,Y^{(2)}_{2,s}-\widetilde{dl}\,Y^{(2)}_{2,s+1})\Big\},
\label{ee2}
\end{equation}
\begin{equation}
\widetilde{dl}\,b^{(1)}_{1,s}-\widetilde{dl}\,B^{(1)}_{1,s}=
\frac{1}{\Sigma}
(p\,\widetilde{dl}\,Y^{(1)}_{1,s}-\widetilde{dl}\,Y^{(1)}_{1,s+1}).
\qquad\qquad\qquad\qquad\qquad
\qquad\qquad\,\,
\label{ee3}
\end{equation}
Similarly for $p>1$ ($\omega>0$) 
\begin{equation}
\begin{split}
\widetilde{dl}\,d^{(3)}_{3,s}-\widetilde{dl}\,D^{(3)}_{3,s}=
\frac{1}{\Sigma^3-2\Sigma}&\Big\{
(\Sigma^2-1)(\frac{1}{p}\widetilde{dl}\,Y^{(3)}_{3,s}-
\widetilde{dl}\,Y^{(3)}_{3,s+1})\\
&\Sigma(\frac{1}{p}\widetilde{dl}\,Y^{(3)}_{2,s}
-\widetilde{dl}\,Y^{(3)}_{2,s+1})+
(\frac{1}{p}\widetilde{dl}\,Y^{(3)}_{1,s}
-\widetilde{dl}\,Y^{(3)}_{1,s+1})\Big\},
\end{split}
\label{ff1}
\end{equation}
\begin{equation}
\widetilde{dl}\,d^{(2)}_{2,s}-\widetilde{dl}\,D^{(2)}_{2,s}=
\frac{1}{\Sigma^2-1}\Big\{
\Sigma(\frac{1}{p}\widetilde{dl}\,Y^{(2)}_{2,s}
-\widetilde{dl}\,Y^{(2)}_{2,s+1})+
(\frac{1}{p}\widetilde{dl}\,Y^{(2)}_{1,s}
-\widetilde{dl}\,Y^{(2)}_{1,s+1})\Big\},
\label{ff2}
\end{equation}
\begin{equation}
\widetilde{dl}\,d^{(1)}_{1,s}-\widetilde{dl}\,D^{(1)}_{1,s}=
\frac{1}{\Sigma}
(\frac{1}{p}\widetilde{dl}\,Y^{(1)}_{1,s}-\widetilde{dl}\,Y^{(1)}_{1,s+1}).
\qquad\qquad\qquad\qquad\qquad
\qquad\qquad\,\,
\label{ff3}
\end{equation}
Here 
\begin{equation}
\Sigma=p+\frac{1}{p}=\frac{1}{\tilde s(\omega)}
\end{equation}
is used to express various components of the Fourier space kernels 
$\tilde M^{(r)}_{ab}$ using the general formula
\begin{equation}
\tilde M_{ab}(\omega)=\frac{\cosh(k-|a-b|)\mu-\cosh(k-a-b)\mu}
{2\sinh k\mu\sinh\mu},\qquad\mu=\frac{\omega}{g}
\end{equation}
with $k=r+1$.

Next we solve the set of equations (\ref{aa1}-\ref{ff3}) in the following 
sense. We want to write down equations, which (after going back to rapidity
space) allow us to determine the unknown functions $b^{(r)}_{a,s}$,
$d^{(r)}_{a,s}$ for $r=1,2,3$, $a=1,\dots,r$ in terms of 
$Y_{a,s}=Y^{(3)}_{a,s}$, $a=1,2,3$, which serve as \lq\lq input'' from
the TBA equations of the central nodes. We have $6+3+3+6+3=21$ equations
(both for the $p>1$ and for the $p<1$ cases) for the $6+6$ unknowns and the
$6+3$ Y-functions ($Y^{(r)}_{a,s+1}$, $r=1,2,3$ and $Y^{(r)}_{a,s}$, $r=1,2$),
which have to be eliminated from the equations. We see that we have just the 
right number of equations that allow us to obtain the NLIE integral 
equations first in Fourier space and then in rapidity space. 

We note that the counting works similarly in the case of a general SU$(k)$
TBA system. From the functional equations (\ref{FEr1}-\ref{FEr4}) we get

$\binom{k}{2}+\binom{k-1}{2}+\binom{k-1}{2}+\binom{k}{2}=2(k-1)^2$

\noindent
equations, which, together with $k-1$ \lq\lq halfplane'' relations form
a total of $(k-1)(2k-1)$ equations. This number exatly matches the sum
of the number of unknowns ($2\times\binom{k}{2}=k(k-1)$) and the number
of Y-functions to be eliminated ($\binom{k}{2}+\binom{k-1}{2}=(k-1)^2$).

Solving for the unknowns in our case in Fourier space we obtain 
\begin{eqnarray}
\widetilde{dl}  b_{a,s}^{(r)} &=& \sum_{p=1}^{3} \, 
\sum_{a'=1}^{p} \, (\widetilde{K}_{bB})_{aa'}^{(rp)} 
 \, \widetilde{dl}  B_{a',s}^{(p)}+\sum_{p=1}^{3} \, 
\sum_{a'=1}^{p} \, 
(\widetilde{K}_{bD})_{aa'}^{(rp)}  
\, \widetilde{dl}  D_{a',s}^{(p)}+\sum_{a'=1}^{3} \, 
(\widetilde{K}_{bY})_{aa'}^{(r)}  
\, \widetilde{dl}  Y_{a',s},  \nonumber\\
\widetilde{dl} d_{a,s}^{(r)} &=& \sum_{p=1}^{3} \, 
\sum_{a'=1}^{p} \, (\widetilde{K}_{dB})_{aa'}^{(rp)} 
 \, \widetilde{dl}  B_{a',s}^{(p)}+\sum_{p=1}^{3} \, 
\sum_{a'=1}^{p} \, 
(\widetilde{K}_{dD})_{aa'}^{(rp)}  
\, \widetilde{dl}  D_{a',s}^{(p)}+\sum_{a'=1}^{3} \, 
(\widetilde{K}_{dY})_{aa'}^{(r)}  
\, \widetilde{dl}  Y_{a',s}.\nonumber \\
\label{Kkernels}
\end{eqnarray}
The matrices of Fourier space kernels $\tilde K_{bB}$, $\tilde K_{bD}$ etc. 
are listed in appendix \ref{kernels}.

We still have to express $Y_{a,s+1}=Y^{(3)}_{a,s+1}$ in terms of 
the NLIE variables in order to close the set of equations for the central 
nodes. We achieve this goal by combining the Y-system equations for
$y_{a,s}=y^{(3)}_{a,s}$ with (\ref{dd1}-\ref{dd3}):
\begin{equation}
\tilde{dl}  y_{a,s} = \tilde{s} \, \tilde{dl} \! \left( B_{a,s}^{(3)-} 
\, D_{a,s}^{(3)+} \, \frac{b^{(3)}_{a+1,s}}{B^{(3)}_{a+1,s}} \, 
\frac{d^{(3)}_{a-1,s}}{D^{(3)}_{a-1,s}}\right)+\tilde{s} \, \, 
\tilde{dl}Y_{a,s-1} +\tilde{s} \, \tilde{dl}  \!
\left(\frac{y_{a+1,s}}{ Y_{a+1,s}} \, \frac{y_{a-1,s}}{Y_{a-1,s}}
\right).  
\label{FdNLIE}
\end{equation}
We understand that terms appearing on the RHS of (\ref{FdNLIE}) with 
$a$-type indices \lq\lq out of range'' ($\not=1,2,3$) must be omitted.

In order to be able to transform the equations (\ref{Kkernels}) 
back into rapidity space (where the kernel multiplications become 
convolutions), the Fourier form of the kernels should satisfy the following 
requirements:
\begin{itemize}
\item they must be continuous at $\omega=0$,
\item they must tend to zero exponentially as $\omega \to \pm \infty$.
\end{itemize}
The $\omega=0$ condition is to ensure that kernels decay as $1/u^2$ at infinity
and we can check (by comparing the $\omega\to\pm0$ limits of the 
representations valid for positive and negative frequencies) that it is 
satisfied by our kernels. 

The second requirement (which is necessary to ensure that the inverse Fourier
transformation exists) is not automatically satisfied for all matrix elements 
of the kernel matrices listed in appendix~\ref{kernels}. 
Fortunately the problem can be solved by a simple redefinition of the NLIE 
variables. It can be shown that if we redefine our NLIE functions by shifting
their arguments appropriately: 
\begin{eqnarray}
b_{a,s}^{(r)}(u) &\rightarrow& \mathfrak{b}_{a,s}^{(r)}(u)=
b_{a,s}^{(r)}(u+(i/g) \,(r-3+\, \gamma_{a}^{(r)})),
\qquad r=1,2,3 \quad a=1,..,r  \nonumber \\
d_{a,s}^{(r)}(u) &\rightarrow& \mathfrak{d}_{a,s}^{(r)}(u)=
d_{a,s}^{(r)}(u+ \, (i/g)\, \eta_{a}^{(r)}),
\qquad r=1,2,3 \quad a=1,..,r \nonumber \\
\quad y_{a,s}(u) &\rightarrow& \mathfrak{y}_{a,s}(u)=
y_{a,s}(u+ \, (i/g)\, \epsilon_{a}),\quad \qquad a=1,2,3 
\nonumber
\end{eqnarray}
then the kernels entering the NLIE of these redefined variables do satisfy the
requirements imposed above provided the shift parameters satisfy the 
inequalities:
\begin{equation}
-\frac12<\gamma_{1}^{(2)}<\gamma_{2}^{(3)}
<\epsilon_2<\eta_{2}^{(3)}<\eta_{2}^{(2)}<\frac12,
\label{ineq1}
\end{equation}
\begin{equation}
-\frac12<\gamma_{1}^{(3)}<\epsilon_1<\eta_{1}^{(3)}<\eta_{1}^{(2)}
<\eta_{1}^{(1)}
<\gamma_{1}^{(1)}<\gamma_{2}^{(2)}<\gamma_{3}^{(3)}
<\epsilon_3<\eta_{3}^{(3)}<\frac12.
\label{ineq2}
\end{equation}
The NLIE for the new variables in rapidity space takes the form:
\begin{eqnarray}
\ln  \mathfrak{b}_{a,s}^{(r)} &=& \sum_{p=1}^{3} \, \sum_{a'=1}^{p} \, 
({G}_{bB})_{aa'}^{(rp)} 
 \star \ln  \mathfrak{B}_{a',s}^{(p)}+\sum_{p=1}^{3} \, \sum_{a'=1}^{p} \, 
({G}_{bD})_{aa'}^{(rp)}  \star  \ln \mathfrak{D}_{a',s}^{(p)}+
\sum_{a'=1}^{3} \, ({G}_{bY})_{aa'}^{(r)}  \star 
\ln  \mathfrak{Y}_{a',s} \nonumber \\
\ln \mathfrak{d}_{a,s}^{(r)} &=& \sum_{p=1}^{3} \, \sum_{a'=1}^{p} \, 
({G}_{dB})_{aa'}^{(rp)} 
\star \ln  \mathfrak{B}_{a',s}^{(p)}+\sum_{p=1}^{3} \, \sum_{a'=1}^{p} \, 
({G}_{dD})_{aa'}^{(rp)}  
\star \ln  \mathfrak{D}_{a',s}^{(p)}+\sum_{a'=1}^{3} \, 
({G}_{dY})_{aa'}^{(r)}  
\star \ln  \mathfrak{Y}_{a',s}, \nonumber \\
\label{mathfrakNLIE}
\end{eqnarray}
where we have used the notation 
${\mathfrak B}^{(p)}_{a,s}=1+{\mathfrak b}^{(p)}_{a,s}$,
${\mathfrak D}^{(p)}_{a,s}=1+{\mathfrak d}^{(p)}_{a,s}$,
${\mathfrak Y}_{a,s}=1+{\mathfrak y}_{a,s}$ and the kernels are rapidity 
space representations of the appropriately modified Fourier kernels 
(\ref{KbB+}-\ref{KdY-}):
\begin{equation} 
\begin{split}
(\tilde{G}_{bB})_{aa'}^{(rr^\prime)}(\omega)&=
(\tilde{K}_{bB})_{aa'}^{(rr^\prime)}(\omega) \, 
p^{r-r^\prime+\gamma_{a}^{(r)}-\gamma_{a'}^{(r^\prime)}}, \\ 
(\tilde{G}_{bD})_{aa'}^{(rr^\prime)}(\omega)&=
(\tilde{K}_{bD})_{aa'}^{(rr^\prime)}(\omega) \, 
p^{r-3+\gamma_{a}^{(r)}-\eta_{a'}^{(r^\prime)}}, 
\label{FourierG}
\end{split}
\end{equation}
etc. The rapidity space version of (\ref{FdNLIE}) in the new variables becomes
\begin{eqnarray}
\ln  \mathfrak{y}_{a,s} &=& {s}^{[-1+\epsilon_a-\gamma_a^{(3)}]}
\star \ln \! \mathfrak{B}_{a,s}^{(3) } \, +
s^{[1+\epsilon_a-\eta_a^{(3)}]} \star \ln \!
\mathfrak{D}_{a,s}^{(3)} \,+ s^{[\epsilon_a-\gamma_{a+1}^{(3)}]} \star
\ln \! \left(
\frac{\mathfrak{b}_{a+1,s}^{(3)}}{\mathfrak{B}_{a+1,s}^{(3)}} \right) \,
\nonumber \\
&+& s^{[\epsilon_a-\eta_{a-1}^{(3)}]} \star \ln \! \left(
\frac{\mathfrak{d}_{a-1,s}^{(3)}}{\mathfrak{D}_{a-1,s}^{(3)}}\right) +
{s}^{[\epsilon_a]}
 \star \ln{Y}_{a,s-1} +{s}^{[\epsilon_a-\epsilon_{a-1}]} \star \ln \!
\left(\frac{\mathfrak y_{a-1,s}}{\mathfrak Y_{a-1,s}} \right)  \,  
\label{FdNLIE1}\\
&+&{s}^{[\epsilon_a-\epsilon_{a+1}]} \star \ln \!
\left(\frac{\mathfrak y_{a+1,s}}{\mathfrak Y_{a+1,s}} \right).  \nonumber
\end{eqnarray}
We note that (\ref{FdNLIE1}) is well-defined since (\ref{ineq1}) and
(\ref{ineq2}) imply
\begin{equation}
\epsilon_a-\gamma^{(3)}_a>0,\qquad
\epsilon_a-\eta^{(3)}_a<0,\qquad a=1,2,3.
\end{equation}

\section{Linearized equations}

In this section we compute the leading (${\cal O}(\varepsilon^2)$) corrections
to the asymptotic solution corresponding to the ground state of the
$\gamma$-deformed model. These corrections have been calculated recently
\cite{TBA2ndLusch} using the TBA equations. In this paper their contribution 
to the ground state energy was also calculated and it was found to be in 
agreement with the NLO L\"uscher formula. Here we calculate these corrections 
from the ground state NLIE and show that they are identical to the ones 
obtained from the TBA integral equations directly. This agreement is a useful 
analytical evidence of the equivalence of the two approaches.

\subsection{Linearization of the right-wing SU$(2)$ problem}

We start with linearizing the NLIE equations corresponding to the right-wing
nodes. Of course, the equivalence of the SU$(2)$ type TBA equations with
the corresponding NLIE equations is well known. Here we consider this problem 
not only for completeness but also because the logic of the calculation in 
this case is similar to what we will follow in the technically much more 
complicated case of the upper nodes.

We will write for any function $f$
\begin{equation}
f=f^o(1+f^{(1)}+\dots),
\end{equation}
where $f^o$ is the asymptotic limit of the function and $f^{(1)}$ is the
corresponding ${\cal O}(\varepsilon^2)$ correction. With this notation
\begin{equation}
\widetilde{dl}\frac{f}{f^o}=\widetilde{f^{(1)\prime}}+\dots
\end{equation}
and for the corresponding $F=1+f$ we have
\begin{equation}
F^o=1+f^o,\qquad\qquad
\widetilde{dl}\frac{F}{F^o}=\frac{f^o}{F^o}\widetilde{f^{(1)\prime}}+\dots
\end{equation}
The NLIE integral equations of subsection 7.1 are based on the NLIE functional 
relations (\ref{fe}) and (\ref{half1}-\ref{half2}). The same relations are also
satisfied by the asymptotic limit of the NLIE functions and therefore
the NLIE integral equations are also valid if written for the ratios
of type $f/f^o$ (or equivalently in Fourier space for the differences
of the logarithmic derivatives). Introducing the notations
\begin{equation}
\tilde a_m=\widetilde{b_m^{(1)\prime}},\qquad
\tilde f_m=\widetilde{d_m^{(1)\prime}},\qquad
\tilde \xi_m=\widetilde{X_m^{(1)\prime}}
\end{equation}
and
\begin{equation}
\frac{b^o_m}{B^o_m}=\beta^o_m,\qquad\quad
\frac{d^o_m}{D^o_m}=\delta^o_m,\qquad\quad p_o={\rm e}^{-\frac{|\omega|}{g}}
\end{equation}
we can write the linearization of (\ref{ps1}-\ref{pg1}) in Fourier space as
\begin{equation}
\tilde a_m=p_o\tilde s(\beta^o_m\tilde a_m-\delta^o_m\tilde f_m)+
p\,\tilde s\,\tilde\xi_m,\qquad
\tilde f_m=p_o\tilde s(\delta^o_m\tilde f_m-\beta^o_m\tilde a_m)+
\frac{1}{p}\,\tilde s\,\tilde\xi_m
\label{tildaf}
\end{equation}
and from (\ref{Xm1}) we have
\begin{equation}
\tilde\xi_{m+1}=\frac{1}{p}\beta^o_m\tilde a_m+p\,\delta^o_m\tilde f_m.
\label{tildexim1}
\end{equation}
We first solve (\ref{tildaf}):
\begin{equation}
\tilde a_m=\frac{(p-p_o\delta^o_m)\tilde\xi_m}
{\frac{1}{p_o}+p_o(1-\beta^o_m-\delta^o_m)},\qquad\quad
\tilde f_m=\frac{(\frac{1}{p}-p_o\beta^o_m)\tilde\xi_m}
{\frac{1}{p_o}+p_o(1-\beta^o_m-\delta^o_m)},\qquad\quad
\end{equation}
then use the result in (\ref{tildexim1}) and find
\begin{equation}
\tilde\xi_{m+1}=\frac{\beta^o_m\delta^o_m+\frac{1}{p_o^2}(\beta^o_m\delta^o_m
-\beta^o_m-\delta^o_m)}{(\beta^o_m+\delta^o_m-1)-\frac{1}{p_o^2}}
\,p_o\tilde\xi_m.
\label{rawres}
\end{equation}
Recalling the definition of $q$-numbers
\begin{equation}
[N]_q=\frac{q^N-q^{-N}}{q-1/q}
\end{equation}
we rewrite the asymptotic limits (\ref{6.54})
\begin{equation}
\beta^o_m=\frac{1}{q}\,\frac{[m+1]_q}{[m+2]_q},\qquad\quad
\delta^o_m=q\,\frac{[m+1]_q}{[m+2]_q},\qquad\quad
\end{equation}
and analogously
\begin{equation}
x^o_m=[m]_q\,[m+2]_q,\qquad\quad
X^o_m=([m+1]_q)^2.
\end{equation}
The recursion relation (\ref{rawres}) becomes
\begin{equation}
\tilde\xi_{m+1}=\frac{[m+1]_q}{[m+2]_q}\,p_o\,
\frac{[m+1]_q-\frac{1}{p_o^2}\,[m+3]_q}{[m]_q-\frac{1}{p_o^2}[m+2]_q}\,
\tilde\xi_m.
\end{equation}
We note that
\begin{equation}
\tilde\xi_m=\widetilde{X^{(1)\prime}_m}=\frac{x^o_m}{X^o_m}
\widetilde{x^{(1)\prime}_m}
=\frac{[m]_q\,[m+2]_q}{([m+1]_q)^2}\,\widetilde{x^{(1)\prime}_m}
\end{equation}
and using this relation it is easy to see that our result follows from
eq. (4.68) of ref. \cite{TBA2ndLusch}, which in our notation reads
\begin{equation}
\widetilde{x_m^{(1)}}={\rm const.}\,p_o^{m+2}\left(\frac{[m+1]_q}{[m+2]_q}
-\frac{1}{p_o^2}\,\frac{[m+1]_q}{[m]_q}\right),
\end{equation}
where the complicated constant is $m$-independent.

\subsection{Linearization of the upper Y-system}

We saw above that the final result of the NLIE linearization procedure is a 
recursion relation among ${\cal O}(\varepsilon^2)$ coefficients of TBA
Y-functions. In the next subsection we will go through this procedure
for the upper nodes and will show that they agree with the relations obtained
directly from TBA. Although the latter problem has been solved in 
ref.~\cite{TBA2ndLusch}, in this subsection we reproduce the derivation and 
present the relations in a form easily comparable to the NLIE results.

We start here by recalling the upper node (SU$(4)$ problem) Y-system equations:
\begin{eqnarray}
y^+_{1,s}y^-_{1,s}&=&Y_{1,s+1}\,Y_{1,s-1}\,\frac{y_{2,s}}{Y_{2,s}},\\
y^+_{2,s}y^-_{2,s}&=&Y_{2,s+1}\,Y_{2,s-1}\,
\frac{y_{1,s}}{Y_{1,s}}\,\frac{y_{3,s}}{Y_{3,s}},\\
y^+_{3,s}y^-_{3,s}&=&Y_{3,s+1}\,Y_{3,s-1}\,\frac{y_{2,s}}{Y_{2,s}}.
\end{eqnarray}
The asymptotic solution\footnote{(\ref{asyy}) is an {\it exact} solution of
the Y-system relations, it is only the relations $Y^o_{a,s}=1+y^o_{a,s}$
which are not exactly satisfied for $a=2$.} 
can be recalled from subsection 5.4.1
\begin{equation}
y^o_{1,s}=y^o_{3,s}=s^2-1,\qquad Y^o_{1,s}=Y^o_{3,s}=s^2,\qquad
y^o_{2,s}=Y^o_{2,s}=\frac{1}{\Psi_s},
\label{asyy}
\end{equation}
where
\begin{equation}
\Psi_s=s^2\,\frac{(1-q)^2(1-\dot q)^2}{q\dot q}
\,\frac{\psi^{[s]}}{\psi^{[-s]}}.
\end{equation}
The new feature of the asymptotic solution here is that for $a=2$ it is
${\cal O}(1/\varepsilon^2)$ and for this reason if we introduce the ${\cal O}
(\varepsilon^2)$ coeffiecients $z_{a,s}$, $Z_{a,s}$ by
\begin{equation}
y_{a,s}=y^o_{a,s}(1+z_{a,s}+\dots),\qquad\quad
Y_{a,s}=Y^o_{a,s}(1+Z_{a,s}+\dots)
\end{equation}
we find
\begin{equation}
a=1,3:\qquad Z_{a,s}=r_s\,z_{a,s},\qquad\quad r_s=\frac{s^2-1}{s^2}
\end{equation}
and
\begin{equation}
a=2:\qquad\ \ Z_{2,s}=z_{2,s}+\Psi_s.
\end{equation}

As in the previous subsection, the ratios
\begin{equation}
x_{a,s}=\frac{y_{a,s}}{y^o_{a,s}},\qquad\quad
X_{a,s}=\frac{Y_{a,s}}{Y^o_{a,s}},\qquad\quad
\end{equation}
are exactly solving the Y-system equations, which can be rewritten as TBA 
integral equations as follows.
\begin{eqnarray}
\ln\,x_{1,s}&=&s\star\{\ln\,X_{1,s+1}+\ln\,X_{1,s-1}
+\ln\,x_{2,s}-\ln\,X_{2,s}\},\\
\ln\,x_{2,s}&=&s\star\{\ln\,X_{2,s+1}+\ln\,X_{2,s-1}
+\ln\,x_{1,s}+\ln\,x_{3,s}-\ln\,X_{1,s}-\ln\,X_{3,s}\},\\
\ln\,x_{3,s}&=&s\star\{\ln\,X_{3,s+1}+\ln\,X_{3,s-1}
+\ln\,x_{2,s}-\ln\,X_{2,s}\}.
\end{eqnarray}
Going to Fourier space after linearizing the above system we get
\begin{eqnarray}
\Sigma\,\tilde z_{1,s}&=&r_{s+1}\,\tilde z_{1,s+1}+r_{s-1}\,\tilde z_{1,s-1}
-\tilde\Psi_s,\\
\Sigma\,\tilde z_{2,s}&=&\tilde z_{2,s+1}+\tilde z_{2,s-1}
+\tilde\Psi_{s+1}+\tilde\Psi_{s-1}+(1-r_s)(\tilde z_{1,s}+\tilde z_{3,s}),
\label{tildez2s}\\
\Sigma\,\tilde z_{3,s}&=&r_{s+1}\,\tilde z_{3,s+1}+r_{s-1}\,\tilde z_{3,s-1}
-\tilde\Psi_s.
\end{eqnarray}

\subsubsection{$\tilde z_{1,s}$, $\tilde z_{3,s}$ problem}

Since the linearized equations are separated, we first study the $a=1,3$ cases,
which are of the form
\begin{equation}
\Sigma\,\tilde z_s=r_{s+1}\,\tilde z_{s+1}+r_{s-1}\,\tilde z_{s-1}
-\tilde\Psi_s.
\label{13type}
\end{equation}
Let us introduce a few building blocks in terms of which we will write
the solution.
\begin{equation}
\begin{split}
{\cal N}_s&=sp_o^2-s-2,\\
\alpha_s&=p_o^s\,\frac{s+1}{s(s+2)}\,{\cal N}_s=p_o^s\,\frac{s+1}{s(s+2)}\,
[s(p_o^2-1)-2],\\
\beta_s&=\alpha_{-s}=p_o^{-s}\,\frac{s-1}{s(s-2)}\,[s(p_o^2-1)+2]
\end{split}
\end{equation}
and further
\begin{equation}
f_k=\frac{p_o^4}{(p_o^2-1)^3}\beta_{k+2}r_{k+1}\tilde\Psi_{k+1},\qquad
g_k=\frac{p_o^2}{(p_o^2-1)^3}\alpha_k r_{k+1}\tilde\Psi_{k+1}.
\end{equation}
We note that the quantities introduced above satisfy the following identities.
\begin{equation}
\begin{split}
\Sigma\alpha_{s-1}&=r_{s+1}\alpha_s+r_{s-1}\alpha_{s-2},\\
\beta_s\alpha_{s-1}-\beta_{s+1}\alpha_{s-2}&=\frac{s(s-1)}{(s-2)(s+1)}\,
\frac{(p_o^2-1)^3}{p_o^3},\\
\Sigma\beta_{s+1}&=r_{s-1}\beta_s+r_{s+1}\beta_{s+2}.
\end{split}\label{iden3}
\end{equation}

The main observation that allows one to present the recursion relations in
a simple form is that the second order difference equations (\ref{13type})
are equivalent to the first order difference equations
\begin{equation}
b_s=\frac{\tilde z_{s+1}}{\alpha_s}-\frac{\tilde z_s}{\alpha_{s-1}}
\label{bzz}
\end{equation}
provided the new function $b_s$ satisfies
\begin{equation}
r_{s+1}\alpha_s b_s-r_{s-1}\alpha_{s-2}b_{s-1}=\tilde\Psi_s.
\label{bss1}
\end{equation}
This can be easily shown using the identities (\ref{iden3}), which are
also useful to verify that the general solution of (\ref{bss1}) is of the
form
\begin{equation}
b_s=p_o\left(\frac{\beta_{s+2}}{\alpha_s}-
\frac{\beta_{s+1}}{\alpha_{s-1}}\right)\,\left(A_2-\sum_{r=1}^{s-1}g_r\right),
\end{equation}
where $A_2$ is an arbitrary \lq\lq integration'' constant. It is also easy
to write down the general solution of (\ref{bzz}):
\begin{equation}
\tilde z_s=\frac{\alpha_{s-1}}{p_o}\,\left(\sum_{r=1}^{s-1}f_r-A_1\right)+p_o
\beta_{s+1}\left(A_2-\sum_{r=1}^{s-1}g_r\right)
\end{equation}
containing an other arbitrary constant $A_1$. This is the general solution
of (\ref{13type}), but we are interested in particular solutions satisfying
the requirement $\lim_{s\to\infty}\tilde z_s=0$.  This is a boundary condition
(at infinity) and requires \cite{TBA2ndLusch}
\begin{equation}
A_2=\sum_{r=1}^\infty g_r.
\end{equation}
(The other integration constant $A_1$ remains arbitrary. It is a complicated
expression that can be determined \cite{TBA2ndLusch} from the coupling of the 
upper nodes to the rest of the AdS/CFT Y-functions.)

For our purposes (comparison to the recursion relations coming from the 
linearized NLIE) the relations (\ref{bzz}) are sufficient. After imposing
the boundary condition they are uniquely determined and can be rewritten in
the form
\begin{equation}
\tilde Z_{1,s+1}={\cal A}_s\tilde Z_{1,s}+{\cal B}_s,\qquad
\tilde Z_{3,s+1}={\cal A}_s\tilde Z_{3,s}+{\cal B}_s,
\label{BZZ}
\end{equation}
where
\begin{equation}
{\cal A}_s=\frac{r_{s+1}\alpha_s}{r_s\alpha_{s-1}}=\frac{sp_o}{s+1}\,
\frac{{\cal N}_s}{{\cal N}_{s-1}},\qquad
{\cal B}_s=r_{s+1}\alpha_sb_s=-\frac{sp_o}{{\cal N}_{s-1}}\sum_{r=s}^\infty
\frac{p_o^{r-s}}{s+1}{\cal N}_r\tilde\Psi_{r+1}.
\label{AsBs}
\end{equation}

\subsubsection{$\tilde z_{2,s}$ problem}

Having found the solution for $\tilde z_{a,s}$ for $a=1,3$ we now turn to
(\ref{tildez2s}), which we rewrite as
\begin{equation}
\Sigma\tilde Z_{2,s}=\tilde Z_{2,s+1}+\tilde Z_{2,s-1}+\Sigma
\tilde\Psi_s+(1-r_s)(\tilde z_{1,s}+\tilde z_{3,s}).
\end{equation}
Again, it is easy to verify that this is equivalent to the first order
equation
\begin{equation}
\tilde Z_{2,s+1}=p_o\tilde Z_{2,s}+{\cal C}_s(\tilde Z_{1,s}+\tilde Z_{3,s})
+{\cal D}_s,\qquad
{\cal C}_s=-\frac{p_o^2}{(s+1){\cal N}_{s-1}},
\label{Z2rec}
\end{equation}
provided ${\cal D}_s$ satisfies
\begin{equation}
{\cal D}_s-\frac{1}{p_o}{\cal D}_{s-1}+\left(p_o+\frac{1}{p_o}\right)
\tilde\Psi_s-\frac{2{\cal B}_{s-1}}{(s-1){\cal N}_{s-1}}=0.
\label{Ds}
\end{equation}

Similarly to what we saw above for the $a=1,3$ cases, if we find a 
\lq\lq good'' solution ${\cal D}^{(g)}_s$ (satisfying the boundary condition 
at infinity), the general solution for (\ref{Ds}) will be 
\begin{equation}
{\cal D}_s={\cal D}^{(g)}_s+d_op_o^{-s},
\end{equation}
where $d_o$ is arbitrary, however, imposing the boundary condition requires
$d_o=0$ again, making the \lq\lq good'' solution unique. The general solution
of (\ref{Z2rec}) for $\tilde Z_{2,s}$ contains a term $\zeta_op_o^s$ with 
arbitrary constant $\zeta_o$, but in the next section we will only need the
relation (\ref{Z2rec}) itself, which is unique.

\subsection{Linearization of the NLIE equations for the upper nodes}

Here again it is useful to start from the NLIE equations written for
the ratios of functions divided by their asymptotic values. Let us introduce
the shorthand notations\footnote{Note that the index $s$ is arbitrary, but 
fixed in our present considerations. To simplify the notation, $s$ is omitted
from most of our formulas.}
\begin{equation}
\begin{split}
\hat b_{i,r}&=\widetilde{dl}\,b^{(r)}_{i,s}-\widetilde{dl}\,b^{(r)o}_{i,s},\\
\hat d_{i,r}&=\widetilde{dl}\,d^{(r)}_{i,s}-\widetilde{dl}\,d^{(r)o}_{i,s},\\
\hat p_i&=\widetilde{dl}\,Y^{(3)}_{i,s}-\widetilde{dl}\,Y^{(3)o}_{i,s},
\end{split}
\qquad\qquad
\begin{split}
\hat B_{i,r}&=\widetilde{dl}\,B^{(r)}_{i,s}-\widetilde{dl}\,B^{(r)o}_{i,s},\\
\hat D_{i,r}&=\widetilde{dl}\,D^{(r)}_{i,s}-\widetilde{dl}\,D^{(r)o}_{i,s},\\
\hat Y_i&=\widetilde{dl}\,Y^{(3)}_{i,s+1}-\widetilde{dl}\,Y^{(3)o}_{i,s+1}.
\end{split}
\end{equation}

The NLIE equations for these variables are of the same form as (\ref{Kkernels})
\begin{eqnarray}
\hat b_{i,r}&=&\sum_{j,w}(\tilde K_{bB})^{(rw)}_{ij}\hat B_{j,w}
+\sum_{j,w}(\tilde K_{bD})^{(rw)}_{ij}\hat D_{j,w}
+\sum_j(\tilde K_{bY})^{(r)}_{ij}\hat p_j,\label{hatbir}\\
\hat d_{i,r}&=&\sum_{j,w}(\tilde K_{dB})^{(rw)}_{ij}\hat B_{j,w}
+\sum_{j,w}(\tilde K_{dD})^{(rw)}_{ij}\hat D_{j,w}
+\sum_j(\tilde K_{dY})^{(r)}_{ij}\hat p_j,\label{hatdir}\\
\hat Y_i&=&\sum_{j,w}(\tilde K_{YB})^{(w)}_{ij}\hat B_{j,w}
+\sum_{j,w}(\tilde K_{YD})^{(w)}_{ij}\hat D_{j,w}
+\sum_j(\tilde K_{YY})_{ij}\hat p_j.\label{hatYi}
\end{eqnarray}
The kernels occurring in (\ref{hatbir}) and (\ref{hatdir}) are listed in 
appendix C. New kernels appear in (\ref{hatYi}). These can be calculated
by substituting (\ref{hatbir}) and (\ref{hatdir}) into (\ref{dd1}-\ref{dd3}).

Let us recall the asymptotic solutions found in subsection 5.4.1:
\begin{eqnarray}
b^{(3)o}_{1,s}&=&b^{(3)o}_{3,s}=d^{(3)o}_{1,s}=d^{(3)o}_{3,s}=s,\\
B^{(3)o}_{1,s}&=&B^{(3)o}_{3,s}=D^{(3)o}_{1,s}=D^{(3)o}_{3,s}=s+1,\\
b^{(3)o}_{2,s}&=&B^{(3)o}_{2,s}=-\frac{1}{sC_{2,s}},\\
d^{(3)o}_{2,s}&=&D^{(3)o}_{2,s}=-\frac{1}{sC_{1,s}},\\
b^{(2)o}_{1,s}&=&B^{(2)o}_{1,s}=\frac{1}{(s-\Delta)C_{2,s-1}},\\
b^{(2)o}_{2,s}&=&d^{(2)o}_{1,s}=\Delta-s,\qquad\qquad
B^{(2)o}_{2,s}=D^{(2)o}_{1,s}=\Delta+1-s,\\
d^{(2)o}_{2,s}&=&D^{(2)o}_{2,s}=\frac{1}{(s-\Delta)C_{1,s}},\\
b^{(1)o}_{1,s}&=&d^{(1)o}_{1,s}=s-1-\Delta,\qquad\qquad
B^{(1)o}_{1,s}=D^{(1)o}_{1,s}=s-\Delta.
\end{eqnarray}
Here
\begin{equation}
C_{1,s}=(1-q)(1-\dot q)\,\psi^{[s]},\qquad\quad
C_{2,s}=\frac{(1-q)(1-\dot q)}{q\dot q}\,\frac{1}{\psi^{[-s]}}.
\end{equation}
Both functions are ${\cal O}(\varepsilon)$ and
\begin{equation}
\Psi_s=s^2\,C_{1,s}\,C_{2,s}\sim{\cal O}(\varepsilon^2).
\end{equation}

A serious complication as compared to the cases discussed so far is that
some of the asymptotic solutions above are ${\cal O}(1/\varepsilon)$. This
implies that the leading corrections to the asymptotic solution are 
(relatively) ${\cal O}(\varepsilon)$. We will call this order 1/2. 
We are interested in the corrections ${\cal O}(\varepsilon^2)$ (order 1),
which are unfortunately NLO corrections in this expansion.

Let us denote our variables generically by $x$ and write $X=1+x$. 
($x=b^{(r)}_{i,s}$, $X=B^{(r)}_{i,s}$,~etc.) Define the expansion coefficients
by
\begin{equation}
\begin{split}
x&=x^o(1+x^{(1/2)}+x^{(1)}+\dots),\qquad 
\ \ \ \ x^{(1/2)}\sim{\cal O}(\varepsilon),
\qquad x^{(1)}\sim{\cal O}(\varepsilon^2),\\
X&=X^o(1+X^{(1/2)}+X^{(1)}+\dots),\qquad X^{(1/2)}\sim{\cal O}(\varepsilon),
\qquad X^{(1)}\sim{\cal O}(\varepsilon^2).
\end{split}
\label{expa}
\end{equation}
The ratio $x_c=x^o/X^o$ is a constant for those variables for which $x^o$
is ${\cal O}(1)$ (constant) and $x_c=1$ for the cases where $x^o$ is
${\cal O}(1/\varepsilon)$. Accordingly,
\begin{equation}
\begin{split}
x^o\sim{\cal O}(1)&:X^{(1/2)}=x_cx^{(1/2)},
\ \ \ \ \qquad X^{(1)}=x_cx^{(1)},\\
x^o\sim{\cal O}(1/\varepsilon)&:X^{(1/2)}=\frac{1}{x^o}+x^{(1/2)},\qquad 
X^{(1)}=x^{(1)}.
\end{split}
\end{equation}
Up to ${\cal O}(\varepsilon^2)$,
\begin{equation}
\widetilde{dl}\,x-\widetilde{dl}\,x^o=\tilde x^{(1/2)}+\tilde x^{(1)}-
\tilde x^{(2/2)}+\dots,\qquad x^{(2/2)}=x^{(1/2)}x^{(1/2)\prime}.
\end{equation}
Here we have used the Fourier transforms
\begin{equation}
x^{(1/2)\prime}\rightarrow\tilde x^{(1/2)},\qquad
x^{(1)\prime}\rightarrow\tilde x^{(1)},\qquad
x^{(1/2)}x^{(1/2)\prime}\rightarrow\tilde x^{(2/2)}
\end{equation}
and we will use the analogously defined objects for $X$.

After substituting the expansion (\ref{expa}) into the NLIE equations
the problem can be solved order by order. We find it convenient to treat
$x^{(1/2)}$ as our independent variable at order 1/2 and write 
\begin{equation}
x^o\sim{\cal O}(1):X^{(1/2)}=x_cx^{(1/2)},\qquad\quad
x^o\sim{\cal O}(1/\varepsilon):X^{(1/2)}=\frac{1}{x^o}+x^{(1/2)},
\end{equation}
while for order 1 we introduce the independent variable $x^{(1m)}$ and write
\begin{equation}
X^{(1)\prime}-X^{(2/2)}=x_cx^{(1m)},\qquad
x^{(1)\prime}-x^{(2/2)}=x^{(1m)}+x^{(1d)},
\end{equation}
where
\begin{equation}
\begin{split}
x^o\sim{\cal O}(1)&:x^{(1d)}=(x_c-1)x^{(1/2)}x^{(1/2)\prime},\\
x^o\sim{\cal O}(1/\varepsilon)&:x^{(1d)}=\frac{1}{x^o}
\left(\frac{1}{x^o}\right)^\prime+\left(\frac{1}{x^o}x^{(1/2)}\right)^\prime.
\end{split}
\end{equation}

Let us start at order 1/2 and introduce the shorthand notation
\begin{equation}
\begin{split}
\tilde a_{i,r}&=\widetilde{b^{(r)(1/2)\prime}_{i,s}},\\
\tilde f_{i,r}&=\widetilde{d^{(r)(1/2)\prime}_{i,s}},
\end{split}
\qquad\quad
\begin{split}
\tilde A_{i,r}&=\widetilde{B^{(r)(1/2)\prime}_{i,s}},\\
\tilde F_{i,r}&=\widetilde{D^{(r)(1/2)\prime}_{i,s}}
\end{split}
\end{equation}
and
\begin{equation}
c_{1,s}=\widetilde{C^\prime_{1,s}},\qquad\qquad
c_{2,s}=\widetilde{C^\prime_{2,s}}.
\end{equation}
At order 1/2 we have to make the substitutions
\begin{equation}
\hat b_{i,r}\rightarrow \tilde a_{i,r},\quad
\hat d_{i,r}\rightarrow \tilde f_{i,r},\quad
\hat B_{i,r}\rightarrow \tilde A_{i,r},\quad
\hat D_{i,r}\rightarrow \tilde F_{i,r},\quad
\hat p_i\rightarrow 0
\end{equation}
in (\ref{hatbir})-(\ref{hatYi}) and use the relations
\begin{equation}
\begin{split}
\tilde A_{1,3}&=\frac{s}{s+1}\,\tilde a_{1,3},\\
\tilde A_{2,3}&=\tilde a_{2,3}-sc_{2,s},\\
\tilde A_{3,3}&=\frac{s}{s+1}\,\tilde a_{3,3},\\
\tilde A_{1,2}&=\tilde a_{1,2}+(s-\Delta)pc_{2,s},\\
\tilde A_{2,2}&=\frac{s-\Delta}{s-\Delta-1}\,\tilde a_{2,2},\\
\tilde A_{1,1}&=\frac{s-\Delta-1}{s-\Delta}\,\tilde a_{1,1},
\end{split}
\qquad\quad
\begin{split}
\tilde F_{1,3}&=\frac{s}{s+1}\,\tilde f_{1,3},\\
\tilde F_{2,3}&=\tilde f_{2,3}-sc_{1,s},\\
\tilde F_{3,3}&=\frac{s}{s+1}\,\tilde f_{3,3},\\
\tilde F_{1,2}&=\frac{s-\Delta}{s-\Delta-1}\,\tilde f_{1,2},\\
\tilde F_{2,2}&=\tilde f_{2,2}+(s-\Delta)c_{1,s},\\
\tilde F_{1,1}&=\frac{s-\Delta-1}{s-\Delta}\,\tilde f_{1,1}.
\end{split}
\end{equation}
We have to solve the set of linear equations obtained by these substitutions
from (\ref{hatbir}) and (\ref{hatdir}) for the unknowns $\tilde a_{i,r}$, 
$\tilde f_{i,r}$. The source terms of the equations are proportional to 
$c_{1,s}$ or $c_{2,s}$. These Fourier space source functions can be 
characterized by the following properties.
\begin{equation}
c_{1,s}(\omega)=0,\quad\omega>0,\qquad\quad
c_{2,s}(\omega)=0,\quad\omega<0
\end{equation}
\begin{equation}
c_{1,s}(\omega)=p^sa_1(\omega),\qquad\quad
c_{2,s}(\omega)=p^{-s}a_2(\omega).
\end{equation}
The resulting solution is also proportional to $c_{1,s}$, $c_{2,s}$. We have
\begin{equation}
\tilde a_{i,r}(\omega),\ \tilde f_{i,r}(\omega)\sim\left\{
\begin{split}
&c_{2,s}(\omega)\qquad\omega>0,\\
&c_{1,s}(\omega)\qquad\omega<0.
\end{split}\right.
\end{equation}
Substituting the solution for $\tilde a_{i,r}$ and $\tilde f_{i,r}$ into
(\ref{hatYi}) we find $\hat Y_i=0$, as expected (to this order), 
since the expansion of the Y-functions starts at ${\cal O}(\varepsilon^2)$, 
order 1.

We now turn to order 1. We make the substitutions
\begin{equation}
\begin{split}
\hat B_{j,w}&\rightarrow\left(b^{(w)}_{j,s}\right)_c
\widetilde{\left(b^{(w)}_{j,s}\right)^{(1m)}},\\
\hat D_{j,w}&\rightarrow\left(d^{(w)}_{j,s}\right)_c
\widetilde{\left(d^{(w)}_{j,s}\right)^{(1m)}},
\end{split}
\qquad
\begin{split}
\hat b_{i,r}&\rightarrow\widetilde{\left(b^{(r)}_{i,s}\right)^{(1m)}}
+\widetilde{\left(b^{(r)}_{i,s}\right)^{(1d)}},\\
\hat d_{i,r}&\rightarrow\widetilde{\left(d^{(r)}_{i,s}\right)^{(1m)}}
+\widetilde{\left(d^{(r)}_{i,s}\right)^{(1d)}}
\end{split}
\end{equation}
and
\begin{equation}
\hat p_j\rightarrow \tilde Z_{j,s},\qquad\quad
\hat Y_i\rightarrow \tilde Z_{i,s+1}.
\end{equation}
Recall that we treat the variables of type $\widetilde{x^{(1m)}}$ as our 
independent variables and solve (\ref{hatbir}) and (\ref{hatdir}) for them
in terms of the sources: the variables $\tilde Z_{j,s}$ and the variables of 
type $\widetilde{x^{(1d)}}$. This solution is then substituted
into (\ref{hatYi}) and the result is that $\tilde Z_{i,s+1}$ is expressed
in terms of the sources. Due to the linearity of the problem
we can consider the two sources separately. 

First we solve the problem 
corresponding to the sources $\tilde Z_{j,s}$. Substituting this part of the
solution to (\ref{hatYi}) we find that the homogeneous parts (terms 
proportional to ${\cal A}_s$ and $p_o$, ${\cal C}_s$) of (\ref{BZZ}) and 
(\ref{Z2rec}) are reproduced.

The case of the inhomogeneous terms is much more complicated. Here 
$\widetilde{x^{(1d)}}$ type terms act as sources for the solution.
These are known from the order 1/2 solution. More precisely, the functions
\begin{equation}
a_{i,r}=b^{(r)(1/2)}_{i,s},\qquad
f_{i,r}=d^{(r)(1/2)}_{i,s}
\label{airfir}
\end{equation}
can be considered as known from the order 1/2 solution, since the order
1/2 solution is given in terms of the Fourier transform of their derivatives:
\begin{equation}
\tilde a_{i,r}=\widetilde{a^\prime_{i,r}},\qquad
\tilde f_{i,r}=\widetilde{f^\prime_{i,r}}.
\end{equation}
The inhomogeneous case sources are given in terms of the functions 
(\ref{airfir}): 
\begin{eqnarray}
b^{(3)(1d)}_{1,s}&=&-\frac{1}{s+1}a_{1,3}a^\prime_{1,3},\\
b^{(3)(1d)}_{2,s}&=&(a_{2,3}-sC_{2,s})(a_{2,3}-sC_{2,s})^\prime
-a_{2,3}a^\prime_{2,3},\\
b^{(3)(1d)}_{3,s}&=&-\frac{1}{s+1}a_{3,3}a^\prime_{3,3},\\
b^{(2)(1d)}_{1,s}&=&(a_{1,2}+(s-\Delta)C_{2,s-1})
(a_{1,2}+(s-\Delta)C_{2,s-1})^\prime-a_{1,2}a^\prime_{1,2},\\
b^{(2)(1d)}_{2,s}&=&\frac{1}{s-\Delta-1}a_{2,2}a^\prime_{2,2},\\
b^{(1)(1d)}_{1,s}&=&-\frac{1}{s-\Delta}a_{1,1}a^\prime_{1,1},
\end{eqnarray}
and
\begin{eqnarray}
d^{(3)(1d)}_{1,s}&=&-\frac{1}{s+1}f_{1,3}f^\prime_{1,3},\\
d^{(3)(1d)}_{2,s}&=&(f_{2,3}-sC_{1,s})(f_{2,3}-sC_{1,s})^\prime
-f_{2,3}f^\prime_{2,3},\\
d^{(3)(1d)}_{3,s}&=&-\frac{1}{s+1}f_{3,3}f^\prime_{3,3},\\
d^{(2)(1d)}_{1,s}&=&\frac{1}{s-\Delta-1}f_{1,2}f^\prime_{1,2},\\
d^{(2)(1d)}_{2,s}&=&(f_{2,2}+(s-\Delta)C_{1,s})
(f_{2,2}+(s-\Delta)C_{1,s})^\prime-f_{2,2}f^\prime_{2,2},\\
d^{(1)(1d)}_{1,s}&=&-\frac{1}{s-\Delta}f_{1,1}f^\prime_{1,1}.
\end{eqnarray}

We have to compute the Fourier transform of the functions listed above. The
generic structure of this calculation is as follows. Given the Fourier 
transform of the derivative of some function ${\cal F}$,
\begin{equation}
\widetilde{\cal F}_1(\omega)=
\int_{-\infty}^\infty{\rm d}u\,{\rm e}^{iu\omega}\,
{\cal F}^\prime(u)=f_p(\omega)+f_m(\omega),
\end{equation}
where
\begin{equation}
f_p(\omega)=0,\quad\omega<0,\qquad\qquad
f_m(\omega)=0,\quad\omega>0,
\end{equation}
find the Fourier transform
\begin{equation}
\widetilde{\cal F}_2(\omega)=
\int_{-\infty}^\infty{\rm d}u\,{\rm e}^{iu\omega}\,
{\cal F}(u){\cal F}^\prime(u).
\end{equation}
In all cases
\begin{equation}
f_p(\omega)\sim c_{2,s}(\omega)\sim a_2(\omega),\qquad
f_m(\omega)\sim c_{1,s}(\omega)\sim a_1(\omega).
\end{equation}
$\widetilde{\cal F}_2(\omega)$ is given by the integrals
\begin{equation}
\left\{
\begin{split}
&\frac{i\omega}{4\pi}\int_0^\omega\frac{{\rm d}\nu}{\nu(\omega-\nu)}
f_p(\nu)f_p(\omega-\nu)+
\frac{i\omega}{2\pi}\int_{-\infty}^0\frac{{\rm d}\nu}{\nu(\omega-\nu)}
f_m(\nu)f_p(\omega-\nu),\quad\ \ \omega>0,\\
&\frac{i\omega}{4\pi}\int_{-\omega}^0\frac{{\rm d}\nu}{\nu(\omega-\nu)}
f_m(\nu)f_m(\omega-\nu)+
\frac{i\omega}{2\pi}\int_0^\infty\frac{{\rm d}\nu}{\nu(\omega-\nu)}
f_p(\nu)f_m(\omega-\nu),\quad\omega<0.
\end{split}\right.
\end{equation}
Thanks to the linearity of the problem, it is sufficient to solve the
problem for the integrands as sources, moreover for each type of integrand
separately. We have four types of contributions:
\begin{equation}
\left.\begin{split}
f_pf_p\ \ {\rm type\ terms}&\sim c_{2,s}(\nu)c_{2,s}(\omega-\nu)\\
f_mf_p\ \ {\rm type\ terms}&\sim c_{1,s}(\nu)c_{2,s}(\omega-\nu)
\end{split}\right\}\qquad\omega>0,
\end{equation}
\begin{equation}
\left.\begin{split}
f_mf_m\ \ {\rm type\ terms}&\sim c_{1,s}(\nu)c_{1,s}(\omega-\nu)\\
f_pf_m\ \ {\rm type\ terms}&\sim c_{2,s}(\nu)c_{1,s}(\omega-\nu)
\end{split}\right\}\qquad\omega<0.
\end{equation}
We can introduce the notation for the $f_mf_p$ part of the function
$G(\omega)$
\begin{equation}
G_{(mp)}(\omega)=\frac{i\omega}{2\pi}\int_{-\infty}^0
\frac{{\rm d}\nu}{\nu(\omega-\nu)}\,G^{(I)}_{(mp)}(\omega,\nu)c_{1,s}(\nu)
c_{2,s}(\omega-\nu)
\end{equation}
and similarly for $G_{(pm)}(\omega)\sim G^{(I)}_{(pm)}(\omega,\nu)$, etc.

The formula for the Fourier transform of the derivative of $\Psi_s$ has
a similar structure:
\begin{equation}
-i\omega\tilde\Psi_s(\omega)=
\left\{
\begin{split}
&\frac{is^2\omega}{2\pi}\int_{-\infty}^0\frac{{\rm d}\nu}{\nu(\omega-\nu)}
c_{1,s}(\nu)c_{2,s}(\omega-\nu),\quad\ \ \omega>0,\\
&\frac{is^2\omega}{2\pi}\int_0^\infty\frac{{\rm d}\nu}{\nu(\omega-\nu)}
c_{2,s}(\nu)c_{1,s}(\omega-\nu),\quad\ \ \ \omega<0.
\end{split}\right.
\end{equation}
Let us concentrate on the $s$ dependence of the integrand in the above formula:
\begin{equation}
{\rm integrand\ of\ }\tilde\Psi_s(\omega)\sim
\left\{
\begin{split}
&s^2\left(\frac{k^2}{p}\right)^sa_1(\nu)a_2(\omega-\nu),\quad\ \ \omega>0,\\
&s^2\left(\frac{p}{k^2}\right)^sa_2(\nu)a_1(\omega-\nu),\quad\ \ \ \ \omega<0.
\end{split}\right.
\end{equation}
Here we introduced the notation
\begin{equation}
k={\rm e}^{\frac{\nu}{g}}.
\end{equation}
Using the simple $s$-dependence of the integrand, the summation in 
(\ref{AsBs}) can be performed and we find for the corresponding integrands
\begin{equation}
{\cal B}_{s(mp)}^{(I)}={\cal B}_{s(o)},
\label{intBs}
\end{equation}
where 
\begin{equation}
\begin{split}
{\cal B}_{s(o)}&=-\frac{sp_o^2k_o^2}{[(s-1)p_o^2-s-1](1-p_o^2k_o^2)^3}
\Big\{p_o^2[
s(s-1)p_o^4k_o^4-2(s^2-1)p_o^2k_o^2\\
&+s(s+1)]-[s(s+1)p_o^4k_o^4-2s(s+2)p_o^2k_o^2+(s+1)(s+2)]\Big\}
\end{split}
\label{bsres}
\end{equation}
and $k_o=k$. ${\cal B}^{(I)}_{s(pm)}$ is given by the same formula, but
with $k_o=1/k$. 

If we now solve the inhomogeneous part of the order 1 problem at the level
of integrands and substitute the solution to (\ref{hatYi}) we find that all
$f_pf_p$ and $f_mf_m$ type contributions vanish and the $f_mf_p$ ($\omega>0$)
and $f_pf_m$ ($\omega<0$) contributions to $\tilde Z_{1,s+1}$ and
$\tilde Z_{3,s+1}$ can be given by integrands that are
precisely the same as (\ref{bsres}). Analogously, the $f_mf_p$ and
$f_pf_m$ type contributions to the integrand of $\tilde Z_{2,s+1}$ are of
the form  
\begin{equation}
{\cal D}^{(I)}_{s(mp)}={\cal D}_{s(o)} {\rm \ \ with\ \ }k_o=k,\qquad\quad 
{\cal D}^{(I)}_{s(pm)}={\cal D}_{s(o)} {\rm \ \ with\ \ }k_o=1/k, 
\end{equation}
where
\begin{equation}
\begin{split}
{\cal D}_{s(o)}&=\frac{p_ok_o^2}{[(s-1)p_o^2-s-1](1-p_o^2k_o^2)^3}
\Big\{k_o^4 p_o^8 s^3-k_o^4 p_o^8 s^2-2 k_o^4 p_o^6 s^2-k_o^4 p_o^4 s^3\\
&-k_o^4 p_o^4 s^2
-2 k_o^2 p_o^6 s^3+3 k_o^2 p_o^6 s-k_o^2 p_o^6
+4 k_o^2 p_o^4 s^2+2 k_o^2 p_o^4 s
+2 k_o^2 p_o^2 s^3+4 k_o^2 p_o^2 s^2\\
&+k_o^2 p_o^2 s -k_o^2 p_o^2+p_o^4 s^3
+p_o^4 s^2-p_o^4 s-p_o^4-2 p_o^2 s^2-2 p_o^2 s-s^3-3 s^2-3 s-1\Big\}
\end{split}
\label{dsres}
\end{equation}
and the coefficient functions satisfy the recursion relation
\begin{equation}
{\cal D}_{s(o)}-\frac{{\cal D}_{s-1(o)}}{p_o^2k_o^2}
+\left(p_o+\frac{1}{p_o}\right)s^2-
\frac{2{\cal B}_{s-1(o)}}{(s-1)p_ok_o^2[(s-1)p_o^2-s-1]}=0,
\end{equation}
which is equivalent to (\ref{Ds}), taking into account the $(p_ok_o^2)^s$
factor in $c_{1,s}c_{2,s}$.
Note that the arbitrary constant $\Delta$ is absent from the final formulas
(\ref{bsres}) and (\ref{dsres}), which is an important check on the overall
consistency of our results.

\section{Summary}

In this paper we derived an alternative finite NLIE description for
the AdS/CFT spectral problem, which we call hybrid-NLIE. The term
hybrid-NLIE was first used in \cite{nlie22} referring to the
property of the equations that semi-infinite parts of the infinite
Y-system are resummed by appropriate NLIE functions
which are coupled to the rest (unsummed part) of the Y-functions. 
Our equations differ in various aspects from the recently published 
finite FiNLIE \cite{GKLV11} formulation of the spectral problem.

The main differences (apart from the obvious differences in
the derivation as well as the construction of NLIE unknowns) 
are as follows: our NLIE is defined on the mirror sheet, while in
ref. \cite{GKLV11} also the \lq\lq magic sheet" is important. Our equations 
have a more conventional form than those of ref. \cite{GKLV11} but at 
the price of having more variables than in the FiNLIE. 
The unknowns of \cite{GKLV11} are discontinuities
along (short) cuts of the \lq\lq magic sheet", while our unknowns are complex
functions on the whole real line.  In \cite{GKLV11} the derivation is an
appropriate generalization of the methods of \cite{GKV08},
\cite{KL10} and it is based on the Wronskian solutions of the
T-system \cite{ujKazi}, \cite{Wronskian} on the \lq\lq magic sheet". In our
derivation we remain on the mirror sheet and work in the spirit of
\cite{KBP}, the very first NLIE paper in the literature, and its
generalizations \cite{Suz98},\cite{nlie21},\cite{nlie22}, where the
NLIE originates from the TQ-relations of the integrable model under
consideration. Here because of the involved nesting structure of the
problem the derivation is based on a set of hierarchical B\"acklund
equations of the corresponding T-system.

The starting point of the derivation of the hybrid-NLIE was the
quasi-local formulation of the mirror TBA \cite{BH10}. In this reformulation
of the simplified mirror TBA equations \cite{AF09d} two SU$(2)$ and an 
SU$(4)$ type semi-infinite sub Y-systems are coupled by the
quasi-local TBA equations to the central Y-functions. 
Our approach to get the hybrid-NLIE description was to
transform the SU$(N)$ type semi-infinite sub Y-systems into
hybrid-NLIEs. This required to derive two SU$(2)$ type
and an SU$(4)$ type hybrid-NLIEs.
The SU$(2)$ type hybrid-NLIEs have already been derived in \cite{RyoHybrid},
and in this paper we completed the derivation by constructing the
missing SU$(4)$ type hybrid-NLIE.
The derivation of these hybrid-NLIEs proceeds in three steps: finding the
proper set of unknown variables, the functional relations they satisfy and 
their analytic properties.

To each semi-infinite SU$(N)$ type sub Y-system of the whole AdS/CFT 
Y-system, 
there corresponds an infinite SU$(N)$ type sub T-system. The rank 
(w.r.t. SU$(N)$) of these sub T-systems  can be reduced via subsequent 
B\"acklund transformations, connecting
the T-systems corresponding to neighboring levels of the nesting procedure.
The unknown functions are constructed from the B\"acklund equations
corresponding to this nested hierarchy. Every unknown is a simple 
multiplicative expression composed of T-functions of neighboring levels and
their inverses. This multiplicative structure allowed us
to derive the functional relations connecting the NLIE unknowns and 
the Y-functions at each level of the  nesting. 
Another advantage of this construction is that analyticity information 
is available for the NLIE unknowns if the analytic properties of the
T-functions are known at all levels of the nesting.

The analyticity information on the nested T-functions consists of 2 pieces. 
First of all, in order to be able to derive the NLIE for a given state, 
we have to know the analyticity domains where the T-functions are free of 
discontinuities and are close to the asymptotic solution. Furthermore we need 
to know (qualitatively) the positions of their point-like singularities 
(poles and zeroes). 
The information on the analyticity domains were extracted from the known 
analyticity properties of Y-functions at the highest level of the nesting 
and the explicit construction of B\"acklund transformations allowed us to 
determine the analyticity domains at lower levels of the nesting as well. The 
qualitative information on point-like singularities is encoded in the 
asymptotic solution that was calculated in section~5 for an arbitrary state in 
the undeformed model and for the  ground state of the $\gamma$-deformed  
theory.

The knowledge of the functional relations satisfied by the NLIE unknowns, 
of their analyticity domains together with their qualitative singularity 
structure given by the asymptotic solution makes it possible to determine the 
form of the hybrid-NLIE of AdS/CFT for any excited state  of the theory in a 
quite straightforward manner\footnote{For excited states the hybrid-NLIE
has to be supplemented by quantization conditions. The most important
of these are the exact Bethe equations, but they pose no extra problem
since only the central Y-functions appear in the formulas.}. 
However, for our present purposes in this paper 
it was sufficient to give the explicit form of the hybrid-NLIE  only for the 
simplest nontrivial state, namely the ground state of the $\gamma$-deformed 
theory. This state has non-zero energy, has the simplest singularity structure
and therefore it is ideal for describing the structure of the hybrid-NLIE and 
to test it analytically in the large volume (small coupling) limit.

The hybrid-NLIE equations can be grouped into four sets of equations.
Three of them correspond to the left-wing and right-wing SU$(2)$ type 
hybrid-sub-NLIEs 
and the SU$(4)$ type hybrid-sub-NLIE. These three hybrid-sub-NLIEs are joined 
together to a closed set of equations by the fourth group of equations: the 
so-called central node quasi-local TBA equations.
We list below all four groups of equations specified for the the ground state 
of the $\gamma$-deformed theory.

The first two groups of equations are formed by the left and right
wing SU$(2)$ hybrid-sub-NLIEs. These equations  tell us how the
SU$(2)$ NLIE functions couple to each other and the rest of
the Y-functions. They are obtained from
(\ref{7.18},\ref{7.19}) with $m=1$ to get the truncation with a minimal
number of unknowns:
\begin{eqnarray}  \! \! \! \! \! \! \! \! \! \! \! \! \! \! \! \! \! \!
\ln \,\mathfrak{b}_1^{(\alpha)}&=&H\star \ln
\,\mathfrak{B}_1^{(\alpha)}-H^{[-2\eta]}\star
\ln \,\mathfrak{D}_1^{(\alpha)}+s^{[1-\eta]}\star \ln \,
(1+Y^{(\alpha)}_{1\vert w})
+Cb_1^{(\alpha)} ,\\
 \! \! \! \! \! \! \! \! \! \! \! \! \! \! \! \! \! \!\ln
\,\mathfrak{d}_1^{(\alpha)}&=&H\star \ln
\,\mathfrak{D}_1^{(\alpha)}-H^{[2\eta]}\star \ln
\,\mathfrak{B}_1^{(\alpha)}+s^{[\eta-1]}\star \ln
\,(1+Y^{(\alpha)}_{1\vert w})+Cd_1^{(\alpha)}, \\
 \! \! \! \! \! \! \! \! \! \! \! \! \! \! \! \! \! \!
Y^{(\alpha)}_{1\vert w}&=&\exp \left\{\! s^{[\eta-1]} \star
\ln\mathfrak{B}_1^{(\alpha)} +s^{[1-\eta]} \star
\ln\mathfrak{D}_1^{(\alpha)} \! + \!
\ln\left[\frac{1-\frac{1}{Y_-^{(\alpha)}}}{1-\frac{1}{Y_+^{(\alpha)}}}\right]
\! \hat\star\ s \right\}, \qquad \alpha=\pm,
\end{eqnarray}
where the Fourier form of the kernel $H(u)$ is given by
(\ref{7.15}), $0<\eta<1$ is a small shift parameter, and
the $+$ and $-$ values of the index $^{(\alpha)}$ refer to the right and left 
wing SU$(2)$ NLIEs, respectively.

$\mathfrak{b}_1^{(\alpha)},\mathfrak{d}_1^{(\alpha)}$ together with 
$Y^{(\alpha)}_{1\vert w}$ constitute the set of unknowns of the SU$(2)$ type 
hybrid-sub-NLIEs. We recall that
$\mathfrak{B}_1^{(\alpha)}=1+\mathfrak{b}_1^{(\alpha)}$,
$\mathfrak{D}_1^{(\alpha)}=1+\mathfrak{d}_1^{(\alpha)}$. 
The integration constants given in terms of the deformation parameters of the 
$\gamma$-deformed theory are as follows:
$Cd_1^{(+)}=-Cb_1^{(+)}=i \frac{\gamma_3+\gamma_2}{2}L$ 
and $Cd_1^{(-)}=-Cb_1^{(-)}=i \frac{\gamma_3-\gamma_2}{2}L$.

The next group of equations form the upper SU$(4)$ hybrid-NLIE. They are 
given by (\ref{mathfrakNLIE}) and (\ref{FdNLIE1}) taken at $s=3$:
\begin{eqnarray}
\ln  \mathfrak{b}_{a,3}^{(r)} &=& \sum_{p=1}^{3} \, \sum_{a'=1}^{p} \, 
({G}_{bB})_{aa'}^{(rp)}
\star \ln  \mathfrak{B}_{a',3}^{(p)}+\sum_{p=1}^{3} \, \sum_{a'=1}^{p} \,
({G}_{bD})_{aa'}^{(rp)}  \star  \ln \mathfrak{D}_{a',3}^{(p)}+\sum_{a'=1}^{3}
 \, ({G}_{bY})_{aa'}^{(r)}  \star
\ln  \mathfrak{Y}_{a',3} \nonumber \\
\ln \mathfrak{d}_{a,3}^{(r)} &=& \sum_{p=1}^{3} \, \sum_{a'=1}^{p} \, 
({G}_{dB})_{aa'}^{(rp)}
\star \ln  \mathfrak{B}_{a',3}^{(p)}+\sum_{p=1}^{3} \, \sum_{a'=1}^{p} \,
({G}_{dD})_{aa'}^{(rp)}
\star \ln  \mathfrak{D}_{a',3}^{(p)}+\sum_{a'=1}^{3} \, ({G}_{dY})_{aa'}^{(r)}
\star \ln  \mathfrak{Y}_{a',3} \nonumber \\
\ln  \mathfrak{y}_{a,3} &=& {s}^{[-1+\epsilon_a-\gamma_a^{(3)}]}
\star \ln \! \mathfrak{B}_{a,3}^{(3) } \, +
s^{[1+\epsilon_a-\eta_a^{(3)}]} \star \ln \!
\mathfrak{D}_{a,3}^{(3)} \,+ s^{[\epsilon_a-\gamma_{a+1}^{(3)}]} \star
\ln \! \left(
\frac{\mathfrak{b}_{a+1,3}^{(3)}}{\mathfrak{B}_{a+1,3}^{(3)}} \right) \,
 \\
&+& s^{[\epsilon_a-\eta_{a-1}^{(3)}]} \star \ln \! \left(
\frac{\mathfrak{d}_{a-1,3}^{(3)}}{\mathfrak{D}_{a-1,3}^{(3)}}\right) +
{s}^{[\epsilon_a]}
 \star \ln{Y}_{a,2} -{s}^{[\epsilon_a-\epsilon_{a-1}]} \star \ln \!
\left(\frac{\mathfrak Y_{a-1,3}}{\mathfrak y_{a-1,3}} \right)  \, \nonumber \\
&-&{s}^{[\epsilon_a-\epsilon_{a+1}]} \star \ln \!
\left(\frac{\mathfrak Y_{a+1,3}}{\mathfrak y_{a+1,3}} \right). \nonumber
\end{eqnarray}
Here the first two equations are for the $6+6$ NLIE variables
$\mathfrak{b}_{a,3}^{(r)}$ and $\mathfrak{d}_{a,3}^{(r)}$, $r=1,2,3$, 
$a=1,\dots,r$ and the kernels appearing here are given in appendix~C.
In the third equation $Y_{a,2}$ are given by $(\ref{iden2})$ taken at $s=2$
and the additional variables are $(\mathfrak y_{1,3},\mathfrak y_{2,3},
\mathfrak y_{3,3})=
(Y_{2|vw}^{(+)\, [\epsilon_1]}),1/Y_3^{[\epsilon_2]},Y_{2|vw}^{(-)\, 
[\epsilon_3]})$.
Here it is understood that terms with $a$-type indices \lq\lq out of range''
are omitted. 
Recall that $\mathfrak{B}_{a,3}^{(p)}=1+\mathfrak{b}_{a,3}^{(p)}$ etc. 

The last group of equations is given by the central part of the quasi-local 
TBA equations:
\begin{equation}
\ln Y_2=-s \star \ln \left(1+\frac{1}{Y_1}\right)-s^{[-\epsilon_2]} \star 
\ln \mathfrak Y_{2,3}+
\sum_{\alpha=\pm} \, \ln \left(1+\frac{1}{Y_{1|vw}^{(\alpha)}}\right) \star s,
\end{equation}
\begin{equation}
\ln Y_{1|vw}^{(+)}=s^{[-\epsilon_1]} \star \ln \mathfrak Y_{1,3} -s
\star \ln (1+Y_2)+
\ln \frac{1-Y_{-}^{(+)}}{1-Y_{+}^{(+)}} \, \hat{\star} \, s,
 \end{equation}
 \begin{equation}
 \ln Y_{1|vw}^{(-)}=s^{[-\epsilon_3]} \star \ln \mathfrak Y_{3,3} -s
  \star \ln (1+Y_2)+
 \ln \frac{1-Y_{-}^{(-)}}{1-Y_{+}^{(-)}} \, \hat{\star} \, s,
\end{equation}
\begin{equation}
\frac{Y_-^{(\alpha)}}{Y_+^{(\alpha)}}= \exp\left\{-L_1 \star K_{1y}
-\Omega(K_{Qy}) \right\},
\end{equation}
\begin{equation}
\begin{split}
Y^{(\alpha)}_+Y^{(\alpha)}_-= &\exp\Bigg\{
2\ln\left[\frac{1+Y^{(\alpha)}_{1\vert vw}}{1+Y^{(\alpha)}_{1\vert
w}} \right]\star s + L_1\star\left[
-K_1+2K^{11}_{xv}\star s\right]  \\
& -\Omega(K_Q)+2 \, \Omega(K^{Q1}_{xv}\star s) \Bigg\},
\end{split} 
\end{equation}
\begin{equation}
\begin{split}
\ln Y_1=&-L\tilde{\cal E}_1+
\sum_{\alpha=\pm} \, r_1^{(\alpha)}\star s\ \hat\star\ K_{y1}\\
&- \sum_{\alpha=\pm} \left( \ln\left[\frac{1-Y_-^{(\alpha)}}
{1-Y_+^{(\alpha)}}\right]
\ \hat\star\ s\star K^{11}_{vwx}
+{\mathscr L}_-^{(\alpha)}\ \hat\star\ K^{y1}_-+{\mathscr L}_+^{(\alpha)}
\ \hat\star\ K^{y1}_+ \right)  \\
&\qquad+ L_1\star K^{11}_{{\alg{sl}(2)}}
+\Omega(K^{Q1}_{{\alg{sl}(2)}})+2 \,\Omega(s\star K^{Q-1\,1}_{vwx}),
\end{split}
\end{equation}
where $\Omega({\cal K}_Q)$ is a linear functional of a vector kernel 
${\cal K}_Q$ given by (\ref{final}) and depends also linearly on the
logarithmic expressions (\ref{jeloles}) of the central 
Y-functions: $Y_1,Y_2,Y_{1|vw}^{(\alpha)},
Y_{1|w}^{(\alpha)},Y_{\pm}^{(\alpha)}$. The explicit forms of the 
kernels of this group of equations are listed in appendix~A.

The ground state problem consists of these four sets of equations
supplemented by requirements on the large $u$ behavior,
the boundary conditions (\ref{largeuvw}) and (\ref{largeupm}). 
This problem turned out to be an ideal analytical 
testing ground for the correctness of the NLIE equations. In section~8 
we have shown that the hybrid-NLIE results for the wrapping corrections up to
2nd order are in agreement with those of earlier TBA computations 
\cite{TBA2ndLusch}.
The equations for excited states and their numerical solution will be
discussed in a future publication.

 \vspace{1cm}
{\tt Acknowledgements}

\noindent 
This investigation was supported by the Hungarian National Science Fund OTKA (under K 77400).

\appendix

\section{Notations and TBA kernels}
\label{notations}

In this paper we adopted the definitions and conventions of ref. \cite{Arutyunov:2009ax}. 
For completeness, in this appendix we collect these definitions and give a list
of all kernel functions used  in the paper.

We use the notation $f^{\pm}(u)=f(u\pm \frac{i}{g})$ for any
function $f$ and in general $f^{[a]}(u)=f(u+\frac{i}{g} a)$. We will
also use $w^\pm=w\pm\frac{i}{g}$ for $w$ some parameter.

Most of the kernels and also the asymptotic solution of the Y-system is expressed
in terms of the function $x(u)$:
\begin{equation}
x(u)=\frac12 (u-i\sqrt{4-u^2}), \qquad \mbox{Im}\, x(u)<0,
\end{equation}
which maps the $u$-plane with cuts $[-\infty,-2] \cup [2,\infty]$ onto the physical region
of the mirror theory, and the function $x_s(u)$
\begin{equation}
x_s(u)=\frac{u}{2} \left(1+\sqrt{1-\frac{4}{u^2}}  \right), \qquad |x_s(u)|\geq 1,
\end{equation}
which maps the $u$-plane with the short cut $[-2,2]$ onto the
physical region of the string theory. Both functions satisfy the
identity $x(u)+\frac{1}{x(u)}=u$ and they are related by
$x(u)=x_s(u),$ and $x(u)=1/x_s(u)$ in the lower and upper halves of
the complex plane respectively.

The momentum $\tilde{p}^Q$ and the energy $\tilde{\cal{E}}_Q$ of a
mirror $Q$-particle are expressed in terms of $x(u)$ as follows
\begin{eqnarray}
 \tilde{p}_Q=g x\big(u-\frac{i}{g}Q\big)-g
x\big(u+\frac{i}{g}Q\big)+i Q\, ,
~~~~~\tilde{\cal{E}}_Q=\log\frac{x\big(u-\frac{i}{g}Q\big)}{x\big(u+\frac{i}{g}Q\big)}\,.
 \end{eqnarray}
Two different types of convolutions appear in the quasi-local TBA
equations. These are:
\begin{eqnarray}
\nonumber &&f \star {\cal K}(v) \equiv \int_{-\infty}^\infty\, {\rm d}u\,
f(u) \, { \cal K}(u,v)\,, \quad f \, {\, \hat{\star} \,}{\cal K}(v)
\equiv \int_{-2}^2\, {\rm d}u\, f(u) \, {\cal K}(u,v)\,. \label{convs}
\end{eqnarray}
In addition, we also use the standard definition
for convolutions with kernel functions depending on rapidity differences
only:
\begin{equation}
k\star f(v)=\int_{-\infty}^\infty{\rm d}u\,k(v-u)f(u).
\end{equation}
This definition is convenient because it is equivalent to ordinary
multiplication in Fourier space. It was used for example in (\ref{sol})
and also for the NLIE equations in section 7. 

The kernels and kernel vectors entering the mirror TBA equations can
be grouped into two sets. The kernels from the first group are
functions of only the difference of the rapidities, while kernels
form the other group are not of difference type.

We start with listing kernels depending on a single variable:
\begin{alignat}{2}
s (u) & = \frac{1}{2 \pi i} \, \frac{d}{du} \log t^-(u)= {g \over 4\cosh {\pi g u \over 2}}\,,\quad t(u)=\tanh[
\frac{\pi g}{4} u ]\,,
\nonumber \\
K_Q (u) &= \frac{1}{2\pi i} \, \frac{d}{du} \, \log S_Q(u) = \frac{1}{\pi} \, \frac{g\, Q}{Q^2 + g^2
u^2}\,,\quad S_Q(u)= \frac{u - \frac{iQ}{g}}{u + \frac{i Q}{g}} \,, \nonumber\\
K_{MN}(u) &= \frac{1}{2\pi i} \, \frac{d}{du} \, \log
S_{MN}(u)=K_{M+N}(u)+K_{N-M}(u)+2\sum_{j=1}^{M-1}K_{N-M+2j}(u)\,,\nonumber\\
S_{MN}(u) &=S_{M+N}(u)S_{N-M}(u)\prod_{j=1}^{M-1}S_{N-M+2j}(u)^2 =S_{NM}(u)\,. \label{sKQ}
\end{alignat}
The fundamental building block of kernels which are not of difference type is:
\begin{eqnarray}
 K(u,v) = \frac{1}{2 \pi i} \, \frac{d}{du} \, \log S (u,v) = \frac{1}{2 \pi i} \,
\frac{ \sqrt{4-v^2}}{\sqrt{4-u^2}}\, {1\over u-v} \,,\ \  S(u,v)=\frac{x(u) - x(v)}{x(u) x(v) - 1}\,.~~~
\label{Kuv}
 \end{eqnarray}
Using the kernels $K(u,v)$ and $K_Q(u-v)$ it is possible to define a series of kernels which are connected
to the fermionic $Y_{\pm}^{(\alpha)}$-functions.
They are:
\begin{eqnarray}
K_{Qy}(u,v)&=&K(u-\frac{i}{g}Q,v)-K(u+\frac{i}{g}Q,v)\,, \label{KQy}\\
K^{Qy}_\mp(u,v)&=&{1\over 2}\Big( K_Q(u-v) \pm  K_{Qy}(u,v)\Big) \label{KQypm}
\end{eqnarray}
and
\begin{eqnarray}
K_{yQ}(u,v)&=&K(u,v+{i\over g}Q)-K(u,v-{i\over g}Q), \label{KyQ}  \\
K^{yQ}_\pm(u,v) &=& {1\over 2}\Big(K_{yQ}(u,v)\mp K_Q(u-v)\Big)\, . \label{KyQpm}
\end{eqnarray}
The kernels entering the right hand sides of
(\ref{YpxYm}),(\ref{hybrid}) are
\begin{eqnarray}
\nonumber
K_{xv}^{QM}(u,v) &=&{1\over 2\pi i}{d\over du}\log S_{xv}^{QM}(u,v)\,,\\
\nonumber S_{xv}^{QM}(u,v) &=&
\frac{x(u-i{Q \over g })-x(v+i{M \over g})}{x(u+i{Q \over g })-x(v+i{M
\over g})}\, \frac{x(u-i{Q \over g })-x(v-i{M \over g})}{x(u+i{Q \over g
})-x(v-i{M \over g})}\, \frac{x(u+i{Q \over g })}{x(u-i{Q \over g
})}~~~~\\ &\times
&\prod_{j=1}^{M-1}\frac{u-v-\frac{i}{g}(Q-M+2j)}{u-v+\frac{i}{g}(Q-M+2j)} \label{Sxv}
\end{eqnarray}
and 
\begin{eqnarray}
\nonumber
K_{vwx}^{QM}(u,v) &=&{1\over 2\pi i}{d\over du}\log S_{vwx}^{QM}(u,v)\,,\\
\nonumber S_{vwx}^{QM}(u,v) &=&
\frac{x(u-i{Q \over g })-x(v+i{M \over g})}{x(u-i{Q \over g })-x(v-i{M\over g})}\,
 \frac{x(u+i{Q \over g })-x(v+i{M \over g})}{x(u+i{Q \over g})-x(v-i{M \over g})}\,
 \frac{x(v-i{M \over g })}{x(v+i{M \over g})}~~~~
\\ &\times
&\prod_{j=1}^{Q-1}\frac{u-v-\frac{i}{g}(M-Q+2j)}{u-v+\frac{i}{g}(M-Q+2j)}\,. \label{Svwx}
\end{eqnarray}
The equations for the momentum carrying node (\ref{hybrid}) contain
the dressing phase, an important building block of the
${\alg{sl}(2)}$ S-matrix of the model \cite{AFrev}. It is of the
form
\begin{eqnarray}
\label{Ssl2}
S_{{\alg{sl}(2)}}^{QM}(u,v)= S_{QM}(u-v)^{-1} \,
\Sigma_{QM}(u,v)^{-2}\,,
\end{eqnarray}
where  $\Sigma_{QM}$ is the improved dressing factor \cite{dresscross}.
The corresponding ${\alg{sl}(2)}$ and dressing kernels are defined in the
usual way
\begin{eqnarray}
K_{\alg{sl}(2)}^{QM}(u,v)= \frac{1}{2\pi i}\frac{d}{du}\log
S_{\alg{sl}(2)}^{QM}(u,v) \,,\quad K_{QM}^{\Sigma}(u,v)=\frac{1}{2\pi
i}\frac{d}{du}\log \Sigma_{QM}(u,v)\,.~~~~
\end{eqnarray}
The asymptotic solution and the source terms in the excited state
generalization of the TBA equations involve the ${\alg{sl}(2)}$
S-matrix analytically continued to the physical region in the first
argument.
\begin{eqnarray}\nonumber
S_{\alg{sl}(2)}^{1_*M}(u,v)&=&{1\over
S_{1M}(u-v)\Sigma_{1_*M}(u,v)^2}\,.~~~~
\end{eqnarray}
Explicit expressions for the improved dressing factors $\Sigma_{QM}(u,v)$ and $\Sigma_{1_*M}(u,v)$
can be found in section 6 of ref. \cite{dresscross}.

In the quasi-local TBA formulation the vector $\delta {\cal K}_Q$
defined by (\ref{kerid0}) must be known.
Since in the expression (\ref{final}) and thus in the quasi-local TBA formulation
 $\delta {\cal K}_Q$ with $Q \geq 2$ appear only, we
list them for this set of the indexes.
They are given by:
\begin{eqnarray}
\delta K_{Q} &=& 0, \qquad \qquad \quad \delta K_{Qy} = 0,\qquad \qquad \delta K^{Q1}_{xv} = 0,
\qquad \qquad Q\geq 2,
\label{dKQszeruek} \\
\delta(s \star K^{Q-1, 1}_{vwx}) &=& \delta_{Q,2} \, \, s \, \star s \, \hat{\star} \, K_{y1}
, \qquad \qquad
\delta K^{Q 1}_{\alg{sl}(2)} = - \delta_{Q,2} \, \, s, \qquad \qquad \ Q \geq 2. \label{dKsl2}
\end{eqnarray}
Using (\ref{dKQszeruek}-\ref{dKsl2}) and (\ref{final}) all the
$\Omega({\cal K}_Q)$ terms can be explicitly evaluated. The
substitution of these expressions into the equations
(\ref{TBAmvw}-\ref{hybrid}) completes the quasi-local form of the
mirror TBA equations.

\section{Asymptotic solution}
\label{AsySol}

The asymptotic solutions of the vertical SU$(4)$ and the horizontal
SU$(2)$ parts of the T-hook were given at all levels of the nesting
in section 5 and 6. The solution is built from the asymptotic
Q-functions of the left (L) and right (R) SU$(2|2)$ T-systems to
which the T-hook T-system splits in the  asymptotic limit. In this
appendix we collect the T-functions and the basic building elements
such as the Q-functions of the relevant SU$(2|2)$ fat-hook solution
\cite{GKV09}. We consider $N$ magnon states with magnon rapidities
$u_j$ and introduce the following functions:
\begin{equation}
R_m(u)=\prod_{j=1}^{N} \, \frac{x(u)-x_j^+}{({x_j^+})^{\frac12}}, \qquad B_m(u)=\prod_{j=1}^{N} \,
\frac{\frac{1}{x(u)}-x_j^+}{({x_j^+})^{\frac12}},    \label{RB_m}
\end{equation}
\begin{equation}
R_p(u)=\prod_{j=1}^{N} \, \frac{x(u)-x_j^-}{({x_j^-})^{\frac12}},
\qquad B_p(u)=\prod_{j=1}^{N} \,
\frac{\frac{1}{x(u)}-x_j^-}{({x_j^-})^{\frac12}}, \label{RB_p}
\end{equation}
where $x_j^{\pm}=x_s(u_j\pm \frac{i}{g})$. These functions satisfy the relation
\begin{equation}
R_m^+(u) \, B_m^+(u)=R_p^-(u) \, B_p^-(u)=(-1)^N \, Q(u),
\end{equation}
with $Q(u)=\prod_{j=1}^{N} (u-u_j)$.
For general states auxiliary Bethe roots will also appear in the formulae. To take into account
their contribution as well, we need to introduce the following functions:
\begin{equation}
R_l(u)=\prod_{j=1}^{K_l} \,
\frac{x(u)-y_{l,j}}{(y_{l,j})^{\frac12}}, \qquad
B_l(u)=\prod_{j=1}^{K_l} \,
\frac{\frac{1}{x(u)}-y_{l,j}}{(y_{l,j})^{\frac12}}, \quad
Q_l(u)=\prod_{j=1}^{K_l} (u-u_{l,j}), \quad l=1,2,3, \label{RBl}
\end{equation}
where $y_{l,j}=x(u_{l,j})$ and they satisfy the relation
\begin{equation}
R_l(u) \, B_l(u)=(-1)^{K_l} \, Q_l(u), \qquad l=1,2,3.  \label{RBQl}
\end{equation}
The sets $\{u_{l,j}\}_{l=1,2,3}$ form the 3 family of Bethe roots
corresponding to the 3 levels of the SU$(2|2)$ nested Bethe
Ansatz. Choosing the $\alg{sl}(2)$ grading for the reference state
 they satisfy the asymptotic SU$(2|2)$ Bethe equations \cite{GKV09}:
\begin{equation} 
\frac{Q_2^-(u_{1,j})}{Q_2^+(u_{1,j})} \frac{B_p (u_{1,j})}{B_m (u_{1,j})}=1, \qquad j=1,.., K_1
\label{BA1}
\end{equation}
\begin{equation}
\frac{Q_2^{++}(u_{2,j})}{Q_2^{--}(u_{2,j})}  \frac{Q_1^-(u_{2,j})}{Q_1^+(u_{2,j})}
\frac{Q_3^-(u_{2,j})}{Q_3^+(u_{2,j})}=-1, \qquad j=1,.., K_2
\label{BA2}
\end{equation}
\begin{equation}
\frac{Q_2^-(u_{3,j})}{Q_2^+(u_{3,j})} \frac{R_p(u_{3,j})}{R_m(u_{3,j})}=1, \qquad j=1,.., K_3.
\label{BA3}
\end{equation}
Roughly speaking each type of Bethe root corresponds to a zero
of a Q-function of the system. There are nine
$Q^{(k,m)}|_{m=0,1,2}^{k=0,1,2}$ Q-functions in an SU$(2|2)$
problem. They are not all independent, but connected by the
so-called QQ-relations \cite{KSZ07}:
\begin{equation}
Q^{(k,m)}\,Q^{(k+1,m+1)++}-Q^{(k+1,m+1)} \, Q^{(k,m)++}=Q^{(k,m+1)}\, Q^{(k+1,m)++}, \label{QQ}
\end{equation}
which express the fact that starting from different reference states in the Bethe ansatz description
leads to the same final result for the eigenvalues of T-functions.

The ${\cal W}$ quantum characteristic function \cite{Beisert06} and
its inverse, are generators of the T-functions in the symmetric and
antisymmetric representations respectively,
\begin{equation}
{\cal W}=\sum_{s=0}^{\infty} {\bf t}_{1,s}^{o [s-1]} \, e^{2 \, s \,
\partial_u}, \qquad {\cal W}^{-1}=\sum_{a=0}^{\infty} (-1)^a \, {\bf
t}_{a,1}^{o [a-1]} \, e^{2 \, a \, \partial_u}.
\end{equation}
They were explicitly given in the $\alg{sl}(2)$ grading for the SU$(2|2)$ asymptotic solution in \cite{GKV09}
 and  can be expressed in terms of the $Q^{(k,m)}$ functions as follows:
\begin{eqnarray}
{\cal W}=\left(1-\frac{Q^{(2,2)--}}{Q^{(2,2)}}\frac{Q^{(2,1)++}}{Q^{(2,1)}} \, e^{2 \, \partial_u}  \right)
\left(1-\frac{Q^{(2,1)++}}{Q^{(2,1)}}\frac{Q^{(1,1)--}}{Q^{(1,1)}} \, e^{2 \, \partial_u}  \right)^{-1} \nonumber \\
\left(1-\frac{Q^{(1,1)++}}{Q^{(1,1)}}\frac{Q^{(0,1)--}}{Q^{(0,1)}}
\, e^{2 \, \partial_u}  \right)^{-1}
\left(1-\frac{Q^{(0,0)++}}{Q^{(0,0)}}\frac{Q^{(0,1)--}}{Q^{(0,1)}}
\, e^{2 \, \partial_u}  \right). \label{WQ}
\end{eqnarray}
{Comparing}{\footnote {Fitting the conventions appropriately}} the
expression for ${\cal W}$ given in \cite{GKV09}, to (\ref{WQ}) and
using the QQ-relations (\ref{QQ}) the 9 Q-functions of the SU$(2|2)$
fat-hook can be obtained in the ${\alg sl(2)}$ grading, namely as
functions of the Bethe roots of equations (\ref{BA1}-\ref{BA3}). Using
the function $\Omega$ defined by 
\begin{equation}
\frac{\Omega^+}{\Omega^-}=\frac{R_p}{R_m},
\end{equation} 
the nine Q-functions take the form:
\begin{equation} \label{Q2.}
Q^{(2,2)}=\frac{(-1)^N \, Q}{\Omega^2}, \qquad
Q^{(2,1)}=\frac{Q_1^-}{\Omega \, \Omega^{--}}, \qquad
Q^{(2,0)}=\frac{-{\cal F}^{(0) --}}{\Omega^2}
\frac{Q_1^{[-3]}}{Q_3^{[-3]}} \left( \frac{R_p^-}{R_m^-}\right)^2
\frac{R_p^{[-3]}}{R_m^{[-3]}},
\end{equation}
\begin{eqnarray}
Q^{(1,0)}&=&\frac{-1}{\Omega^{--}} \frac{Q_2^{--}}{Q_3^-} \left(1-\frac{Q_2}{Q_2^{--}} \frac{R_m^-}{R_p^-} \right),
\qquad Q^{(1,1)}=\frac{Q_2}{\Omega}, \qquad \nonumber \\
Q^{(1,2)}&=&\frac{(-1)^N \, Q^{++}}{\Omega^{++}} \frac{Q_2}{Q_1^+}
\left(1-\frac{Q}{Q^{++}}\frac{Q_2^{++}}{Q_2}\frac{R_p^+}{R_m^+}
\right), \label{Q1.}
\end{eqnarray}
\begin{equation} \label{Q0.}
Q^{(0,0)}=1, \qquad Q^{(0,1)}=Q_3^+, \qquad Q^{(0,2)}=  -(-1)^N \,
{\cal G}^{(0) ++} \, Q^{[4]} \, \frac{Q_3^{[3]}}{Q_1^{[3]}}
\frac{R_m^{[3]}}{R_p^{[3]}},
\end{equation}
where are ${\cal{F}}^{(0)}$ and ${\cal{G}}^{(0)}$ are given by
\cite{BH9}:
\begin{equation}
{\cal{F}}^{(0)}=-\frac{Q_3^{-}}{Q_3^{+}}+\frac{R_m^+}{R_p^+} \,
\left( \frac{Q_2^{++} \, Q_3^-}{Q_2 \, Q_3^+} + \frac{Q_2^{--} \, Q_1^+}{Q_2 \, Q_1^-} \right)-
\frac{R_m^- \, R_m^+}{R_p^- \, R_p^+} \, \frac{Q_1^+}{Q_1^-}, \label{F0}
\end{equation}
\begin{equation}
{\cal{G}}^{(0)}=-\frac{Q_3^{-}}{Q_3^{+}}+\frac{B_m^+}{B_p^+} \,
\left( \frac{Q_2^{++} \, Q_3^-}{Q_2 \, Q_3^+} + \frac{Q_2^{--} \, Q_1^+}{Q_2 \, Q_1^-} \right)-
\frac{B_m^- \, B_m^+}{B_p^- \, B_p^+} \, \frac{Q_1^+}{Q_1^-}.     \label{G0}
\end{equation}
Expanding ${\cal W}^{-1}$ the explicit form of the T-functions
corresponding to the antisymmetric representations can be obtained
\cite{BH9}:
\begin{eqnarray}
{\bf t}_{a,1}^o&=&(-1)^{a} \, \frac{Q_3^{[-a]} \,
Q_1^{[a]}}{Q^{[a+1]}} \, \frac{B_p^{[a]}}{B_m^{[a]}} \, \left \{
\frac{Q^{[1+a]}}{Q_1^{[a]} \, Q_3^{[a]}} \,
\frac{B_m^{[a]}}{B_p^{[a]}}+ \frac{Q^{[1-a]}}{Q_3^{[-a]} \,
Q_1^{[-a]}} \, \frac{B_m^{[-a]}}{B_p^{[-a]}}
\right. \nonumber \\
&+& \Theta(a-2) \, \sum_{n=0}^{a-2} \, \frac{Q^{[a-1-2n]}}{Q_3^{[a-2-2n]} \, Q_1^{[a-2-2n]}} \,
\left( \frac{B_m^{[a-2-2n]}}{B_p^{[a-2-2n]}}+\frac{R_m^{[a-2-2n]}}{R_p^{[a-2-2n]}} \right) \label{Ta10} \\
&-& \Theta(a-1) \, \sum_{n=0}^{a-1}
\left. \frac{Q^{[a-1-2n]}}{Q_3^{[a-2-2n]} \, Q_1^{[a-2n]} } \, \left(
\frac{Q_1^{[a-2n]} \, Q_2^{[a-3-2n]}}{ Q_1^{[a-2-2n]} \, Q_2^{[a-1-2n]}}+
\frac{ Q_3^{[a-2-2n]} \, Q_2^{[a+1-2n]}}{ Q_3^{[a-2n]} \, Q_2^{[a-1-2n]}}  \right)\right\}, \nonumber
\end{eqnarray}
where $\Theta(x)$ is the unitstep function such that $\Theta(0)=1$.
The functions defined in (\ref{RB_m}), (\ref{RB_p}), (\ref{RBl}) have
their only discontinuities along the real line, and the B-type
functions are analytical continuations of the R-type functions
through the real cut line, this is why in spite of the seemingly
complicated discontinuity structure of (\ref{Ta10}), it can be shown
that ${\bf t}_{a,1}^o$ is a $(-a,a)$ function.

The T-functions in the symmetric representations can be obtained by
expanding ${\cal W}$:
\begin{eqnarray}
{\bf t}_{1,s}^{o}&=&\frac{1}{Q_1^{[-s]} \, Q_3^{[s]}} \,
\prod_{j=1}^{s-1} \frac{R_m^{[2j-s]}}{R_p^{[2j-s]}} \, \left \{
Q_2^{[-s-1]} \, Q_2^{[s+1]} \, \frac{R_m^{[s]}}{R_p^{[s]}} \,
\sum_{k=0}^{s} \, F_{s,k}
\right. \nonumber \\
&-&
\Theta(s-1) \, \left( Q_2^{[s-1]} \, Q_2^{[-s-1]} \, \sum_{k=0}^{s-1} \, F_{s,k}
-Q_2^{[1-s]} \, Q_2^{[s+1]} \, \frac{B_m^{[-s]}}{B_p^{[-s]}} \, \frac{R_m^{[s]}}{R_p^{[s]}} \,
\sum_{k=1}^{s} F_{s,k} \right) \nonumber \\
&+& \left. \Theta(s-2) \, Q_2^{[1-s]} \, Q_2^{[s-1]} \, \frac{B_m^{[-s]}}{B_p^{[-s]}} \, \sum_{k=1}^{s-1} \, F_{s,k}
\right \},    \label{T1s0}
\end{eqnarray}
where
\begin{equation}
F_{s,k}=\frac{Q_1^{[2k-s]} \, Q_3^{[2k-s]}}{Q_2^{[2k-1-s]} \, Q_2^{[2k+1-s]} }. \label{Fsk}
\end{equation}
Finally the form of the T-functions in the $(1,1)$ gauge on the
interior boundaries of the fat-hook can be read off from the
boundary conditions (\ref{TQ1}-\ref{TQ4}), (\ref{5.10}-\ref{5.15})
and using (\ref{Q2.}),(\ref{Q0.}) their form can be given
explicitly:
\begin{equation}
{\bf t}_{a,2}^o=\left(  {\cal{F}}^{(0)} \, \frac{R_m^-}{R_p^-} \,
\frac{1}{Q^{[-2]}} \, \frac{Q_1^-}{Q_3^-} \right)^{[a]} \, \left(
{\cal{G}}^{(0)} \, \frac{R_m^+}{R_p^+} \, Q^{[2]} \,
\frac{Q_3^+}{Q_1^+} \right)^{[-a]}, \qquad a \geq 2 \label{Ta20}
\end{equation}
\begin{equation}
{\bf t}_{2,s}^o=\left(  {\cal{F}}^{(0)} \, \frac{R_p^-}{R_m^-} \,
\frac{Q_1^-}{Q_3^-} \right)^{[s]} \, \left( {\cal{G}}^{(0)} \,
\frac{R_m^+}{R_p^+} \, \frac{Q_3^+}{Q_1^+} \right)^{[-s]}  \, \left(
\prod_{k=1}^{s-1} \frac{R_m^{[2k+1-s]}}{R_p^{[2k+1-s]}} \right)^2,
\qquad s \geq 2. \label{T2s0}
\end{equation}

In the treatment of the horizontal SU$(2)$ wing of the problem the
T-functions were used in a "cut-free" gauge. Their relation to the
$(1,1)$ gauge expressions (\ref{T1s0}) and (\ref{T2s0}) is given by:
\begin{equation}
\tilde{t}^{1}_{s}={\bf t}^o_{1,s} \, \prod_{j=1}^{s-1}
\frac{R_p^{[2j-s]}}{R_m^{[2j-s]}}, \qquad\qquad
\tilde{t}^{2}_{s}={\bf t}^o_{2,s} \, \left(\prod_{j=1}^{s-1}
\frac{R_p^{[2j+1-s]}}{R_m^{[2j+1-s]}}\right)^2\,. 
\label{T1s0tilde}
\end{equation}
It can be seen from the above explicit formulas that 
$\tilde{t}^{1}_{s}$ are $(-s,s)$ functions and
$\tilde{t}^{2}_{s}$ are of type $(1-s,s-1)$.

In order to complete the asymptotic solution of AdS/CFT we should
discuss the asymptotic behavior of massive Y-functions. The
asymptotic solution for the massive nodes on the AdS/CFT Y-system is
given by
\begin{equation}
{\mathbf y}^o_{a,0}=\eta_a{\mathbf t}^R_{a,1}{\mathbf t}^L_{a,1},
\end{equation}
where the prefactor is a solution of the discrete Laplace equation
and takes the form:
\begin{equation} \label{B.25}
\eta_a=\left(\frac{x^{[a]}}{x^{[-a]}}\right)^{J_{\rm eff}}{\cal D}_a
\frac{\phi^{[-a]}}{\phi^{[a]}}
\end{equation}
with
\begin{equation} \label{B.26}
\phi=\frac{B_{1,L}B_{1,R}}{B_{3,L}B_{3,R}},\qquad {\cal
D}_a=\prod_{k=0}^{a-1}{\cal D}_1^{[a-1-2k]}\equiv
\prod\limits_{j=1}^N S_{\alg{sl}(2)}^{a 1_*}(u,u_j).
\end{equation} 
The parameter $J_{eff}$ is an effective length composed of the
$J$-charge and the numbers of the auxiliary Bethe roots by the
formula $J_{eff}=J+\frac{K_{3,L}-K_{1,L}+K_{3,R}-K_{1,R}}{2}$. The
factor ${\cal{D}}_1 $ is the product of the dressing phases of
fundamental magnons in the mirror-physical channel ${\cal{D}}_1(u)
\equiv \prod\limits_{j=1}^N S_{\alg{sl}(2)}^{1 1_*}(u,u_j),$
\cite{BH9}.

The analytical properties of the dressing phase and the asymptotic
T-functions imply that the ${\mathbf y}^o_{a,0}$ functions are of
type $(-a,a)$ and they decay with a high power of $u$ at 
infinity \cite{BH9}.

 The requirement that on the
physical sheet $1+{\mathbf y}^o_{a,0}$ is zero at the positions of
magnon rapidities leads to the  Beisert-Staudacher asymptotic Bethe
equations \cite{BS}:

\begin{equation}
\left( \frac{x_s^{+}}{x_s^{-}} \right)^{J_{eff}}  S_{\alg{sl}(2)} \,
\,\frac{B_{1,L}^{-}}{R_{1,L}^{+}} \, \frac{R_{3,L}^{-}}{B_{3,L}^{+}}
\,\frac{B_{1,R}^{-}}{R_{1,R}^{+}} \, \frac{R_{3,R}^{-}}{B_{3,R}^{+}}
\bigg|_{u_k} =-1, \qquad k=1,..,N, \label{Beisert}
\end{equation}
where for short we introduced the notation
\begin{equation}
S_{\alg{sl}(2)}(u) =\prod\limits_{j=1}^N S_{\alg{sl}(2)}^{1_*
1_*}(u,u_j) 
\end{equation}
for the dressing factor in the physical-physical region.

\section{NLIE kernels}
\label{kernels}

All the kernels appearing in the equations (\ref{Kkernels}) for 
$b_{a,s}^{(r)}$ and $d_{a,s}^{(r)}$  are composed as sums of their
negative and positive frequency parts. For instance in Fourier
space: 
$$ (\tilde{K}_{bB})_{aa'}^{(rp)}
(\omega)=\Theta(\omega) \, (\tilde{K}^{(+)}_{bB})_{aa'}^{(rp)}
(\omega)+\Theta(-\omega) \, (\tilde{K}^{(-)}_{bB})_{aa'}^{(rp)}
(\omega),$$
$(\tilde{K}^{(\pm)}_{bB})_{aa'}^{(rp)} (\omega)$ are the positive and negative frequency parts respectively
and $\Theta(\omega)$ is the Heaviside function.

Here we list below the negative and positive frequency parts of the kernels.
Recall that $p=e^{\frac{\omega}{g}}$ and introduce two important functions
that eliminate the denominator parts of the kernels.
\begin{equation}
\tilde{N}_{+}(p)=\frac{1}{1+2p^2+p^8} \, \quad \omega \geq 0 , \qquad \tilde{N}_{-}(p)=\frac{1}{1+2p^{-2}+p^{-8}} \,
\quad \omega \leq 0.
\end{equation}

Then the kernels $\tilde{K}_{bB}, \, \tilde{K}_{bD}, \,
\tilde{K}_{dB}, \, \tilde{K}_{dD}$ in (\ref{Kkernels}) are arranged
into $6\times 6$ matrices by identifying the 6 index pairs
$(a,r)_{r=1,...,3, a=1,..,r}$ to the indexes of the $6\times 6$
matrices as follows: $(1,3) \to 1, \quad (2,3) \to 2, \quad (3,3)
\to 3, \quad (1,2) \to 4, \quad (2,2) \to 5, \quad (1,1) \to 6.$
Based on the same identification the kernels $\tilde{K}_{bY}, \,
\tilde{K}_{dY}$ are represented as $6\times 3$ matrices.

Then the positive and negative parts of the kernels in (\ref{Kkernels}) take the form in Fourier
 space as follows:
{ \tiny \begin{equation} \label{KbB+}
\frac{\tilde{K}_{bB}^{(+)}(\omega)}{\tilde{N}_{+}(p)}=\left(
\begin{array}{cccccc}
 1-p^2 & p^7+p & -p^8+p^6-2 p^2+2 & -p^6-p^2-2 & p^7+2
   p-\frac{1}{p} & -p^2-\frac{1}{p^2} \\
 p^3+p & -p^6-p^4 & p^7+p & p^3+2 p+\frac{1}{p} & -p^6-p^4-p^2-1
   & p-p^5 \\
 p^4-p^2 & p^3+p & 1-p^2 & p^4-1 & 2 p & p^4+1 \\
 -p^8-p^4-2 p^2 & p^9+p^7+p^3+p & -p^8+p^6-p^2+1 &
   -p^6-p^4-p^2-1 & 2 p^7+2 p & -p^4-p^2-\frac{1}{p^2}-1 \\
 2 p^3 & -p^8-p^6-p^4-p^2 & p^9+2 p^3-p & 2 p^3+2 p &
   -p^6-p^4-p^2+1 & -p^5+p^3+p+\frac{1}{p} \\
 -p^8-p^4 & p^9-p^5 & p^6+p^2 & -p^8-p^6-p^4-p^2 &
   p^9+p^7-p^5+p^3 & -2 p^4
\end{array}
\right),
\end{equation} }
{ \tiny \begin{equation} \label{KbB-}
\frac{\tilde{K}_{bB}^{(-)}(\omega)}{\tilde{N}_{-}(p)}=\left(
\begin{array}{cccccc}
 1-\frac{1}{p^2} & \frac{1}{p^3}+\frac{1}{p} &
   \frac{1}{p^4}-\frac{1}{p^2} &
   -\frac{1}{p^8}-\frac{1}{p^4}-\frac{2}{p^2} & \frac{2}{p^3} &
   -\frac{1}{p^8}-\frac{1}{p^4} \\
 \frac{1}{p^7}+\frac{1}{p} & -\frac{1}{p^6}-\frac{1}{p^4} &
   \frac{1}{p^3}+\frac{1}{p} &
   \frac{1}{p^9}+\frac{1}{p^7}+\frac{1}{p^3}+\frac{1}{p} &
   -\frac{1}{p^8}-\frac{1}{p^6}-\frac{1}{p^4}-\frac{1}{p^2} &
   \frac{1}{p^9}-\frac{1}{p^5} \\
 -\frac{1}{p^8}+\frac{1}{p^6}-\frac{2}{p^2}+2 &
   \frac{1}{p^7}+\frac{1}{p} & 1-\frac{1}{p^2} &
   -\frac{1}{p^8}+\frac{1}{p^6}-\frac{1}{p^2}+1 &
   \frac{1}{p^9}+\frac{2}{p^3}-\frac{1}{p} &
   \frac{1}{p^6}+\frac{1}{p^2} \\
 -\frac{1}{p^6}-\frac{1}{p^2}-2 & \frac{1}{p^3}+p+\frac{2}{p} &
   \frac{1}{p^4}-1 &
   -\frac{1}{p^6}-\frac{1}{p^4}-\frac{1}{p^2}-1 &
   \frac{2}{p^3}+\frac{2}{p} &
   -\frac{1}{p^8}-\frac{1}{p^6}-\frac{1}{p^4}-\frac{1}{p^2} \\
 \frac{1}{p^7}-p+\frac{2}{p} &
   -\frac{1}{p^6}-\frac{1}{p^4}-\frac{1}{p^2}-1 & \frac{2}{p} &
   \frac{2}{p^7}+\frac{2}{p} &
   -\frac{1}{p^6}-\frac{1}{p^4}-\frac{1}{p^2}+1 &
   \frac{1}{p^9}+\frac{1}{p^7}-\frac{1}{p^5}+\frac{1}{p^3} \\
 -p^2-\frac{1}{p^2} & \frac{1}{p}-\frac{1}{p^5} &
   \frac{1}{p^4}+1 & -\frac{1}{p^4}-p^2-\frac{1}{p^2}-1 &
   -\frac{1}{p^5}+\frac{1}{p^3}+p+\frac{1}{p} & -\frac{2}{p^4}
\end{array}
\right),
\end{equation} }

{ \tiny \begin{equation}
\frac{\tilde{K}_{bD}^{(+)}(\omega)}{\tilde{N}_{+}(p)}=\left(
\begin{array}{cccccc}
 p^2-1 & -p^7-p & p^8-p^6+2 p^2-2 & -p^8-2 p^2+1 &
   p^7-p^5+p-\frac{1}{p} & -p^4-1 \\
 -p^3-p & p^6+p^4 & -p^7-p & p^7+p^5+p^3+p & -p^8-p^6-p^2-1 &
   p^3-p^7 \\
 p^2-p^4 & -p^3-p & p^2-1 & -2 p^2 & p^7+p^3+2 p & p^6+p^2 \\
 p^8+p^4+2 p^2 & -p^9-p^7-p^3-p & p^8-p^6+p^2-1 & -2 p^8-2 p^2 &
   p^9-p^5+p^3-\frac{1}{p} & -p^6-p^4-p^2-1 \\
 -2 p^3 & p^8+p^6+p^4+p^2 & -p^9-2 p^3+p & p^7+p^5+p^3-p & -2
   p^8-2 p^2 & -p^7+p^5+p^3+p \\
 p^8+p^4 & p^5-p^9 & -p^6-p^2 & -p^8+p^6+p^4+p^2 &
   -p^7-p^5-p^3-p & -p^8-2 p^6-2 p^2-1
\end{array}
\right),
\end{equation} }

{ \tiny \begin{equation}
\frac{\tilde{K}_{bD}^{(-)}(\omega)}{\tilde{N}_{-}(p)}=\left(
\begin{array}{cccccc}
 \frac{1}{p^2}-1 & -\frac{1}{p^3}-\frac{1}{p} &
   \frac{1}{p^2}-\frac{1}{p^4} & -\frac{2}{p^2} &
   \frac{1}{p}-\frac{1}{p^5} & -\frac{1}{p^6}-\frac{1}{p^2} \\
 -\frac{1}{p^7}-\frac{1}{p} & \frac{1}{p^6}+\frac{1}{p^4} &
   -\frac{1}{p^3}-\frac{1}{p} &
   \frac{1}{p^7}+\frac{1}{p^5}+\frac{1}{p^3}+\frac{1}{p} &
   -\frac{1}{p^4}-\frac{2}{p^2}-1 & \frac{1}{p^7}-\frac{1}{p^3}
   \\
 \frac{1}{p^8}-\frac{1}{p^6}+\frac{2}{p^2}-2 &
   -\frac{1}{p^7}-\frac{1}{p} & \frac{1}{p^2}-1 &
   -\frac{1}{p^8}-\frac{2}{p^2}+1 &
   \frac{1}{p^7}+\frac{1}{p^3}+\frac{2}{p} & \frac{1}{p^4}+1 \\
 \frac{1}{p^6}+\frac{1}{p^2}+2 & -\frac{1}{p^3}-p-\frac{2}{p} &
   1-\frac{1}{p^4} & -\frac{2}{p^2}-2 &
   -\frac{1}{p^5}-\frac{1}{p^3}+p+\frac{1}{p} &
   -\frac{1}{p^6}-\frac{1}{p^4}-\frac{1}{p^2}-1 \\
 -\frac{1}{p^7}+p-\frac{2}{p} &
   \frac{1}{p^6}+\frac{1}{p^4}+\frac{1}{p^2}+1 & -\frac{2}{p} &
   \frac{1}{p^5}+\frac{1}{p^3}-p+\frac{1}{p} & -\frac{2}{p^2}-2
   & \frac{1}{p^7}+\frac{1}{p^5}-\frac{1}{p^3}+\frac{1}{p} \\
 p^2+\frac{1}{p^2} & \frac{1}{p^5}-\frac{1}{p} &
   -\frac{1}{p^4}-1 &
   \frac{1}{p^6}+\frac{1}{p^4}-\frac{1}{p^2}+1 &
   -\frac{1}{p^5}-\frac{1}{p^3}-p-\frac{1}{p} &
   -\frac{1}{p^6}-p^2-\frac{2}{p^2}-2
\end{array}
\right),
\end{equation} }

{ \tiny \begin{equation}
\frac{\tilde{K}_{bY}^{(+)}(\omega)}{\tilde{N}_{+}(p)}=\left(
\begin{array}{ccc}
 p^8+3 p^2 & p^3+p & p^4-p^6 \\
 -p^3-p & p^8+p^6 & p^7+p^5 \\
 p^2-p^4 & -p^5-p^3 & p^8-p^6 \\
 p^8+p^4+2 p^2 & p^5+2 p^3+p & p^4-p^8 \\
 -2 p^3 & p^8+p^6-p^4-p^2 & 2 p^7 \\
 p^8+p^4 & p^7+2 p^5+p^3 & -p^8+2 p^6+p^4
\end{array}
\right),
\end{equation} }

{ \tiny \begin{equation}
\frac{\tilde{K}_{bY}^{(-)}(\omega)}{\tilde{N}_{-}(p)}=\left(
\begin{array}{ccc}
 \frac{1}{p^6}+\frac{2}{p^2}+1 & \frac{1}{p^5}+\frac{1}{p^3} &
   \frac{1}{p^4}-\frac{1}{p^2} \\
 -\frac{1}{p^7}-\frac{1}{p^5} & -\frac{1}{p^6}+\frac{2}{p^2}+1 &
   \frac{1}{p^3}+\frac{1}{p} \\
 \frac{1}{p^6}-\frac{1}{p^4} & -\frac{1}{p^3}-\frac{1}{p} &
   1-\frac{1}{p^2} \\
 \frac{1}{p^4}+\frac{2}{p^2}+1 &
   \frac{1}{p^5}+\frac{2}{p^3}+\frac{1}{p} & \frac{1}{p^4}-1 \\
 -\frac{2}{p^5} & -\frac{1}{p^6}-\frac{1}{p^4}+\frac{1}{p^2}+1 &
   \frac{2}{p} \\
 -\frac{1}{p^4}+\frac{2}{p^2}+1 & \frac{1}{p^3}+p+\frac{2}{p} &
   \frac{1}{p^4}+1
\end{array}
\right),
\end{equation} }

{ \tiny \begin{equation}
\frac{\tilde{K}_{dB}^{(+)}(\omega)}{\tilde{N}_{+}(p)}=\left(
\begin{array}{cccccc}
 p^2-1 & -p^7-p & p^8-p^6+2 p^2-2 & p^6+p^2+2 & -p^7-2
   p+\frac{1}{p} & p^2+\frac{1}{p^2} \\
 -p^3-p & p^6+p^4 & -p^7-p & -p^3-2 p-\frac{1}{p} &
   p^6+p^4+p^2+1 & p^5-p \\
 p^2-p^4 & -p^3-p & p^2-1 & 1-p^4 & -2 p & -p^4-1 \\
 -2 p^2 & p^7+p^5+p^3+p & -p^8-2 p^2+1 & -2 p^2-2 &
   p^5+p^3+p-\frac{1}{p} & p^6+p^4-p^2+1 \\
 p-p^5 & -p^4-2 p^2-1 & p^7+p^3+2 p & -p^5-p^3+p+\frac{1}{p} &
   -2 p^2-2 & -p^5-p^3-p-\frac{1}{p} \\
 -p^6-p^2 & p^7-p^3 & p^4+1 & -p^6-p^4-p^2-1 & p^7+p^5-p^3+p &
   -p^6-2 p^2-\frac{1}{p^2}-2
\end{array}
\right),
\end{equation} }

{ \tiny \begin{equation} \label{KdB-}
\frac{\tilde{K}_{dB}^{(-)}(\omega)}{\tilde{N}_{-}(p)}=\left(
\begin{array}{cccccc}
 \frac{1}{p^2}-1 & -\frac{1}{p^3}-\frac{1}{p} &
   \frac{1}{p^2}-\frac{1}{p^4} &
   \frac{1}{p^8}+\frac{1}{p^4}+\frac{2}{p^2} & -\frac{2}{p^3} &
   \frac{1}{p^8}+\frac{1}{p^4} \\
 -\frac{1}{p^7}-\frac{1}{p} & \frac{1}{p^6}+\frac{1}{p^4} &
   -\frac{1}{p^3}-\frac{1}{p} &
   -\frac{1}{p^9}-\frac{1}{p^7}-\frac{1}{p^3}-\frac{1}{p} &
   \frac{1}{p^8}+\frac{1}{p^6}+\frac{1}{p^4}+\frac{1}{p^2} &
   \frac{1}{p^5}-\frac{1}{p^9} \\
 \frac{1}{p^8}-\frac{1}{p^6}+\frac{2}{p^2}-2 &
   -\frac{1}{p^7}-\frac{1}{p} & \frac{1}{p^2}-1 &
   \frac{1}{p^8}-\frac{1}{p^6}+\frac{1}{p^2}-1 &
   -\frac{1}{p^9}-\frac{2}{p^3}+\frac{1}{p} &
   -\frac{1}{p^6}-\frac{1}{p^2} \\
 -\frac{1}{p^8}-\frac{2}{p^2}+1 &
   \frac{1}{p^7}+\frac{1}{p^5}+\frac{1}{p^3}+\frac{1}{p} &
   -\frac{2}{p^2} & -\frac{2}{p^8}-\frac{2}{p^2} &
   \frac{1}{p^7}+\frac{1}{p^5}+\frac{1}{p^3}-\frac{1}{p} &
   -\frac{1}{p^8}+\frac{1}{p^6}+\frac{1}{p^4}+\frac{1}{p^2} \\
 \frac{1}{p^7}-\frac{1}{p^5}-p+\frac{1}{p} &
   -\frac{1}{p^8}-\frac{1}{p^6}-\frac{1}{p^2}-1 &
   \frac{1}{p^7}+\frac{1}{p^3}+\frac{2}{p} &
   \frac{1}{p^9}-\frac{1}{p^5}+\frac{1}{p^3}-p &
   -\frac{2}{p^8}-\frac{2}{p^2} &
   -\frac{1}{p^7}-\frac{1}{p^5}-\frac{1}{p^3}-\frac{1}{p} \\
 -\frac{1}{p^4}-1 & \frac{1}{p^3}-\frac{1}{p^7} &
   \frac{1}{p^6}+\frac{1}{p^2} &
   -\frac{1}{p^6}-\frac{1}{p^4}-\frac{1}{p^2}-1 &
   -\frac{1}{p^7}+\frac{1}{p^5}+\frac{1}{p^3}+\frac{1}{p} &
   -\frac{1}{p^8}-\frac{2}{p^6}-\frac{2}{p^2}-1
\end{array}
\right),
\end{equation} }

{ \tiny \begin{equation}
\frac{\tilde{K}_{dD}^{(+)}(\omega)}{\tilde{N}_{+}(p)}=\left(
\begin{array}{cccccc}
 1-p^2 & p^7+p & -p^8+p^6-2 p^2+2 & p^8+2 p^2-1 &
   -p^7+p^5-p+\frac{1}{p} & p^4+1 \\
 p^3+p & -p^6-p^4 & p^7+p & -p^7-p^5-p^3-p & p^8+p^6+p^2+1 &
   p^7-p^3 \\
 p^4-p^2 & p^3+p & 1-p^2 & 2 p^2 & -p^7-p^3-2 p & -p^6-p^2 \\
 2 p^2 & -p^7-p^5-p^3-p & p^8+2 p^2-1 & -p^6-p^4-p^2+1 & 2 p^7+2
   p & p^8+p^6-p^4+p^2 \\
 p^5-p & p^4+2 p^2+1 & -p^7-p^3-2 p & 2 p^3+2 p & -p^6-p^4-p^2-1
   & -p^7-p^5-p^3-p \\
 p^6+p^2 & p^3-p^7 & -p^4-1 & -p^6+p^4+p^2+1 &
   -p^5-p^3-p-\frac{1}{p} & -2 p^4
\end{array}
\right),
\end{equation} }

{ \tiny \begin{equation}
\frac{\tilde{K}_{dD}^{(-)}(\omega)}{\tilde{N}_{-}(p)}=\left(
\begin{array}{cccccc}
 1-\frac{1}{p^2} & \frac{1}{p^3}+\frac{1}{p} &
   \frac{1}{p^4}-\frac{1}{p^2} & \frac{2}{p^2} &
   \frac{1}{p^5}-\frac{1}{p} & \frac{1}{p^6}+\frac{1}{p^2} \\
 \frac{1}{p^7}+\frac{1}{p} & -\frac{1}{p^6}-\frac{1}{p^4} &
   \frac{1}{p^3}+\frac{1}{p} &
   -\frac{1}{p^7}-\frac{1}{p^5}-\frac{1}{p^3}-\frac{1}{p} &
   \frac{1}{p^4}+\frac{2}{p^2}+1 & \frac{1}{p^3}-\frac{1}{p^7}
   \\
 -\frac{1}{p^8}+\frac{1}{p^6}-\frac{2}{p^2}+2 &
   \frac{1}{p^7}+\frac{1}{p} & 1-\frac{1}{p^2} &
   \frac{1}{p^8}+\frac{2}{p^2}-1 &
   -\frac{1}{p^7}-\frac{1}{p^3}-\frac{2}{p} & -\frac{1}{p^4}-1
   \\
 \frac{1}{p^8}+\frac{2}{p^2}-1 &
   -\frac{1}{p^7}-\frac{1}{p^5}-\frac{1}{p^3}-\frac{1}{p} &
   \frac{2}{p^2} & -\frac{1}{p^6}-\frac{1}{p^4}-\frac{1}{p^2}+1
   & \frac{2}{p^3}+\frac{2}{p} &
   -\frac{1}{p^6}+\frac{1}{p^4}+\frac{1}{p^2}+1 \\
 -\frac{1}{p^7}+\frac{1}{p^5}+p-\frac{1}{p} &
   \frac{1}{p^8}+\frac{1}{p^6}+\frac{1}{p^2}+1 &
   -\frac{1}{p^7}-\frac{1}{p^3}-\frac{2}{p} &
   \frac{2}{p^7}+\frac{2}{p} &
   -\frac{1}{p^6}-\frac{1}{p^4}-\frac{1}{p^2}-1 &
   -\frac{1}{p^5}-\frac{1}{p^3}-p-\frac{1}{p} \\
 \frac{1}{p^4}+1 & \frac{1}{p^7}-\frac{1}{p^3} &
   -\frac{1}{p^6}-\frac{1}{p^2} &
   \frac{1}{p^8}+\frac{1}{p^6}-\frac{1}{p^4}+\frac{1}{p^2} &
   -\frac{1}{p^7}-\frac{1}{p^5}-\frac{1}{p^3}-\frac{1}{p} &
   -\frac{2}{p^4}
\end{array}
\right),
\end{equation} }

{ \tiny \begin{equation}
\frac{\tilde{K}_{dY}^{(+)}(\omega)}{\tilde{N}_{+}(p)}=\left(
\begin{array}{ccc}
 1-p^2 & -p^3-p & p^6-p^4 \\
 p^3+p & -p^6+2 p^2+1 & -p^7-p^5 \\
 p^4-p^2 & p^5+p^3 & p^6+2 p^2+1 \\
 2 p^2 & -p^7-p^5+p^3+p & -2 p^6 \\
 p^5-p & p^6+2 p^4+p^2 & p^5+2 p^3+p \\
 p^6+p^2 & p^5+2 p^3+p & -p^6+2 p^4+p^2
\end{array}
\right),
\end{equation} }

{ \tiny \begin{equation} \label{KdY-}
\frac{\tilde{K}_{dY}^{(-)}(\omega)}{\tilde{N}_{-}(p)}=\left(
\begin{array}{ccc}
 \frac{1}{p^8}-\frac{1}{p^6} & -\frac{1}{p^5}-\frac{1}{p^3} &
   \frac{1}{p^2}-\frac{1}{p^4} \\
 \frac{1}{p^7}+\frac{1}{p^5} & \frac{1}{p^8}+\frac{1}{p^6} &
   -\frac{1}{p^3}-\frac{1}{p} \\
 \frac{1}{p^4}-\frac{1}{p^6} & \frac{1}{p^3}+\frac{1}{p} &
   \frac{1}{p^8}+\frac{3}{p^2} \\
 \frac{2}{p^6} &
   \frac{1}{p^7}+\frac{1}{p^5}-\frac{1}{p^3}-\frac{1}{p} &
   -\frac{2}{p^2} \\
 \frac{1}{p^3}-\frac{1}{p^7} & \frac{1}{p^4}+\frac{2}{p^2}+1 &
   \frac{1}{p^7}+\frac{1}{p^3}+\frac{2}{p} \\
 -\frac{1}{p^6}+\frac{2}{p^4}+\frac{1}{p^2} &
   \frac{1}{p^5}+\frac{2}{p^3}+\frac{1}{p} &
   \frac{1}{p^6}+\frac{1}{p^2}
\end{array}
\right).
\end{equation} }

It can be seen from the Fourier representations
(\ref{KbB+}-\ref{KdY-}),(\ref{FourierG}) that the rapidity space
representations of all matrix elements of the kernels in
(\ref{mathfrakNLIE}) can be expressed as linear combinations of the
two denominator functions $N_{\pm}(u)$ with appropriately shifted
arguments. For example let us consider the upper left corner matrix
elements in (\ref{KbB+},\ref{KbB-}). They correspond to
$(G_{bB})_{11}^{(33)}(u)$ and yield the expressions in rapidity
space as follows:
\begin{equation} \label{G3311}
(G_{bB})_{11}^{(33)}(u)=(K_{bB})_{11}^{(33)}(u)=N_{+}(u)-N_{+}(u+\frac{2
\, i}{g})+N_{-}(u)-N_{-}(u-\frac{2 \, i}{g}),
\end{equation}
where $N_{\pm}(u)$ are given as inverse Fourier transforms of $\tilde{N}_{\pm}(e^{\omega/g})$:
\begin{equation}
N_{+}(u)=\int_{0}^{\infty} \, \frac{d \omega}{2 \, \pi} \, e^{-i \, \omega \, u} \,
\tilde{N}_{+}(e^{\omega/g}), \quad \,
N_{-}(u)=\int_{-\infty}^{0} \, \frac{d \omega}{2 \, \pi} \, e^{-i \, \omega \, u} \,
\tilde{N}_{-}(e^{\omega/g}), \quad \,
N_{+}(u)=N_{-}(-u)
\end{equation}
and can be expressed in terms of the incomplete beta function{\footnote{ The incomplete beta function
is defined by its integral representation: $B_z(a,b)=\int_{0}^{z} dt \, t^{a-1} \,(1-t)^{b-1}$}}
$B_z(a,b)$ as follows.
Denote by $a_1,a_2,a_3,a_4$ the four zeroes of the denominator polynomial $1+2 x+x^4$.
They take the form:
\begin{eqnarray}
a_1&=&-1 \nonumber \\
a_2&=&\frac{1}{3} \left(1-\frac{2}{\sqrt[3]{3 \sqrt{33}-17}}+\sqrt[3]{3
   \sqrt{33}-17}\right)\simeq  -0.543689    \nonumber \\
a_3&=&\frac{1}{3}+\frac{1+i \sqrt{3}}{3 \sqrt[3]{3
   \sqrt{33}-17}}-\frac{1}{6} \left(1-i \sqrt{3}\right)
   \sqrt[3]{3 \sqrt{33}-17}\simeq 0.771845 \, + 1.11514  \,i \nonumber \\
a_4&=&\frac{1}{3}+\frac{1-i \sqrt{3}}{3 \sqrt[3]{3
   \sqrt{33}-17}}-\frac{1}{6} \left(1+i \sqrt{3}\right)
   \sqrt[3]{3 \sqrt{33}-17}\simeq 0.771845 \,- 1.11514 \,i    \label{a1234}
\end{eqnarray}
Then $N_{+}(u)=N_{-}(-u)$ is given by:
{ \begin{eqnarray} \label{Npu}
N_{+}(u)&=&
\frac{g}{4 \, \pi} \frac{a_1^{-1-\frac{1}{2} i g u} B_{a_1}\left(\frac{i g
   u}{2}+1,0\right)}{  (a_1-a_2) (a_1-a_3)
   (a_1-a_4)}- \frac{g}{4 \, \pi} \frac{a_2^{-1-\frac{1}{2} i g u} B_{a_2}\left(\frac{i g
   u}{2}+1,0\right)}{  (a_1-a_2) (a_2-a_3)
   (a_2-a_4)} \nonumber \\
   &+&\frac{g}{4 \, \pi} \frac{a_3^{-1-\frac{1}{2} i g u} B_{a_3}\left(\frac{i g
   u}{2}+1,0\right)}{  (a_1-a_3) (a_2-a_3)
   (a_3-a_4)}-\frac{g}{4 \, \pi} \frac{a_4^{-1-\frac{1}{2} i g u} B_{a_4}\left(\frac{i g
   u}{2}+1,0\right)}{  (a_4-a_1) (a_4-a_2) (a_3-a_4)}.
\end{eqnarray} }


\begin{thebibliography}{999}

\bibitem{adscft} J.~M.~Maldacena,
``The large N limit of superconformal field theories and
supergravity,'' Adv.\ Theor.\ Math.\ Phys.\  {\bf 2} (1998) 231
[Int.\ J.\ Theor.\ Phys.\  {\bf 38} (1999) 1113], [hep-th/9711200];\\
S.~S.~Gubser, I.~R.~Klebanov and A.~M.~Polyakov, ``Gauge theory
correlators from non-critical string theory,''
Phys.\ Lett.\ B {\bf 428} (1998) 105, [hep-th/9802109];\\
E.~Witten, ``Anti-de Sitter space and holography,'' Adv.\ Theor.\
Math.\ Phys.\  {\bf 2} (1998) 253, [hep-th/9802150].
\bibitem{AFrev}
  G.~Arutyunov and S.~Frolov,
  ``Foundations of the $AdS_5 \times S^5$ Superstring. Part I,''
  J.\ Phys.\ A  {\bf 42} (2009) 254003
  [arXiv:0901.4937 [hep-th]].
\bibitem{AJK}
  J.~Ambjorn, R.~A.~Janik and C.~Kristjansen,
  ``Wrapping interactions and a new source of corrections to the spin-chain / string duality,''
 {\slshape   Nucl.\ Phys.\  B }{\bf 736} (2006) 288
  [arXiv:hep-th/0510171].
\bibitem{AF07}
  G.~Arutyunov and S.~Frolov,
  ``On String S-matrix, Bound States and TBA,''
  JHEP {\bf 0712} (2007) 024, hep-th/0710.1568.
\bibitem{AF09a}
  G.~Arutyunov and S.~Frolov,
  ``String hypothesis for the $AdS_5 \times S^5$ mirror,''
  JHEP {\bf 0903} (2009) 152
  [arXiv:0901.1417 [hep-th]].
\bibitem{Bombardelli:2009ns}
  D.~Bombardelli, D.~Fioravanti and R.~Tateo,
  ``Thermodynamic Bethe Ansatz for planar AdS/CFT: a proposal,''
  J.\ Phys.\ A  {\bf 42} (2009) 375401
  [arXiv:0902.3930].
\bibitem{GKKV09}
  N.~Gromov, V.~Kazakov, A.~Kozak and P.~Vieira,
  ``Exact Spectrum of Anomalous Dimensions of Planar N = 4 Supersymmetric
   Yang-Mills Theory: TBA and excited states,''
  Lett.Math.Phys.91:265-287,2010, [arXiv:0902.4458v3 [hep-th]].
\bibitem{AF09b}
  G.~Arutyunov and S.~Frolov,
  ``Thermodynamic Bethe Ansatz for the $AdS_5 \times S^5$ Mirror Model,''
  JHEP {\bf 0905} (2009) 068
  [arXiv:0903.0141 [hep-th]].
\bibitem{AF09d}
  G.~Arutyunov and S.~Frolov,
  ``Simplified TBA equations of the $AdS_5 \times S^5$ mirror model,''
  JHEP 0911:019,2009, [arXiv:0907.2647 [hep-th]].
\bibitem{DT}
P. Dorey, R. Tateo, "Excited states by analytic continuation of TBA
equations," Nucl. Phys. B 482, 639 (1996) [arXiv:hep-th/9607167].
\bibitem{GKV09b}
  N.~Gromov, V.~Kazakov and P.~Vieira,
  ``Exact AdS/CFT spectrum: Konishi dimension at any coupling,''
  Phys.Rev.Lett.104:211601,2010, [arXiv:0906.4240 [hep-th]].
\bibitem{Arutyunov:2009ax}
  G.~Arutyunov, S.~Frolov and R.~Suzuki,
  ``Exploring the mirror TBA,''
  JHEP 1005:031,2010, [arXiv:0911.2224 [hep-th]].
\bibitem{BH-BJ}
J. Balog, \'A. Heged\H{u}s, "The Bajnok-Janik formula and wrapping
corrections", JHEP 1009:107,2010,  [arXiv:1003.4303 [hep-th]].

\bibitem{AFT11}
 G. Arutyunov, S. Frolov, S. J. van Tongeren,
 "Bound States in the Mirror TBA", [arXiv:1111.0564[hep-th]].

\bibitem{GKV09}
  N.~Gromov, V.~Kazakov and P.~Vieira,
  ``Integrability for the Full Spectrum of Planar AdS/CFT,''
  Phys.Rev.Lett.103:131601,2009, [arXiv:0901.3753 [hep-th]].

\bibitem{Z27}
 O. Lunin and J. M. Maldacena, "Deforming field theories with
U(1) $\otimes$ U(1) global symmetry and their gravity duals," JHEP
05 (2005) 033, arXiv:hep-th/0502086.

\bibitem{Z28}
 S. Frolov, "Lax pair for strings in
Lunin-Maldacena background," JHEP 05 (2005) 069,
arXiv:hep-th/0503201.

\bibitem{Z29}
 N. Beisert and R. Roiban, "Beauty and the twist: The Bethe
ansatz for twisted N = 4 SYM," JHEP 08 (2005) 039,
arXiv:hep-th/0505187.

\bibitem{Z30}
 S. A. Frolov, R. Roiban, and A. A.
Tseytlin, "Gauge-string duality for (non)supersymmetric deformations
of N = 4 super Yang-Mills theory," Nucl. Phys. B731 (2005) 1–44,
arXiv:hep-th/0507021.

\bibitem{FrolovTsT}
L. F. Alday, G. Arutyunov and S. Frolov, "Green-Schwarz strings in
TsT-transformed backgrounds," JHEP 0606 (2006) 018 [hep-th/0512253].

\bibitem{Z32}
 S. Ananth, S. Kovacs, and H. Shimada, "Proof
of ultra-violet finiteness for a planar nonsupersymmetric Yang-Mills
theory," Nucl. Phys. B783 (2007) 227–237, arXiv:hep-th/0702020.

\bibitem{Z37}
 N. Reshetikhin, "Multiparameter quantum groups and
twisted quasitriangular Hopf algebras," Lett.Math.Phys. 20 (1990)
331–335.

\bibitem{Z34} C. Ahn, Z. Bajnok, D. Bombardelli, and
R. I. Nepomechie, "Finite-size effect for four-loop Konishi of the
$\beta$-deformed N=4 SYM," Phys.Lett. B693 (2010) 380–385,
arXiv:1006.2209 [hep-th].

\bibitem{Z36}
 N. Gromov and F. Levkovich-Maslyuk, "Y-system and beta-deformed N=4
Super-Yang-Mills," J.Phys.A A44 (2011) 015402, [arXiv:1006.5438
[hep-th]].

 \bibitem{Z33}
  G. Arutyunov, M. de Leeuw, and
S. J. van Tongeren, "Twisting the Mirror TBA," JHEP 1102 (2011) 025,
arXiv:1009.4118 [hep-th].

 \bibitem{Z35}
  C. Ahn, Z. Bajnok, D. Bombardelli,
and R. I. Nepomechie, "Twisted Bethe equations from a twisted
S-matrix," JHEP 1102 (2011) 027, [arXiv:1010.3229 [hep-th]].

\bibitem{TBA2ndLusch}
C.~Ahn, Z.~Bajnok, D.~Bombardelli and R.~I.~Nepomechie,
``TBA, NLO Luscher correction, and double wrapping in twisted AdS/CFT,''
  JHEP {\bf 1112} (2011) 059
  [arXiv:1108.4914 [hep-th]].
 
\bibitem{orbi}
M.~de Leeuw and S.~J.~van Tongeren,
``The spectral problem for strings on twisted $AdS_5 \times S^5$,''
arXiv:1201.1451 [hep-th].

\bibitem{Sieg}
  F.~Fiamberti, A.~Santambrogio, C.~Sieg and D.~Zanon,
  ``Wrapping at four loops in N=4 SYM,''
  Phys.\ Lett.\  B {\bf 666} (2008) 100
  [arXiv:0712.3522 [hep-th]].
\bibitem{Vel}
V.~N.~Velizhanin,
``The four-loop anomalous dimension of the Konishi operator 
in N=4 supersymmetric Yang-Mills theory,''
JETP Lett.\  {\bf 89} (2009) 6 [arXiv:0808.3832 [hep-th]].
 
\bibitem{KL02}
A.~V.~Kotikov and L.~N.~Lipatov, ``DGLAP and BFKL Evolution
Equations in the ${\mathcal{N}}\!=4$ Supersymmetric Gauge Theory,''
Nucl.\ Phys.\  B {\bf 661} (2003) 19 [Erratum-ibid.\  B {\bf 685}
(2004) 405] [arXiv:hep-ph/0208220].
\bibitem{40_5}
  A.~V.~Kotikov, L.~N.~Lipatov, A.~Rej, M.~Staudacher and V.~N.~Velizhanin,
  ``Dressing and Wrapping,''
  J.\ Stat.\ Mech.\  {\bf 0710} (2007) P10003
  [arXiv:0704.3586 [hep-th]].


\bibitem{AFS}
G.~Arutyunov, S.~Frolov and R.~Suzuki, ``Five-loop Konishi from the
Mirror TBA'', JHEP 1004:069,2010,  [arXiv:1002.1711 [hep-th]].
\bibitem{BHxxx}
J. Balog, \'A. Heged\H{u}s, "5-loop Konishi from linearized TBA and
the XXX magnet", JHEP 1006:080,2010, [arXiv:1002.4142 [hep-th]].
\bibitem{JL07}
  R.~A.~Janik and T.~Lukowski,
  ``Wrapping interactions at strong coupling -- the giant magnon,''
 {\slshape   Phys.\ Rev.\  D }{\bf 76} (2007) 126008
  [arXiv:0708.2208 [hep-th]].
\bibitem{BJ08}
  Z.~Bajnok and R.~A.~Janik,
  ``Four-loop perturbative Konishi from strings and finite size effects for multiparticle states,''
 {\slshape   Nucl.\ Phys.\  B }{\bf 807} (2009) 625
  [arXiv:0807.0399 [hep-th]].
\bibitem{Bajnok:2008qj}
  Z.~Bajnok, R.~A.~Janik and T.~Lukowski,
  ``Four loop twist two, BFKL, wrapping and strings,''
  Nucl.\ Phys.\  B {\bf 816} (2009) 376
  [arXiv:0811.4448 [hep-th]].
\bibitem{BJ09}
  Z.~Bajnok, A.~Hegedus, R.~A.~Janik and T.~Lukowski,
  ``Five loop Konishi from AdS/CFT,''
  Nucl. Phys. B {\bf 827} (2010), 426-456,
  arXiv:0906.4062 [hep-th].
\bibitem{Lukowski:2009ce}
  T.~Lukowski, A.~Rej and V.~N.~Velizhanin,
  ``Five-Loop Anomalous Dimension of Twist-Two Operators,''
  Nucl.Phys.B {\bf 831}   (2010), 105-132,
  arXiv:0912.1624 [hep-th].
\bibitem{Gromov}
  N.~Gromov,
  ``Y-system and Quasi-Classical Strings,''
  JHEP 1001:112,2010, [arXiv:0910.3608 [hep-th]].
\bibitem{ujKazi}
N. Gromov, V. Kazakov, Z. Tsuboi, "$PSU(2,2|4)$ character of
quasiclasical AdS/CFT" JHEP 1007:097,2010, [arXiv:1002.3981
[hep-th]].
\bibitem{Frolnum}
S. Frolov, ``Konishi operator at intermediate coupling",
 J.Phys.A{\bf 44}:065401,2011, [arXiv:1006.5032 [hep-th]],\\
S.~Frolov, ``Scaling dimensions from the mirror TBA,''
arXiv:1201.2317 [hep-th].
\bibitem{StrongKonishi}
N. Gromov, D. Serban, I. Shenderovich and D. Volin, "Quantum folded
string and integrability:
 from finite size effects to Konishi dimension", JHEP 1108:046,2011,
  [arXiv:1102.1040[hep-th]], \\
R. Roiban and A. A. Tseytlin, "Semiclassical string computation of
strong-coupling corrections to dimensions of operators in Konishi
multiplet", Nucl.Phys.B848:251-267,2011 ,
[arXiv:1102.1209 [hep-th]],  \\
B.~C.~Vallilo and L.~Mazzucato,
``The Konishi multiplet at strong coupling,''
JHEP {\bf 1112} (2011) 029
[arXiv:1102.1219 [hep-th]], \\
M.~Beccaria and G.~Macorini,
``Quantum folded string in $S^5$ and the Konishi multiplet 
at strong coupling,''
JHEP {\bf 1110} (2011) 040
[arXiv:1108.3480 [hep-th]].

\bibitem{betapert}
F. Fiamberti, A. Santambrogio, C. Sieg, and D. Zanon, "Finite-size
effects in the superconformal â-deformed N = 4 SYM," JHEP 08 057
(2008) [hep-th/0806.2103], \\
 F. Fiamberti, A. Santambrogio, C. Sieg, and D. Zanon,
 "Single impurity operators at critical wrapping
order in the beta-deformed N = 4 SYM," JHEP 08 034 (2009)
[hep-th/0811.4594], \\
F. Fiamberti, A. Santambrogio and C. Sieg, "Superspace methods for
the computation of wrapping effects in the standard and
beta-deformed N=4 SYM,"  [arXiv:1006.3475 [hep-th]].

\bibitem{KBP}
A. Klumper, M.T. Batchelor and P.A. Pearce, "Central charges of the
6- and 19-vertex models with twisted boundary conditions",
 J. Phys. A24 (1991) 3111.

\bibitem{KP91}
P.A. Pearce, A. Kl\"umper: "Finite-size corrections and scaling
dimensions of solvable lattice models: An analytic method ", Phys.
Rev. Lett. {\bf 66}, 974-977 (1991).

\bibitem{Suz98}
J. Suzuki, "Spinons in magnetic chains of arbitrary spins at finite
temperature."  J.Phys.A32:2341-2359,1999.

\bibitem{exs1}
J. Suzuki,  "Excited states nonlinear integral equations for an
integrable anisotropic spin 1 chain,"  J.Phys. A37 (2004)
11957-11970 , [arXiv:hep-th/0410243].

\bibitem{HubbardNLIE}
Fabian H. L. Essler, Holger Frahm, Frank Gohmann, Andreas Klumper
and Vladimir E. Korepin, "\emph{The One-Dimensional Hubbard Model}"
Cambridge University Press 2005, \emph{Chapter 13}.

\bibitem{RacsNLIEk}
A. Hegedus,
 "Nonlinear integral equations for the finite size
effects of RSOS and vertex-models and related quantum field
theories",  Nucl.Phys.B732:463-486,2005,  [hep-th/0507132].

\bibitem{Klumpersl4}
 J. Damerau, A. Klumper,
"Non-linear integral equations for the thermodynamics of the
sl(4)-symmetric Uimin-Sutherland model",
J.Stat.Mech.0612:P12014,2006,  [cond-mat/0610559].

\bibitem{nlie1}
C. Destri, H.J. de Vega, "New thermodynamic Bethe ansatz equations
without strings",
 Phys. Rev. Lett. 69 (1992) 2313-2317; \\
  "Unified approach to thermodynamic Bethe Ansatz and finite size
corrections for lattice models and field theories",
 Nucl. Phys. B438 (1995) 413-454, [arXiv:hep-th/9407117]; \\
"Nonlinear integral equation and excited states scaling functions in
the sine-Gordon model",
Nucl. Phys. B504 (1997) 621. \\
D. Fioravanti, A. Mariottini, E. Quattrini, F. Ravanini, "Excited
State Destri - De Vega Equation for Sine-Gordon and Restricted
Sine-Gordon Models ",
Phys. Lett. B390 (1997) 243-251, [arXiv:hep-th/9608091]. \\
G. Feverati, F. Ravanini, G. Takacs, "Scaling Functions in the Odd
Charge Sector of Sine-Gordon/Massive Thirring Theory", Phys.Lett.
B444 (1998) 442-450, [arXiv:hep-th/9807160]; "Nonlinear Integral
Equation and Finite Volume Spectrum of Sine-Gordon Theory ", Nucl.
Phys. B540 (1999) 543-586, [arXiv:hep-th/9805117].

\bibitem{nlie21}
 C.~Dunning,
 ``Finite size effects and the supersymmetric sine-Gordon models,''
  J.\ Phys.\ A  {\bf 36} (2003) 5463
  [arXiv:hep-th/0210225].\\
\bibitem{nlie22}
  A.~Hegedus,
 ``Finite size effects in the SS model: Two component nonlinear integral
 equations,''
  Nucl.\ Phys.\  B {\bf 679} (2004) 545
  [arXiv:hep-th/0310051].\\
\bibitem{nlie23}
  A.~Hegedus,
 ``Nonlinear integral equations for finite volume excited state energies of
 the O(3) and O(4) nonlinear sigma-models,''
  J.\ Phys.\ A  {\bf 38} (2005) 5345
  [arXiv:hep-th/0412125].\\
\bibitem{nlie24}
 A.~Hegedus, F.~Ravanini and J.~Suzuki,
 ``Exact finite size spectrum in super sine-Gordon model,''
  Nucl.\ Phys.\  B {\bf 763} (2007) 330
  [arXiv:hep-th/0610012].\\
\bibitem{nlie25}
  J.~Balog and A.~Hegedus,
``The finite size spectrum of the 2-dimensional O(3) nonlinear
sigma-model,''
  Nucl.\ Phys.\  B {\bf 829} (2010) 425
  [arXiv:0907.1759].

\bibitem{GKV08}
N. Gromov, V. Kazakov, P. Vieira, "Finite Volume Spectrum of 2D
Field Theories from Hirota Dynamics", JHEP 0912:060, (2009),
[arXiv:0812.5091 [hep-th]].
\bibitem{KL10}
V. Kazakov, S. Leurent, ''Finite Size Spectrum of SU(N) Principal
Chiral Field from Discrete Hirota Dynamics'', [arXiv:1007.1770
[hep-th]].

\bibitem{KNS11}
A. Kuniba, T. Nakanishi, J. Suzuki,
 "T-systems and Y-systems in integrable systems",
[arXiv:1010.1344 [hep-th]].

\bibitem{GKLV11}
N. Gromov, V. Kazakov, S. Leurent, D. Volin,
    "Solving the AdS/CFT Y-system",
    [arXiv:1110.0562 [hep-th]].

\bibitem{Z31}
K. Zoubos, "Review of AdS/CFT Integrability, Chapter IV.2:
Deformations, Orbifolds and Open Boundaries," arXiv:1012.3998
[hep-th].

\bibitem{Tateo1}
A. Cavaglia, D. Fioravanti, R. Tateo,  "Extended Y-system for the
$AdS_5/CFT_4$ correspondence",
 Nucl.Phys. B {\bf 843}, (2011) 302-343, [arXiv:1005.3016 [hep-th]].

\bibitem{BH10}
J. Balog, A. Hegedus, "Quasi-local formulation of the mirror TBA",
[arXiv:1106.2100 [hep-th]].


\bibitem{BH9}
J. Balog, A. Hegedus, "$AdS_5\times S^5$ mirror TBA equations from
Y-system and discontinuity relations", JHEP 1108:095,2011,
[arXiv:1104.4054[hep-th]].

\bibitem{RyoHybrid}
R. Suzuki, ''Hybrid NLIE for the Mirror $AdS_5 \times S^5$'',
J.Phys.A44:235401, 2011,  [arXiv:1101.5165 [hep-th]].

\bibitem{talk}
N. Gromov (presented by V. Kazakov), talk at ``Conference on Integrability
in Gauge and String Theory 2010'', in Nordita Stockholm, June, 2010.
{\tt 
http://agenda.albanova.se/contributionDisplay.py?contribId=258\&confId=1561}

\bibitem{KSZ07}
 V. Kazakov, A. Sorin, A. Zabrodin,
    "Supersymmetric Bethe Ansatz and Baxter Equations from Discrete Hirota
    Dynamics",  Nucl.Phys.B790:345-413,2008 ,
[arXiv:hep-th/0703147 [hep-th]].
\bibitem{KLWZ96}
I. Krichever, O.Lipan, P.Wiegmann, A. Zabrodin
    "Quantum Integrable Systems and Elliptic Solutions of Classical Discrete Nonlinear
    Equations", Commun.Math.Phys. 188 (1997) 267-304,
[arXiv:hep-th/9604080 [hep-th]].
\bibitem{Beisert06}
N. Beisert,  "The Analytic Bethe Ansatz for a Chain with Centrally
Extended su$(2|2)$ Symmetry" J.Stat.Mech.0701:P01017,2007,
[arXiv:nlin/0610017].
\bibitem{dresscross}
G. Arutyunov, S. Frolov, ``The dressing factor and crossing
equations'', J.\ Phys.\ A  {\bf 42}, 425401 (2009) [arXiv:0904.4575
[hep-th]].
\bibitem{BS}
  N.~Beisert and M.~Staudacher,
  ``Long-range PSU(2,2$|$4) Bethe ansaetze for gauge theory and strings,''
  Nucl.\ Phys.\ B {\bf 727}, 1 (2005)
  [hep-th/0504190].
\bibitem{KPtanh}
A. Kl\"umper, P. A. Pearce: "Analytic calculation of scaling
dimensions: Tricritical hard squares and critical hard hexagons"
 J. Stat. Phys. {\bf 64}, 13-76 (1991).
\bibitem{Wronskian}
N. Gromov, V. Kazakov, S. Leurent, Z. Tsuboi,
    "Wronskian Solution for AdS/CFT Y-system,"  JHEP 1101:155,2011
    [arXiv:1010.2720[hep-th]].

 
\end{thebibliography}
\end{document}